\documentclass[prd,aps,lengthcheck,twocolumn,nofootinbib,notitlepage,floatfix,superscriptaddress]{revtex4-2}
\usepackage[utf8]{inputenc}     

\usepackage{amssymb}
\usepackage[tbtags]{amsmath}
\usepackage{indentfirst}
\usepackage{mathtools}
\usepackage{graphicx}            
\usepackage{dcolumn}             
\usepackage{bm}                  
\usepackage{physics}             
\usepackage{subfigure}           
\usepackage[normalem]{ulem}      
\usepackage{orcidlink}

\usepackage{ragged2e}

\usepackage{csquotes}            

\usepackage[makeroom]{cancel}    

\usepackage{comment}             

\usepackage[dvipsnames]{xcolor}  
\usepackage{multirow}            

\usepackage{booktabs}

\allowdisplaybreaks

\usepackage{slashed}

\usepackage{tensor}

\usepackage{enumitem}

\usepackage{hyperref}          

\begin{document}

\title{Charged-current neutrino opacity within the relativistic Hartree-Fock framework for astrophysical simulations of core-collapse supernovae and binary neutron star mergers}

\author{Kamil Soko\l{}owski\orcidlink{0009-0006-8674-4773}}
    \email{kamil.sokolowski@pwr.edu.pl}
    \affiliation{Institute of Theoretical Physics, Wroc\l{}aw University of Science and Technology, Wybrze\.{z}e Wyspia\'{n}skiego 27, 50-370 Wroc\l{}aw, Poland
    }
    \affiliation{
    Institute of Theoretical Physics, University of Wroc\l{}aw, plac Maxa Borna 9, 50-204 Wroc\l{}aw, Poland
    }

\author{Anil Kumar\orcidlink{0000-0001-5833-7595}}
     \email{anil.1@iitj.ac.in}
     \affiliation{Institute of Theoretical Physics, Wroclaw University of Science and Technology, Wybrze\.{z}e Wyspia\'{n}skiego 27, 50-370 Wroc\l{}aw, Poland
     }

\author{Tobias Fischer\orcidlink{0000-0003-2479-344X}}
    \email{bert-tobias.fischer@pwr.edu.pl}
    \affiliation{Institute of Theoretical Physics, Wroclaw University of Science and Technology, Wybrze\.{z}e Wyspia\'{n}skiego 27, 50-370 Wroc\l{}aw, Poland
    }
    \affiliation{Research Center for Computational Physics and Data Processing, Institute of Physics, Silesian University in Opava, Bezručovo nám. 13, CZ-746-01 Opava, Czech Republic}

\date{\today}

\begin{abstract}
    Neutrinos and their weak interactions play a vital role in the physics of core-collapse supernovae and binary neutron star mergers. 
    Their description within astrophysical simulations, including the weak rates, is of pivotal importance not only for the prediction of accurate neutrino fluxes and spectra, including the associated conditions relevant to nucleosynthesis, neutrinos are also responsible for heating and cooling of the stellar plasma as well as the transport of lepton number and entropy. 
    In the present article, we develop an essential improvement of the description of the underlying nuclear medium, necessary for the calculations of charged-current weak rates, with the inclusion of explicitly momentum-dependent nuclear interactions.
    To this end, we introduce the relativistic Hartree-Fock (RHF) approach and the associated momentum-dependent nucleon self-energies. 
    We discuss the resulting neutrino and antineutrino opacities and find large discrepancies comparing the weak rates at the RHF level with those of commonly used relativistic mean-field (RMF) models. In particular, we observe a substantial shift of previously reported large medium-dependent modifications associated with the RMF approach. 
\end{abstract}

\maketitle

\section{\label{sec:introducion} Introduction}
    Neutrinos are abundantly produced in astrophysical environments where high temperatures are encountered, on the order of several tens of MeV, such as core-collapse supernovae and binary neutron star mergers. They are not only confirmed messengers of the central engine that drives massive star explosions---$\bar\nu_e$ were detected from the last nearby event SN1987A by the operating neutrino detectors~\cite{hirata87,bionta87,IMB_Bratton88,BAKSAN1988PhLB205,LSD1987EL3}---neutrinos also contribute to the associated shock revival through heating and cooling processes via a variety of weak reactions. 
    More details on the role of core-collapse supernova phenomenology can be found in Ref.~\cite{Janka2025ARNPS}.
    Furthermore, neutrinos play a crucial role in the nucleosynthesis of both core-collapse supernovae and binary neutron star mergers~\cite{Fischer2024PrPNP}, in particular for the late-time ejection, during which the proton-to-baryon ratio, which is given by the electron fraction $Y_e$, is determined by the flux and spectral differences between $\nu_e$ and $\bar\nu_e$. 
    Neutrinos are also subject to oscillations due to their small but finite rest masses, a phenomenon that necessitates the kinetic approach, unlike the commonly employed transport methods in astrophysical simulations (cf. Ref.~\cite{Tamborra2025NatRP7} and references therein).
    
    It is, therefore, of paramount interest to provide accurate neutrino flux and spectra predictions from simulations of core-collapse supernovae and binary neutron star mergers. This, in turn, requires the consistent treatment of weak processes, in particular those involving a nuclear component and the nuclear medium, which has been realized in one of the pioneering works providing first-principle expressions for weak rates~\cite{reddy_1998}. 
    Charged-current weak rates, treating nucleons at the mean-field level, were later applied to self-consistent simulations of massive star explosions~\cite{martinez_2012,Robert:2012PhRvC86}. These models were based on general relativistic neutrino radiation hydrodynamics in spherical symmetry, featuring three-flavor Boltzmann neutrino transport~\cite{Fischer10,Huedepohl10}.
    The simulations covered the long-term evolution of the deleptonizing proto-neutron star, i.e., the central compact remnant of a supernova explosion, which deleptonizes and later cools via the emission of neutrinos of all flavors on a timescale of several tens of seconds. 
    These models provide the first reliable predictions of the nucleosynthesis associated with massive star explosions, including the neutrino-driven wind, initially exhibiting only a minor neutron excess, making them incompatible with the rapid neutron-capture process and later evolving toward proton-rich conditions. 
    Later improvements of these models by including the full kinematics of the reaction rates and taking into account the weak magnetism, pseudoscalar terms, and $q^2$-dependent nucleon form factors in the hadronic current confirm the initial findings~\cite{Fischer2020PhRvC101,Guo2020PhRvD102}.
    It is important to note that these findings based on spherically symmetric models are insufficient and need to be explored in a multidimensional setup, e.g., it has been demonstrated that proto-neutron star convection has a dominating impact on deleptonization~\cite{roberts_2012,mirizzi16,Janka2025ARNPS,Burrows2026arXiv260209025R}.
    
    The present paper extends these previous developments by treating the nuclear medium beyond the mean-field approach. Therefore, a nuclear equation of state (EOS) is developed based on the relativistic Hartree-Fock (RHF) approach~\cite{serot_1986}. The essential difference, with relevance to the weak rates, is the presence of momentum-dependent interactions, expressed in terms of the nucleon self-energies. These, in turn, change the neutrino-energy dependence of the charged-current weak rates in comparison to the commonly employed relativistic mean-field (RMF) approach~\cite{reddy_1998}, i.e., Hartree mean-field approximation and, furthermore, alter the density dependence of these rates.  
    
    The paper is organized as follows.
    In Sec.~\ref{sec:cc}, we will revisit the weak interaction physics for the charged-current weak rates, followed in Sec.~\ref{sec:HF_approx} with the novel RHF expressions. 
    In Sec.~\ref{sec:RHF_vs_RMF_results}, we provide a detailed evaluation of the neutrino absorption rates at selected conditions encountered in core-collapse supernovae and binary neutron star mergers. 
    In Sec.~\ref{sec:RHF_vs_reduced_RHF}, we perform a comparison between neutrino rates within the RHF approach and the RMF model, which was obtained via the proper reduction of the RHF model. 
    Special emphasis is given to the comparison with the mean-field treatment, for which the essential expressions are provided in Appendix~\ref{app:H_approx}.
    In Sec.~\ref{sec:RMF_delta}, we consider the impact of isospin splitting in the RMF framework with the inclusion of $\delta$ mesons, i.e., different  neutron and proton scalar self-energies, and compare the neutrino opacities with the corresponding RMF models without $\delta$ mesons.
    The manuscript closes with a summary and concluding remarks in Sec.~\ref{sec:summary}. 
    In this paper, we use the following notations: $k_{\rm B}=\hbar=c=1$ and $g^{\mu\nu}=\rm{diag}(+1,-1,-1,-1)$.

\section{Charged-current neutrino opacity}
\label{sec:cc}

\begin{figure}[t!]
        \includegraphics[width=0.45\textwidth]{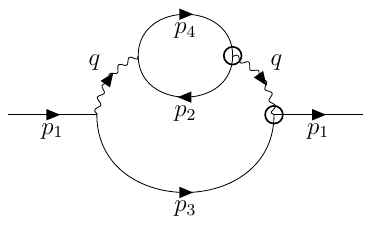}
        \caption{
        \label{fig:neutrino_self_energy}
        Diagrammatic form of the neutrino self-energy, $\Sigma_1^>(E_1,\boldsymbol{p}_1)$, for process~\eqref{eq:cc_process}, which is used for the cutting rules at finite temperature. Solid lines are the fermions and the wavy lines are the exchanged vector bosons $W^+$ (for neutrinos) and $W^-$ (for antineutrinos), with {\em standard} uncircled and circled vertices. We note here that we have the inserted boson self-energy presented by the one-loop with particles 2 and 4 and that cutting the diagram in half results in the usual Feynman diagram for process~\eqref{eq:cc_process} at the tree level for the resulting left-hand side.}
    \end{figure}

    In the following, we focus on the class of charged-current weak processes:
    \begin{equation}
        \nu_1(p_1) + N_2(p_2) \, \longrightarrow \; l_3(p_3) + N_4(p_4) \ ,
    \label{eq:cc_process}
    \end{equation}
    determined by the exchange of $W^\pm$ bosons (wavy lines in Fig.~\ref{fig:neutrino_self_energy}), for neutrinos ($+$) and antineutrinos ($-$), and where $l_3\equiv l$ denotes the leptons ($e$, $\mu$, $\tau$), particle $1$ is the corresponding lepton neutrino, $\nu_1\equiv\nu_l$, and $N_i \ (i=2,4)$ correspond to the initial and final-state baryons. 
    
    In order to calculate the neutrino absorption rate for process~\eqref{eq:cc_process} at finite temperature and density, we apply the real-time formalism and the Kobes-Semenoff rules \cite{bellac_2000,kobes_1986}, which state that at the lowest order in the weak coupling constant and to all orders in the strong coupling constant \cite{bellac_2000}, the interaction (absorption) rate can be obtained from
    \begin{equation}
        \Gamma(E_1)=\frac{1}{2E_1}{\rm Tr}\left\{\slashed p_1\Sigma_1^>(E_1,\boldsymbol{p}_1)\right\} \ , 
        \label{Gamma_general}
    \end{equation}
    where $m_1\approx0$, the slash denotes the Feynman notation, $\slashed{p}=\gamma^\mu p_\mu$, and $\boldsymbol{p}$ denotes the 3-momentum. 
    The neutrino self-energy, $\Sigma_1^>$, presented by Fig.~\ref{fig:neutrino_self_energy}, is obtained by applying the cutting rules at finite temperature following Ref.~\cite{kobes_1986},  
    \begin{align}
        \Sigma_1^>(E_1,\boldsymbol{p}_1)=&\frac{G_F^2C^2}{2}\int\frac{d^4p_2}{(2\pi)^4}\int\frac{d^4p_3}{(2\pi)^4}\,\bar V^{\rm l}_\mu \, G^>_3(p_3)\,V^{\rm l}_\nu \nonumber \\
        &\times \, \left[-\Tr\left\{G_2^<(p_2)\,\bar V_{\rm h}^\mu \,G^>_4(p_4)\,V_{\rm h}^\nu\right\}\right] \ , \label{neutrino_sigma}
    \end{align}
    where $V_\mu$ and $\bar V_\mu=\gamma^0 V_\mu^\dagger\gamma^0$ are the hadron (h) and lepton (l) vertex functions. 
    The Dirac propagators in the real-time formalism are given by
    \begin{align}
        G^>_a(p_a)&=2\pi(\slashed p_a + m_a)\left[1-f_a(p_a^0)\right]\,\Theta\left(p_a^0\right)\,\delta\left(p_a^2-m_a^2\right) \ , \label{G>}\\
        G^<_a(p_a)&=- 2\pi(\slashed p_a + m_a)\,f_a(p_a^0)\,\Theta\left(p_a^0\right)\,\delta\left(p_a^2-m_a^2\right) \ , \label{G<}
    \end{align}
    where \enquote{$>$} and \enquote{$<$} denote flow into and out of the circled vertex, respectively,  $f_a(p_a^0)=\left[1+\exp\{(p_a^0-\mu_a)/T\}\right]^{-1}$ is the Fermi-Dirac distribution function for particle species $a$ with the corresponding chemical potential, $\mu_a$, $\Theta$ denotes the Heaviside step function, and $p_a^\mu=(p_a^0,\boldsymbol{p})$ is the 4-momentum ($a=1,2,3,4$). 
    We note here that when dealing with antiparticles, we have the following replacement: $\slashed p_a+m_a \longrightarrow \slashed p_a-m_a$, in Eqs.~\eqref{G>} and \eqref{G<}. 

    Furthermore, Eq.~\eqref{neutrino_sigma} is obtained for the current-current interaction described by the following Lagrangian~\cite{reddy_1998}:
    \begin{equation}
        \mathcal{L} = \frac{G_F C}{\sqrt{2}}l^\mu j^{cc}_\mu \ ,
    \end{equation} \\
    where $C=\cos\theta_C$ (in agreement with the convention of Ref.~\cite{reddy_1998}, we chose $\sin^2\theta_C=0.053$ for the processes considered in this paper) is the cosine of the Cabibbo up-down quark mixing angle with no exchange of strangeness, for which we have the following leptonic vertex: $V^{\rm l}_\mu=\gamma_\mu \left(1-\gamma_5\right)$, giving rise to the leptonic current, $l^\mu = \bar{\psi}_3 \, V_{\rm l}^\mu \, \psi_1$, and the hadronic vertex, 
    \begin{equation}
        V_{\rm h}^\mu=
        \;
        \gamma^\mu\left(g_\text{V} - \gamma_5\,g_\text{A}\right)+
        F_2\frac{i\sigma^{\mu\alpha}\,q_\alpha}{2M_N}-\frac{G_\text{P}}{M_N}\gamma_5\,q^\mu \ ,
        \label{eq:cc_current}
    \end{equation}
    defining the hadronic current, $j^{cc}_\mu = \bar\psi_4\,V^{\rm h}_\mu\,\psi_2$, where $\psi_a$ and $\bar\psi_a$ are the Dirac spinors for the initial- and final-state fermion and in which we have the following contributions:
    \begin{align}
        {\rm LO} &: \qquad g_\text{V} - \gamma_5 g_\text{A} \ , 
        \label{eq:LO} \\[5pt]
        {\rm WM} &:\qquad F_2\;=\mathcal{Z}_p-\mathcal{Z}_n-1 \ , 
        \label{eq:WM} \\[5pt]
        {\rm PS} &: \qquad  G_\text{P}=\frac{2\,M_N^2\,g_\text{A}}{m_\pi^2-q_\mu^2} \ ,
        \label{eq:PS}
    \end{align}
    with vector and axial-vector coupling constants, $g_\text{V}=1$ and $g_\text{A}=1.27$, respectively, neutron and proton magnetic moments, $\mathcal{Z}_p=2.793$ and $\mathcal{Z}_n=-1.913$, respectively, the pion mass $m_\pi=139.57$~MeV, the vector mass $M_\text{V}=840$~MeV, and the axial-vector mass $M_\text{A}=1$~GeV (cf.~\cite{PDG}). The labels in expressions~\eqref{eq:LO}--\eqref{eq:PS} refer to the hadronic current contributions due to the leading order (LO) and due to the inclusion of weak magnetism (WM) and pseudoscalar terms (PS).
    
    The 4-momentum transfer is denoted as $q^\mu=(q_0,\boldsymbol{q})$ for which we have $q^\mu=p_1^\mu-p_3^\mu=p_4^\mu-p_2^\mu$ and $q_\mu^2=q_0^2-q^2$, with $q\equiv|\boldsymbol{q}|$, and $\sigma^{\mu\nu}=\frac{i}{2}[\gamma^\mu,\gamma^\nu]$. Note further that $M_N=(m_n+m_p)/2$ is the average nucleon mass with bare neutron and proton rest masses, $m_n$ and $m_p$, respectively. Furthermore, weak form factors (FFs) are introduced via the following $q_\mu^2$-dependent terms as follows~\cite{Guo2020PhRvD102}:
    \par
    \begin{subequations}
    \begin{align}
        g_\text{V}&\rightarrow G_\text{V}(q_\mu^2)=\frac{g_\text{V}\left[1-\frac{q_\mu^2(\mathcal{Z}_p-\mathcal{Z}_n)}{4M_N^2}\right]}{\left(1-\frac{q_\mu^2}{4M_N^2}\right)\left(1-\frac{q_\mu^2}{M_\text{V}^2}\right)^2} \ , 
        \\
        g_\text{A}&\rightarrow G_\text{A}(q_\mu^2)=\frac{g_\text{A}}{\left(1-\frac{q_\mu^2}{M_\text{A}^2}\right)^2} \ , 
        \\
        F_2&\rightarrow F_2(q_\mu^2)\,\,=\frac{\mathcal{Z}_p-\mathcal{Z}_n-1}{\left(1-\frac{q_\mu^2}{4M_N^2}\right)\left(1-\frac{q_\mu^2}{M_\text{V}^2}\right)^2} \ ,
        \\
        G_\text{P}&\rightarrow G_\text{P}(q_\mu^2)=\frac{2\,M_N^2\,G_\text{A}(q_\mu^2)}{m_\pi^2-q_\mu^2} \ .
        \label{eq:FF}
    \end{align}
    \label{form_factors}
    \end{subequations}
    
    In the following, we want to provide an expression for the hadron polarization tensor.
    Therefore, we rewrite the bosonic self-energy in the imaginary-time formalism, which is inserted in Eq~\eqref{neutrino_sigma} in the real-time formalism, in the following way~\cite{bellac_2000,roberts_2017}:
    \begin{widetext}
    \begin{align}
        \int\frac{d^4p_2}{(2\pi)^4}{\rm Tr}
        \left\{
        G_2^<(p_2)\,\bar V_{\rm h}^\mu \, G^>_4(p_4)\,V_{\rm h}^\nu
        \right\}
        &=
        2\,{\rm Im}\,T\sum_n\int
        \frac{d^3p_2}{(2\pi)^3}
        \frac
        {{\rm Tr}\left\{G_2(i\omega_n+\mu_2, \boldsymbol{p}_2)\,\bar V_{\rm h}^\mu\, G_4(i\omega_m+i\omega_n+\mu_4, \boldsymbol{p}_2+\boldsymbol{q})\,V_{\rm h}^\nu\right\}}
        {1-\exp\{-(q_0+\Delta\mu)/T\}} \\
        &\equiv
        \frac{2\,{\rm Im}\,\Pi^{\mu\nu}(i\omega_m-\Delta\mu,\boldsymbol{q})}{1-\exp\{-(q_0+\Delta\mu)/T\}} \ , \label{real_vs_imaginary_0}
    \end{align}
    where the sum is over the fermionic Matsubara frequencies, $\omega_n=(2n+1)\pi T$ for $n=\{0,\pm1,...\}$, $\Delta\mu=\mu_2-\mu_4$ is the neutron-proton chemical potential difference, $\omega_m=2m\pi T$ for $m=\{0,\pm1,...\}$ is the bosonic Matsubara frequency, and $\Pi^{\mu\nu}$ denotes the hadron polarization tensor in the imaginary-time formalism,
        \begin{align}
        \Pi_{\mu\nu}(i\omega_m-\Delta\mu,\boldsymbol{q})=&\,T\sum_n\int
        \frac{d^3p_2}{(2\pi)^3}{\rm Tr}\left\{G_2(i\omega_n+\mu_2, \boldsymbol{p}_2)\,\bar V_{\rm h}^\mu\, G_4(i\omega_m+i\omega_n+\mu_4, \boldsymbol{p}_2+\boldsymbol{q})\,V_{\rm h}^\nu\right\} \ , \label{Pi_free}
    \end{align}
    with imaginary-time propagators,
    \begin{equation}
        G_a(i\omega_n+\mu_a,\boldsymbol{p}_a)=\frac{m_a-\slashed p_a}{E_a(\boldsymbol{p}_a)^2-(i\omega_n+\mu_a)^2} \ , \label{G_imaginary}
    \end{equation}
    where here, $\slashed p_a=-(i\omega_n+\mu_a)\gamma^0+\gamma\cdot\boldsymbol{p}$, since in the imaginary-time formalism we switch to the Euclidean spacetime\footnote{$\gamma^4=i\gamma^0$ and $p_a^4=i(i\omega_n+\mu_a)$.}.
    
    The hadron polarization tensor given by Eq.~\eqref{Pi_free} can then be rewritten in terms of the trace with the summation over the Matsubara frequencies by defining the following quantity:
    \begin{align}
        \tilde\Pi_{\mu\nu}(i\omega_m -\Delta\mu,\boldsymbol{q})=&T\sum_n{\rm Tr}\left\{G_2(i\omega_n+\mu_2, \boldsymbol{p}_2)\,\bar V_{\rm h}^\mu\, G_4(i\omega_m+i\omega_n+\mu_4, \boldsymbol{p}_2+\boldsymbol{q})\,V_{\rm h}^\nu\right\} \ ,\label{tilde_Pi_free}
    \end{align}
    such that Eq.~\eqref{real_vs_imaginary_0} becomes
    \begin{align}
        \int\frac{d^4p_2}{(2\pi)^4}{\rm Tr}
        \left\{
        G_2^<(p_2)\,\bar V_{\rm h}^\mu \, G^>_4(p_4)\,V_{\rm h}^\nu
        \right\}
        &=
        2\int\frac{d^3p_2}{(2\pi)^3}\frac{\,{\rm Im}\,\tilde\Pi^{\mu\nu}(i\omega_m-\Delta\mu,\boldsymbol{q})}{1-\exp\{-(q_0+\Delta\mu)/T\}} \ .\label{real_vs_imaginary}
    \end{align}
     
    On the other hand, the left-hand side of Eq.~\eqref{real_vs_imaginary} can be rewritten explicitly as follows:
    \begin{align}
        \int\frac{d^4p_2}{(2\pi)^4}{\rm Tr}
        \left\{
        G_2^<(p_2)\,\bar V_{\rm h}^\mu \, G^>_4(p_4)\,V_{\rm h}^\nu
        \right\}
        =&-\int\frac{d^4p_2}{(2\pi)^2}\Lambda^{\mu\nu}f_2\left(p_2^0\right)\left(1-f_4\left(p_2^0+q_0\right)\right)\,\Theta\left(p_2^0\right)\,\Theta\left(p_2^0+q_0\right) \nonumber \\
        &\times\delta\left(p_2^2-m_2^2\right)\delta\left((p_2+q)^2-m_4^2\right) \nonumber \\
        =&-\int\frac{d^3p_2}{(2\pi)^2}\frac{\Lambda^{\mu\nu}}{4E_2E_4}f_2(E_2)(1-f_4(E_4))\delta\left(q_0+E_2-E_4\right) \ , \label{real_time_Pi_0}
    \end{align}
    with the usage of Eqs.~\eqref{G>} and \eqref{G<}, and with $E_2\equiv E_2(\boldsymbol{p}_2)$ and $E_4\equiv E_4(\boldsymbol{p}_2+\boldsymbol{q})$. 
    In the last line we integrate over $p_2^0$, using the $\Theta$ function to remove the Dirac delta functions, and where the hadron tensor is given as follows:
    \begin{equation}
        \Lambda^{\mu\nu}=\Tr\left\{(\slashed p_2+m_2)\bar V^\mu_{\rm h}(\slashed p_4+m_4)V^\nu_{\rm h}\right\} \ .
        \label{Lambda_theory}
    \end{equation}
    By applying the following identity: $f_2(1-f_4)=(f_2-f_4)/[1-\exp\{-(q_0+\Delta\mu)/T\}]$, we cast Eq.~\eqref{real_time_Pi_0} into the final form,
    \begin{equation}
        \int\frac{d^4p_2}{(2\pi)^4}{\rm Tr}
        \left\{
        G_2^<(p_2)\,\bar V_{\rm h}^\mu \, G^>_4(p_4)\,V_{\rm h}^\nu
        \right\}
        =-\int\frac{d^3p_2}{(2\pi)^2}\frac{\Lambda^{\mu\nu}}{4E_2E_4}
        \frac{f_2(E_2)-f_4(E_4)}{1-\exp\{-(q_0+\Delta\mu)/T\}}
        \delta\left(q_0+E_2-E_4\right) \label{real_time_Pi} \ .
    \end{equation}

    The hadron tensor given in Eq.~\eqref{tilde_Pi_free}, which appears on the right-hand side of Eq.~\eqref{real_vs_imaginary}, needs to be calculated formally. Therefore, we make use of the formalism of Matsubara summation (note that $\exp\{\pm i\omega_m/T\}=1$), which is based on the following prescription \cite{bellac_2000} (cf. Appendix C of Ref.~\cite{roberts_2017}):
    \begin{equation}
        T\sum_n g(p_0)=\sum_i {\rm Res}(g(p_0),z_i)\tilde f(p_0-\mu_2) \ , \label{frequency_sum_identity}
    \end{equation}
    where $\tilde f(x)=[1+\exp\{x/T\}]^{-1}$ and $p_0$ is taken to be equal to $i\omega_n+\mu_2$. 
    We note that $g(p_0)$ is the meromorphic function of $p_0$, which is regular on the vertical line, ${\rm Re}\,p_0=\mu_2$, decreasing faster than $p_0^{-1}$ for $|p_0|\rightarrow\infty$. Note further that ${\rm Res}(g(p_0),z_i)$ corresponds to the residue of $g$ at the point $z_i$, which is a simple and isolated pole at $p_0=z_i$, for which one obtains the following identity: ${\rm Res}(g(p_0),z_i)=\lim_{p_0\to z_i}(p_0-z_i)g(p_0)$. 
    Applying now Eq.~\eqref{G_imaginary} to Eq.~\eqref{tilde_Pi_free}, we obtain the following expression for the hadron polarization tensor
    \footnote{
    \begin{equation*}
        \text{Here we apply the following identity } 1/[(a-b)(a+b)]=1/2a\,[(a-b)^{-1}+(a+b)^{-1}] \ .
    \end{equation*}
    }:
    \begin{align}
        \tilde\Pi_{\mu\nu}(\omega_0,\boldsymbol{q})=&\;T\sum_n\frac{\tilde\Lambda^{\mu\nu}(p_0)}{(E_2-p_0)(E_2+p_0)(E_4-(p_0+\omega_0))(E_4+p_0+\omega_0)} \nonumber \\
        =&\;T\sum_n\frac{\tilde\Lambda^{\mu\nu}(p_0)}{4E_2E_4}\left[\frac{1}{E_2-p_0}+\frac{1}{E_2+p_0}\right]\left[\frac{1}{E_4-(p_0+\omega_0)}+\frac{1}{E_4+p_0+\omega_0}\right] \ , \label{imaginary_sum_poles}
    \end{align}
    and consequently the hadron tensor in the imaginary-time formalism is given by
    \begin{equation}
        \tilde\Lambda^{\mu\nu}(p_0)=\Tr\left\{\left[\gamma^0p_0-\boldsymbol{\gamma}\cdot\boldsymbol{p}_2+m_2\right]\bar V^\mu_{\rm h}\left[\gamma^0(p_0+\omega_0)-\boldsymbol{\gamma}\cdot\boldsymbol{p}_4+m_4\right]V^\nu_{\rm h}\right\} \ ,
    \end{equation}
    where $\omega_0\equiv i\omega_m-\Delta\mu$. 
    From Eq.~\eqref{imaginary_sum_poles}, one can read off that we have four isolated poles at $p_0=E_2$, $p_0=-E_2$, $p_0=E_4-\omega_0$, and $p_0=-E_4-\omega_0$. Using Eq.~\eqref{frequency_sum_identity} and collecting the terms with the same denominator, Eq.~\eqref{imaginary_sum_poles} becomes
    \begin{align}
        \tilde\Pi^{\mu\nu}(\omega_0,\boldsymbol{q})=&\frac{1}{4E_2E_4}\bigg\{\frac{1}{\omega_0+E_2-E_4}\left[\tilde\Lambda^{\mu\nu}(E_2)f_2-\tilde\Lambda^{\mu\nu}(E_4-\omega_0)f_4\right] \nonumber \\
        &+\frac{1}{\omega_0-E_2+E_4}\left[\tilde\Lambda^{\mu\nu}(-E_2)\left(1-\bar f_2\right)-\tilde\Lambda^{\mu\nu}(-E_4-\omega_0)\left(1-\bar f_4\right)\right] \nonumber \\
        &+\frac{1}{\omega_0+E_2+E_4}\left[-\tilde\Lambda^{\mu\nu}(E_2)f_2+\tilde\Lambda^{\mu\nu}(-E_4-\omega_0)\left(1-\bar f_4\right)\right] \nonumber \\
        &+\frac{1}{\omega_0-E_2-E_4}\left[-\tilde\Lambda^{\mu\nu}(-E_2)\left(1-\bar f_2\right)+\tilde\Lambda^{\mu\nu}(E_4-\omega_0)f_4\right]\bigg\} \ , \label{Pi_Matsubara}
    \end{align}
    \end{widetext}
    where $\bar f_a$ denotes the Fermi-Dirac distribution function of the antiparticle with $\bar\mu_a=-\mu_a$.

    With the use of analytic continuation, $\omega_0=i\omega_m-\Delta\mu\rightarrow q_0+i\eta$, we can obtain the real-time polarization tensor, which is required in the calculations of the neutrino absorption rate, and we come naturally back to the Minkowski spacetime. The only term that will contribute in the considered process \eqref{eq:cc_process}, after we take the imaginary part, is the first term in Eq.~\eqref{Pi_Matsubara} with square brackets, as was discussed in Refs.~\cite{roberts_2017,chin_1977} with the kinematic constraint that $q_\mu^2\le(m_2-m_4)^2$. By applying now the identity with the Cauchy principal value\footnote{$(x\pm i\eta)^{-1}=\mathcal{P}x^{-1} \mp i\pi\delta(x)$.}, performing the analytic continuation and taking the imaginary part, we obtain the following final expression for the hadron polarization tensor:
    \begin{align}
        {\rm Im}\;\tilde\Pi^{\mu\nu}(q_0,q)=&-\pi\frac{f_2-f_4}{4E_2E_4}\Lambda^{\mu\nu} 
        \delta(q_0+E_2-E_4) \ , \label{Im_tilde_Pi}
    \end{align}
    where we note that the imaginary part is not taken over $\Lambda^{\mu\nu}$, given by Eq.~\eqref{Lambda_theory}~\cite{roberts_2017}, which is the requirement to connect the real- and imaginary-time expressions of the inserted bosonic self-energy. 
    Implementing the above result into Eq.~\eqref{real_vs_imaginary} together with Eq.~\eqref{real_time_Pi}, we confirm that this relation is true.
    
    By combining Eq.~\eqref{real_vs_imaginary} for bosonic self-energy and Eq.~\eqref{G>} for particle 3 in Eq.~\eqref{Gamma_general} together with Eq.~\eqref{neutrino_sigma}, we obtain the following neutrino absorption rate, i.e., the neutrino opacity being the inverse of the neutrino mean-free path, $\Gamma(E_1)=1/\lambda_1$:
    \begin{align}
        \Gamma(E_1)=&-\frac{G_F^2C^2}{32\pi^3}\int\frac{d^3p_3}{E_1E_3}\int\frac{d^3p_2}{(2\pi)^3}\left[1-f_3(E_3)\right] 
        \nonumber \\
        &\times\frac{L_{\mu\nu}\,{\rm Im}\,\tilde\Pi^{\mu\nu}(q_0,\boldsymbol{q})}{1-\exp\{-(q_0+\Delta\mu)/T\}}  \,
        \label{neutrino_absorption_general}
    \end{align}
    where the lepton tensor,
    \begin{equation}
        L_{\mu\nu} = \Tr\left\{\slashed p_1 \, \bar V^{\rm l}_\mu \, (\slashed p_3 + m_3) \, V^{\rm l}_\nu \right\} \ ,
    \label{lepton_tensor}
    \end{equation}
    does not factor out of the integral over $p_2$ in expression~\eqref{neutrino_absorption_general}, which is the very essence of the RHF approach, differing from commonly employed mean-field models, in which the lepton tensor factorizes~\cite{reddy_1998,roberts_2017}. 
    More details on the decomposition of $L_{\mu\nu}$ will be given below in Sec.~\ref{sec:HF_approx}.

    Finally, Eq.~\eqref{neutrino_absorption_general} can be cast in the following differential form:
    \begin{align}
        \frac{d\,\Gamma(E_1)}{dE_3d\Omega_{13}}=&-\frac{G_F^2C^2}{32\pi^3}\frac{p_3}{E_1}\frac{1-f_3(E_3)}{1-\exp\{-(q_0+\Delta\mu)/T\}} 
        \nonumber \\
        &\quad\quad\times\int\frac{d^3p_2}{(2\pi)^3}L_{\mu\nu}\,{\rm Im}\,\tilde\Pi^{\mu\nu}(q_0,\boldsymbol{q})~,
        \label{neutrino_absorption_general_diff}
    \end{align}
    where we change the integration variable from $p_3$ to $E_3$, where $E_3^2=p_3^2+m_3^2$, and $d\Omega_{13}=d\mu_{13}d\phi_{13}$, where $\mu_{13}=\cos\theta_{13}$ is the cosine of the relative momentum angle between particles $1$ and $3$ 
    (due to the 3-momentum transfer, $\boldsymbol{q}$, $\mu_{13}$ is also the angle between particle $1$ and $\boldsymbol{q}$)
    and $\phi_{13}$ is their azimuthal angle. A detailed discussion of the derivation of the above expressions within the mean-field approximation and how they are related to the free theory is provided in~\cite{roberts_2017}.
    
\section{\label{sec:HF_approx} Neutrino absorption rates within the relativistic Hartree-Fock treatment}
    
    In this section, we lay out the formalism for the relativistic Hartree-Fock approximation~\cite{horowitz_1983,serot_1986} and show how the expression describing the neutrino opacity compared to the RMF approximation differs. Therefore, the nucleons are put on-shell, i.e., $p_a^0=E_a(p_a)$, and the nucleon self-energy is decomposed into scalar $\Sigma_a^S$, timelike $\Sigma_a^0$, and vector $\Sigma_a^V$ components, as follows~\cite{bouyssy_1987}:
    \begin{equation}\label{Sigma_HF}
        \Sigma_a(p_a) = \Sigma_a^S(p_a) + \gamma^0\Sigma_a^0(p_a) + \boldsymbol{\gamma}\cdot\boldsymbol{\hat p}_a\,\Sigma_a^V(p_a) \ ,
    \end{equation}
    where $\boldsymbol{\hat p}_a\equiv\boldsymbol{p}_a/\vert\boldsymbol{p}_a\vert$ is the unit momentum vector and in the following we use $p_a\equiv|\boldsymbol{p}_a|$. We note that momentum dependence is explicitly included, unlike within the RMF approximation, as it appears on account of the specific form of the Fock diagram (see Appendix~\ref{app:Hartree-Fock}). 
    As a consequence, the nucleon masses, $m_a$, momenta, $\boldsymbol{p}_a$, and dispersion relations, $E_a$, are expressed as effective (dressed) quantities as follows:
    \begin{subequations}
    \begin{align}
        m^*_a(p_a) &= m_a+\Sigma_a^S(p_a) \label{m_eff_HF} \ , \\
        \boldsymbol{p}^*_a(p_a) &= \boldsymbol{p}_a+\boldsymbol{\hat p}_a\,\Sigma_a^V(p_a) \label{p_eff_HF} \ , \\
        E^*_a(p_a) &= E_a(p_a)-\Sigma_a^0(p_a)  \label{E_eff_HF} \ ,
    \end{align} \label{eff_HF}%
    \end{subequations}
    where $E^*_a(p_a)=\sqrt{\boldsymbol{p}_a^{*2}(p_a)+m^{*2}_a(p_a)}$. Details about the nonlinear RHF model employed here, based on the exchange of $\sigma$, $\omega$, $\rho$, and $\pi$ mesons, are given in Appendix~\ref{app:Hartree-Fock}, closely following Refs.~\cite{bouyssy_1987,bernardos_1993,savushkin_1997}, including the explicit expressions for the nucleon self-energies. 
    
    Since according to electroweak kinematics, we have $\boldsymbol{p}_4=\boldsymbol{p}_2+\boldsymbol{q}$ and we only integrate over $p_2$ (not $p_4$), $E_4^*(p_4)$ can be expressed as follows:
    \begin{eqnarray}
        && E_4^*(p_2,q) = \Big[|\boldsymbol{p}_2+\boldsymbol{q}|^2 + \Sigma_4^V(|\boldsymbol{p}_2+\boldsymbol{q}|)^2 \nonumber \\
        && \quad + \; 2|\boldsymbol{p}_2+\boldsymbol{q}|\Sigma_4^V(|\boldsymbol{p}_2+ \boldsymbol{q}|) + m_4^{*2}(|\boldsymbol{p}_2+\boldsymbol{q}|)\Big]^{1/2} \ , \qquad
    \end{eqnarray}
    where 
    \begin{align}
        p_4&=|\boldsymbol{p}_4|=|\boldsymbol{p}_2+\boldsymbol{q}| = \sqrt{(\boldsymbol{p}_2+\boldsymbol{q})^2} \nonumber \\
        &=\sqrt{p_2^2+q^2+2p_2q\,\mu_{24}} \ .
    \end{align}
    Consequently, the nucleon self-energies for the particle $4$ depend explicitly on $p_2$, $q$, and $\mu_{24}$, which is the cosine of the momentum angle between the particles $2$ and $4$ 
    (due to the 3-momentum transfer, $\boldsymbol{q}$, $\mu_{24}$ is also the angle between $\boldsymbol{p_2}$ and $\boldsymbol{q}$).
    
    In the following, we introduce the effective 4-momentum transfer, $\tilde q^\mu=(\tilde q_0,\boldsymbol{\tilde q})$, where
    \begin{subequations}
    \begin{align}
        \tilde q_0 =& \; q_0+\Sigma_2^0(p_2)-\Sigma_4^0(p_4) \equiv q_0+\Delta\Sigma^0 \ , \\[5pt]
        \boldsymbol{ \tilde q} =& \; \boldsymbol{q}-\boldsymbol{ \hat p}_2\Sigma_2^V(p_2)+\boldsymbol{ \hat p}_4\Sigma_4^V(p_4) \equiv \boldsymbol{q}-\Delta\boldsymbol{\Sigma}^V
    \end{align}
    \end{subequations}
    with the following 4-vector: $\Delta\Sigma^\mu=\left(\Delta\Sigma^0,-\Delta\boldsymbol{\Sigma}^V\right)$, which gives the relation $\tilde q^\mu=q^\mu+\Delta\Sigma^\mu$. 
    Moreover, $\tilde q^\mu=p_4^{*\mu}-p_2^{*\mu}$, where $p_a^{*\mu}=(E_a^*(p_a),\boldsymbol{p}_a^*(p_a))$.
    For numerical purposes, we provide the explicit form of $\tilde q_\mu^2=\tilde q_0^2-\boldsymbol{ \tilde q}^2$, where
    \begin{subequations}
    \begin{align}
        \tilde q_0^2 =&\left(q_0+\Delta\Sigma^0\right)^2 \ , \label{q0_tilde_HF} \\[5pt]
        \boldsymbol{\tilde q}^2 =& \; \boldsymbol{q}^2+\left(\Delta\boldsymbol{\Sigma}^V\right)^2-2\,\boldsymbol{q}\cdot\Delta\boldsymbol{\Sigma}^V \ , \label{q_tilde_HF}
    \end{align}
    with
    \begin{align}
        \left(\Delta\boldsymbol{\Sigma}^V\right)^2 =& \; \left(\Sigma_2^V\right)^2 + \left(\Sigma_4^V\right)^2 - \frac{2\Sigma_2^V\Sigma_4^V}{p_4}(p_2+q\,\mu_{24}) \ , \\
        \boldsymbol{q}\cdot\Delta\boldsymbol{\Sigma} =& \; q\,\mu_{24}\Sigma_2^V-\frac{p_2\,q\,\mu_{24} + q^2}{p_4}\Sigma_4^V \label{vec_q_delta_Sigma} \ .
    \end{align}
    \end{subequations}

    Following the approach presented in Ref.~\cite{roberts_2017}, how to replace the bare quantities in the free theory with the corresponding ones in the interacting theory in Eq.~\eqref{Im_tilde_Pi}, e.g., within the mean-field approach, we note that the analog also holds within the RHF approximation. 
    Therefore, in order to rewrite the differential absorption rate given by Eq.~\eqref{neutrino_absorption_general_diff} within the RHF approximation, we replace the bare quantities with the effective ones given in \eqref{eff_HF}, make a substitution $q^\mu\rightarrow\tilde q^\mu$, and insert Eq.~\eqref{Im_tilde_Pi}, in order to to obtain
    \begin{align}
        \frac{d\Gamma(E_1)}{dE_3d\Omega_{13}}
        = &
        \;\frac{G_F^2 C^2}{32\pi^2}
        \frac{p_3}{E_1}
        \frac{1-f_3(E_3)}{1-\exp\{-(q_0+\Delta\mu)/T\}} 
        \nonumber \\
        &\times
        \int\frac{d^3p_2}{(2\pi)^3}
        \delta(\tilde q_0+E_2^*(p_2)-E_4^*(p_4))
        \nonumber \\
        &\times
        \left[f_2(E_2)-f_4(E_4)\right]\,
        \frac{L_{\mu\nu}\Lambda^{\mu\nu}}{4E_2^*(p_2)E_4^*(p_4)} \ , 
        \label{eq:diff_absorption_HF}
    \end{align}
    with
    \begin{align}
        \Lambda^{\mu\nu}
        =&
        \Tr\Bigg\{
        \left(\slashed p_2^*(p_2)+m_2^*(p_2)\right)\bigg[\gamma^\mu(G_\text{V}(q_\mu^2) - G_\text{A}(q_\mu^2)\gamma_5) 
        \nonumber \\
        &
        -\;F_2(q_\mu^2)\frac{i\sigma^{\mu\alpha}\tilde q_\alpha}{2{M_N}}+\frac{G_\text{P}(q_\mu^2)}{M_N}\gamma_5\tilde q^\mu\bigg] 
        \nonumber \\
        &\times\;
        \left(\slashed p_4^*(p_4)+m_4^*(p_4)\right)\bigg[\gamma^\nu(G_\text{V}(q_\mu^2) - G_\text{A}(q_\mu^2)\gamma_5) 
        \nonumber \\
        &
        +\;F_2(q_\mu^2)\frac{i\sigma^{\nu\beta}\tilde q_\beta}{2{M_N}}-\frac{G_\text{P}(q_\mu^2)}{M_N}\gamma_5\tilde q^\nu\bigg]\Bigg\} \ , \qquad
        \label{Lambda_HF}
    \end{align}
    where both lepton and hadron tensors appear under the integral over $p_2$ and we note that in the denominator of the term with $F_2(q_\mu^2)$, there should be $M_N$, not $M_p$, as was stated in Ref.~\cite{roberts_2017}. Note further that the single-particle energy, $E_a$, contained in the Fermi-Dirac distributions has to be determined via solutions of Eq.~\eqref{E_eff_HF}, which makes the distributions explicitly depend on the dressed quantities in \eqref{eff_HF}. In comparison with the RMF approach, the essential difference in the RHF treatment arises due to the explicit momentum-dependent nucleon self-energies, which will be shown further below. 
    In the following, we omit the arguments in the effective quantities given in \eqref{eff_HF} for clarity.
    
    In the RHF framework, the contraction of the lepton and hadron tensors can be written as follows~\cite{roberts_2017}:
    \begin{align}
        L_{\mu\nu}\Lambda^{\mu\nu}
        =& 
        \sum_{i\in\mathcal{A}}\mathcal{C}_i\big[L_L \Lambda_L^i + L_Q \Lambda_Q^i - 2L_{M+} \Lambda_{M+}^i 
        \nonumber \\
        &
        -
        2L_{M-} \Lambda_{M-}^i + 2L_{T+} \Lambda_{T+}^i + 2L_{T-} \Lambda_{T-}^i\big] \ ,
        \label{eq:L_Pi_Hartree-Fock}
    \end{align}
    with the following definitions:
    \begin{subequations}
    \begin{align}
        \mathcal{A}=\{&\rm V, A,T,P,VA,VT,AT,VP,AP,TP\} \ , \\
        \intertext{and}
        \mathcal{C}_i=\{&g_\text{V}^2,g_\text{A}^2,F_2^2,G_\text{P}^2,g_\text{V}g_\text{A},g_\text{V}F_2,g_\text{A}F_2,
        \nonumber \\[5pt]
        & g_\text{V}G_\text{P},g_\text{A}G_\text{P},F_2G_\text{P}\} \ ,
    \end{align}\label{eq:AC}%
    \end{subequations}
    where $\mathcal{A}$ contains all vector (V), axial-vector (A), tensor (T), and pseudoscalar (P) components, as well as their combinations with the corresponding coupling constants in $\mathcal{C}_i$, which can be replaced by the form factors, as given by \eqref{form_factors}.
    
    In the following, we choose the 4-momentum transfer, $\tilde q^\mu=(\tilde q_0,0,0,\tilde q)$, aligned with the z-axis and $\tilde n^\mu=(\tilde q,0,0,\tilde q_0)$ orthogonal to $\tilde q^\mu$, where $\tilde q=\sqrt{\boldsymbol{\tilde q}^2}$, with $\boldsymbol{\tilde q}$ given by Eq.~\eqref{q_tilde_HF}. 
    We note that $q^\mu=(q_0,\boldsymbol{q})$, and that $\boldsymbol{\tilde q}$ and $\boldsymbol{q}$ are in general not parallel.
    Using these vectors, we then build the following set of projectors \cite{roberts_2017}. 
    First, the {\em transverse projector} (orthogonal to both $\tilde q^\mu$ and $\tilde n^\mu$) is given as follows:
    \begin{align}
        P_{\mu\nu}^{T+}
        =&\;
        g_{\mu\nu}-\frac{\tilde q_\mu\tilde q_\nu}{\tilde q_\mu^2}-\frac{\tilde n_\mu\tilde n_\nu}{\tilde n_\mu^2} 
        \nonumber \\
        =&\;g_{\mu\nu}+\frac{\tilde n_\mu\tilde n_\nu-\tilde q_\mu\tilde q_\nu}{\tilde q_\mu^2} \ ,
    \end{align}
    where $\tilde n_\mu^2=-\tilde q_\mu^2$, and the {\em longitudinal projectors} along $\tilde q^\mu$ and $\tilde n^\mu$, respectively, are 
    \begin{subequations}
    \begin{align}
        P_{\mu\nu}^{L}&=\frac{\tilde n_\mu\tilde n_\nu}{\tilde n_\mu^2}=-\frac{\tilde n_\mu\tilde n_\nu}{\tilde q_\mu^2}  \ , \\[5pt]
        P_{\mu\nu}^{Q}&=\frac{\tilde q_\mu\tilde q_\nu}{\tilde q_\mu^2}  \ .
    \end{align}
    \end{subequations}
    We further define the following projectors:
    \begin{subequations}
    \begin{align}
        P_{\mu\nu}^{M\pm}
        &=
        \frac{\tilde q_\mu\tilde n_\nu\pm\tilde n_\mu\tilde q_\nu}{\tilde q_\mu^2} \ , \\[5pt]
        P_{\mu\nu}^{T-}
        &=
        \epsilon_{\mu\nu\rho\sigma}\frac{\tilde q^\rho\tilde n^\sigma}{\tilde q_\mu^2} \ .
    \end{align}
    \end{subequations}
    The only nonzero contractions of these projectors are
    \begin{subequations}
    \begin{align}
        P_{\mu\nu}^{L}P^{\mu\nu}_{L}&=1 \ , \\[5pt]
        P_{\mu\nu}^{Q}P^{\mu\nu}_{Q}&=1 \ , \\[5pt]
        P_{\mu\nu}^{M\pm}P^{\mu\nu}_{M\pm}&=-2 \ , \\[5pt]
        P_{\mu\nu}^{T\pm}P^{\mu\nu}_{T\pm}&=2 \ .
    \end{align}
    \end{subequations}
    From these considerations it follows that any rank-2 tensor, $T_{\mu\nu}$, can be decomposed as follows:
    \begin{align}
        T_{\mu\nu}
        =&\;
        T_QP_{\mu\nu}^{Q} + T_LP_{\mu\nu}^{L} + T_{M+}P_{\mu\nu}^{M+} + T_{M-}P_{\mu\nu}^{M-} \nonumber \\[5pt]
        &
        +\;T_{T+}P_{\mu\nu}^{T+} + T_{T-}P_{\mu\nu}^{T-} \ ,
    \end{align}
    where the components are obtained by using the following relations,
    \begin{subequations}
    \begin{align}
        T_Q&=P^{\mu\nu}_{Q}T_{\mu\nu} \ , \\[5pt]
        T_L&=P^{\mu\nu}_{L}T_{\mu\nu} \ , \\[5pt]
        T_{M\pm}&=-\frac{1}{2}P^{\mu\nu}_{{M\pm}}T_{\mu\nu} \ , \\[5pt]
        T_{T\pm}&=\frac{1}{2}P^{\mu\nu}_{{T\pm}}T_{\mu\nu} \ .
    \end{align}
    \end{subequations}

    Let us first consider the lepton tensor given in Eq.~\eqref{lepton_tensor} for the two processes \eqref{eq:cc_process}, $\nu_e + n \rightarrow e^- + p$ and $\bar\nu_e + p \rightarrow e^+ + n$, which takes the following form:
    \begin{align}
        L_{\mu\nu} =& \Tr\left\{\slashed p_1\gamma_\mu(1\mp\gamma_5)(\slashed p_3\pm m_3)\gamma_\nu(1\mp\gamma_5)\right\} \\[5pt]
        =&\;8\big[2p_{1\mu} p_{1\nu}+p_1^\alpha q_\alpha g_{\mu\nu}-(p_{1\mu} q_\nu+p_{1\nu} q_\mu) \nonumber \\[5pt]
        &\pm i\epsilon_{\mu\nu\alpha\beta}p_1^\alpha q^\beta\big] \ , \label{lepton_tensor_explicit}
    \end{align}
    where the upper sign is for neutrinos and the lower sign is for antineutrinos. 
    All decomposed components will depend on the following 4-momentum contractions:
    \begin{subequations}
    \begin{align}
        p_1\cdot q=&\;\frac{q_\mu^2-m_3^2}{2} \ ,
        \label{p1_q_HF} \\[5pt]
        p_1\cdot\tilde q=&\;p_1\cdot q+p_1\cdot\Delta\Sigma \ ,
        \label{p1_q_tilde_HF} \\[5pt]
        p_1\cdot\tilde n=&\;E_1\tilde q-p_1\tilde q_0\tilde\mu_1 \ , 
        \label{p1_n_tilde_HF} \\[5pt]
        \tilde n\cdot q=&\;\tilde q q_0 - \tilde q_0 q\tilde\mu \ ,
        \label{n_tilde_q_HF} \\[5pt]
        q\cdot\tilde q=&\;q_0\tilde q_0-q\tilde q\tilde\mu  \ ,
    \end{align}
    \label{lepton_contractions_HF}%
    \end{subequations}
    along with $\tilde q\cdot\tilde n=0$, where
    \begin{subequations}
    \begin{align}
        p_1\cdot\Delta\Sigma=&\;E_1\Delta\Sigma^0+\frac{\boldsymbol{p}_1\cdot\boldsymbol{p}_2}{p_2}\Sigma_2^V 
        \nonumber \\
        & \qquad\qquad- \frac{\boldsymbol{p}_1\cdot\boldsymbol{p}_2+\boldsymbol{p}_1\cdot\boldsymbol{q}}{p_4}\Sigma_4^V \ , \\[5pt]
        \boldsymbol{p}_1\cdot\boldsymbol{q}=&\;E_1q_0-p_1\cdot q \ , \\[5pt]
        {\boldsymbol{p}_1\cdot\boldsymbol{p}_2}=&\;p_1^1p_2^1+p_1^2p_2^2+p_1^3p_2^3 \ , \\[5pt]
        \tilde\mu_1=&\;-\frac{1}{p_1\tilde q}\left[p_1\cdot\tilde q-E_1\tilde q_0\right] \ , \\[5pt]
        \tilde\mu=&\;\frac{1}{\tilde q q}\left[q^2-\boldsymbol{q}\cdot\Delta\boldsymbol\Sigma\right] \ , 
    \end{align}
    \end{subequations}
    where $\tilde\mu_1$ is the cosine of the angle between particle 1 and effective 3-momentum $\boldsymbol{\tilde q}$, and $\tilde\mu$ is the angle between $\boldsymbol{q}$ and $\boldsymbol{\tilde q}$. 
    We note further that $\boldsymbol{q}\cdot\Delta\boldsymbol\Sigma$ is given by Eq.~\eqref{vec_q_delta_Sigma} 
    and the momentum components of $\boldsymbol{p_1}$ and $\boldsymbol{p_2}$ are easily found in polar coordinates\footnote{ $p_a^1=p_a\sin\theta_a\cos\phi_a$, $p_a^2=p_a\sin\theta_a\sin\phi_a$, $p_a^3=p_a\cos\theta_a$, where we define $\theta_1\equiv\theta_{13}$, $\theta_2\equiv\theta_{24}$, $\phi_1\equiv\phi_{13}$, and $\phi_2\equiv\phi_{24}$.}, which are given relative to $\boldsymbol{q}$.
    Then, the components of the lepton tensor are
    \begin{subequations}
    \begin{align}
        L_L=&\;\frac{8}{\tilde q_\mu^2}\left[-2(p_1\cdot\tilde n)^2+2(p_1\cdot\tilde n)(q\cdot\tilde n)+q_\mu^2(p_1\cdot q)\right] \ ,
        \label{L_L} \\[5pt]
        L_Q=&\;\frac{8}{\tilde q_\mu^2}\left[2(p_1\cdot\tilde q)^2-2(p_1\cdot\tilde q)(q\cdot\tilde q)+q_\mu^2(p_1\cdot q)\right]  \ ,
        \label{L_Q} \\[5pt]
        L_{M+}=&\;\frac{8}{\tilde q_\mu^2}\left[(p_1\cdot\tilde q)(q\cdot\tilde n)+(p_1\cdot\tilde n)((q\cdot\tilde q)-2(p_1\cdot\tilde q))\right] \ ,
        \label{L_M+} \\[5pt]
        L_{M-}=&\;\mp\frac{8i}{\tilde q_\mu^2}e^{\mu\nu\alpha\beta}\tilde n_\mu p_{1\nu}\tilde q_\alpha q_\beta=0 \ , 
        \label{L_M-} \\[5pt]
        L_{T+}=&\;\frac{8}{\tilde q_\mu^2}\big[(p_1\cdot\tilde n)^2-(p_1\cdot\tilde n)(q\cdot\tilde n)+(p_1\cdot\tilde q)((q\cdot\tilde q) \nonumber  \\
        &\;-\;(p_1 \cdot\tilde q))\big] \ , 
        \label{L_T+} \\[5pt]
        L_{T-}=&\;\pm\frac{8i}{\tilde q_\mu^2}\left[(p_1\cdot\tilde n)(q\cdot\tilde q)-(q\cdot\tilde n)(p_1\cdot\tilde q)\right] \ ,
    \label{L_T-}
    \end{align}
    \label{eq:L_components}%
    \end{subequations}
    where the upper sign is for neutrinos and the lower sign is for antineutrinos.
    
    Moving on to the hadron tensor, we first list the following 4-momentum contractions used in the calculations:
    \begin{subequations}
    \begin{align}
        p_2^*\cdot\tilde q=&\;E_2^*\tilde q_0-p_2^*\,\tilde q\,\tilde\mu_2 \ , \\[5pt]
        p_2^*\cdot\tilde n=&\;E_2^*\tilde q-p_2^*\,\tilde q_0 \,\tilde\mu_2 \ ,
    \end{align}
    \label{hadron_contractions_HF}%
    \end{subequations}
    where
    \begin{equation}
        \tilde\mu_2=-\frac{1}{p_2\tilde q}\left[p_2\cdot\tilde q - E_2\tilde q_0\right]
    \end{equation}
    is the cosine of the angle between particle 2 and $\boldsymbol{\tilde q}$ with
    \begin{subequations}
    \begin{align}
        p_2\cdot\tilde q=&\;p_2\cdot q + p_2\cdot\Delta\Sigma \ , \\[5pt]
        p_2\cdot q =&\;E_2q_0 - p_2q\mu_{24} \ , \\[5pt]
        p_2\cdot\Delta\Sigma=&\;E_2\Delta\Sigma^0 + p_2\Sigma_2^V - \frac{p_2^2 + p_2q\mu_{24}}{p_4}\Sigma_4^V \ ,
    \end{align}
    \end{subequations}
    where $E_2$ is found via Eq.~\eqref{E_eff_HF} and we note that $p_2^2=\boldsymbol{p}_2^2$.
    Then, the nonzero components of the hadron tensor given in Eq.~\eqref{Lambda_HF} are as follows:
    \begin{enumerate}
        \item[(1)] Vector:
            \begin{subequations}
            \begin{align}
                \Lambda_Q^\text{V}=&\;\frac{4}{\tilde q_\mu^2}\Big[-\left(m_2^{*2}-m_2^*m_4^*\right)\tilde q_\mu^2+2(p_2^*\cdot\tilde q)^2 
                \nonumber \\
                &+(p_2^*\cdot\tilde q)\tilde q_\mu^2\Big] \ , 
                \label{Lambda_V_Q} \\[5pt]
                \Lambda_L^\text{V}=&-\frac{4}{\tilde q_\mu^2}\Big[\left(m_2^{*2}-m_2^*m_4^*\right)\tilde q_\mu^2+(p_2^*\cdot\tilde q)\tilde q_\mu^2 
                \nonumber \\
                &+2(p_2^*\cdot\tilde n)^2\Big] \ , \\[5pt]
                \Lambda_{T+}^\text{V}=&\;\frac{1}{2}\Big\{8\big[-m_2^{*2}+2m_2^*m_4^*-(p_2^*\cdot\tilde q)\big]-\Lambda_L^\text{V}-\Lambda_Q^\text{V}\Big\} \ , \\[5pt]
                \Lambda_{M+}^\text{V}=&-\frac{4}{\tilde q_\mu^2}\big[2(p_2^*\cdot\tilde q)(p_2^*\cdot\tilde n)+(p_2^*\cdot\tilde n)\tilde q_\mu^2\big] \ .
            \end{align}
            \label{Lambda_V}%
            \end{subequations}
        \item[(2)] Axial-vector:
            \begin{subequations}
            \begin{align}
                \Lambda_Q^\text{A}=&\;\frac{4}{\tilde q_\mu^2}\Big[-\left(m_2^{*2}+m_2^*m_4^*\right)\tilde q_\mu^2+2(p_2^*\cdot\tilde q)^2
                \nonumber \\
                &+(p_2^*\cdot\tilde q)\tilde q_\mu^2\Big] \ , \\[5pt]
                \Lambda_L^\text{A}=&-\frac{4}{\tilde q_\mu^2}\Big[\left(m_2^{*2}+m_2^*m_4^*\right)\tilde q_\mu^2+(p_2^*\cdot\tilde q)\tilde q_\mu^2 
                \nonumber \\
                &+2(p_2^*\cdot\tilde n)^2\Big] \ , \\[5pt]
                \Lambda_{T+}^\text{A}=&\;\frac{1}{2}\Big\{8\left[-m_2^{*2}-2m_2^*m_4^*-(p_2^*\cdot\tilde q)\right]-\Lambda_L^\text{A}-\Lambda_Q^\text{A}\Big\} \ , \\[5pt]
                \Lambda_{M+}^\text{A}=&-\frac{4}{\tilde q_\mu^2}\left[2(p_2^*\cdot\tilde q)(p_2^*\cdot\tilde n)+(p_2^*\cdot\tilde n)\tilde q_\mu^2\right] \ .
            \end{align}
            \label{Lambda_A}%
            \end{subequations}
        \item[(3)] Tensor:
            \begin{subequations}
                \begin{align}
                \Lambda_L^\text{T}=&\;\frac{1}{M_N^2}\Big\{\left(m_2^{*2}+m_2^*m_4^*\right)\tilde q_\mu^2-(p_2^*\cdot\tilde q)\left[\tilde q_\mu^2+2(p_2^*\cdot\tilde q)\right] 
                \nonumber \\
                &+\;2(p_2^*\cdot\tilde n)^2\Big\} \ , \\[5pt]
                \Lambda_{T+}^\text{T}=&\;\frac{1}{2}\bigg\{\frac{1}{M_N^2}\big[\left(m_2^{*2}+3m_2^*m_4^*\right)\tilde q_\mu^2-3(p_2^*\cdot\tilde q)\tilde q_\mu^2 
                \nonumber \\
                &-\;4(p_2^*\cdot\tilde q)^2\big]-\Lambda_L^\text{T}\bigg\} \ .
            \end{align}
            \label{Lambda_T}%
            \end{subequations}
        \item[(4)] Pseudoscalar:
            \begin{equation}
                \Lambda_Q^\text{P}=4\frac{\tilde q_\mu^2}{M_N^2}\big[m_2^{*2}-m_2^*m_4^*+(p_2^*\cdot\tilde q)\big] \ . \label{Lambda_P}
            \end{equation}
        \item[(5)] Vector axial-vector:
            \begin{equation}
                \Lambda_{T-}^\text{VA}=-8i(p_2^*\cdot\tilde n) \ . \label{Lambda_VA}
            \end{equation}
        \item[(6)] Vector tensor:
            \begin{subequations}
            \begin{align}
                \Lambda_L^\text{VT}=&\;-4\Delta(p_2^*\cdot\tilde q)+4\frac{m_2^*}{M_N}\tilde q_\mu^2 \ , \\[5pt]
                \Lambda_{T+}^\text{VT}=&\;-6\Delta(p_2^*\cdot\tilde q)+6\frac{m_2^*}{M_N}\tilde q_\mu^2-\frac{1}{2}\Lambda_L^\text{VT} \ , \\[5pt]
                \Lambda_{M+}^\text{VT}=&\;-2\Delta(p_2^*\cdot\tilde n) \ , \label{Lambda_VT_M_plus}
            \end{align}\label{Lambda_VT}%
            \end{subequations}
            where $\Delta=(m_4^*-m_2^*)/{M_N}$, and we note here that the denominator of $\Delta$ contains $M_N$, not $M_2^*$, as given in the denominator in Ref.~\cite{roberts_2017}.
        \item[(7)] Axial-vector tensor:
            \begin{equation}
                \Lambda_{T-}^\text{AT}=-4i\left(\Delta+2\frac{m_2^*}{M_N}\right)(p_2^*\cdot\tilde n) \ . \label{Lambda_AT}
            \end{equation}\vspace{0.1em}
        \item[(8)] Axial-vector pseudoscalar:
            \begin{subequations}
            \begin{align}
                \Lambda_Q^\text{AP}=&\;8\left[\Delta(p_2^*\cdot\tilde q)-\tilde q_\mu^2\frac{m_2^*}{M_N}\right] \ , \\[5pt]
                \Lambda_{M+}^\text{AP}=&\;-4\Delta(p_2^*\cdot\tilde n) \ . 
                \label{Lambda_AP_M+}
            \end{align}
            \label{Lambda_AP}
            \end{subequations}
    \end{enumerate}
    When considering the process $\bar\nu_e + p \rightarrow e^+ + n$, the only modification required in the hadron tensor [Eq.~\eqref{Lambda_HF}] is the interchange of the labels $2 \leftrightarrow 4$ in the nonzero components.

    Before writing the expression for the neutrino absorption rate within the RHF approximation, we note that the kinematics of the incoming and outgoing leptons can be expressed in terms of the 4-momentum transfer introduced above,
    i.e., the energy and 3-momentum transfer for a given incoming neutrino energy, $E_1$,
    \begin{align}
        q_0=&\;E_1 - E_3 \ , 
        \label{q0} \\
        q=&\;\sqrt{E_1^2 + p_3^2 - 2E_1p_3\mu_{13}} \ , 
        \label{q}
    \end{align}\\
    where $p_3=\sqrt{E_3^2-m_3^2}$. From Eq.~\eqref{q0}, considering the allowed range for $E_3 \in [m_3, \infty)$, one finds that $q_0$ is constrained by $-\infty < q_0 < E_1 - m_3$. Moreover, from Eq.~\eqref{q}, taking the extreme values of $\mu_{13} = \pm 1$, the momentum transfer $q$ satisfies $|E_1 - p_3| < q < E_1 + p_3$.

    Finally, since the contracted 4-momenta given in Eqs.~\eqref{lepton_contractions_HF} and \eqref{hadron_contractions_HF} within the RHF approximation involve all the possible angles and integration variables, i.e., $q_0$ [it originates from $E_3$ via Eq.~\eqref{q0}], $\mu_{13}$, $\mu_{24}$, $\phi_{13}$, $\phi_{24}$, and $p_2$, the neutrino opacity has the following general expression obtained from Eq.~\eqref{eq:diff_absorption_HF}:
    \begin{widetext}
    \begin{align}
        \Gamma(E_1)
        =&\;
        \frac{G_F^2 C^2}{32\pi^2}
        \int_{-\infty}^{E_1-m_3} dq_0
        \frac{p_3}{E_1}
        \frac{1-f_3(E_3)}{1-\exp\{-(q_0+\Delta\mu)/T\}}
        \int_{-1}^1d\mu_{13}
        \int_0^{2\pi}d\phi_{13}
        \int_{-1}^1d\mu_{24}
        \int_0^{2\pi}d\phi_{24}
        \nonumber \\
        &\times
        \int_0^\infty\frac{dp_2}{(2\pi)^3}\,p_2^2 \;
        \delta(\tilde q_0+E_2^*(p_2)-E_4^*(p_4))\left[f_2(E_2)-f_4(E_4)\right]
        \frac{L_{\mu\nu}\Lambda^{\mu\nu}}{4E_2^*(p_2)E_4^*(p_4)} \ .
        \label{eq:total_absorption_HF}
    \end{align}
    \end{widetext}
    The standard form of neutrino opacity within the RMF approximation with the integrals over $q_0$ and $q$ is given in Appendix~\ref{app:H_approx}. The last modification, which one has to apply in case of the process $\bar\nu_e + p \rightarrow e^+ + n$, is the following replacement: $f_3\rightarrow\bar f_3$, i.e., $\mu_3\rightarrow -
    \mu_3$.

\section{Comparing the neutrino opacity within RHF and RMF approaches}\label{sec:RHF_vs_RMF_results}
    In the following, we will evaluate the expressions~\eqref{eq:diff_absorption_HF} and \eqref{eq:total_absorption_HF} numerically. Therefore, we choose two representative conditions, which we select from Ref.~\cite{Guo2020PhRvD102}. 
    We list these conditions in Table~\ref{tab:AB}, henceforth denoted as A and B, in terms of temperature $T$, baryon density $n_{\rm B}$ (in units of the nuclear saturation density $n_0$), electron and muon abundances $Y_e$ and $Y_\mu$ (satisfying the local charge neutrality condition, $Y_p=Y_e+Y_\mu$), and electron and muon chemical potentials $\mu_e$ and $\mu_\mu$, respectively.
    Furthermore, we provide the corresponding thermodynamic chemical potentials, $\mu_n$ and $\mu_p$, and the RMF scalar self-energy, $\Sigma^S$, and zero-component (timelike) self-energies, $\Sigma^0_n$ and $\Sigma^0_p$, distinguishing neutrons and protons, in Table~\ref{tab:cond}. 
    In the following, we distinguish the RHF nuclear model,
    with the nonlinear terms given via the self-interacting $\sigma$ and $\omega$ mesons, 
    from common RMF parametrizations. To this end, we select the TMA \cite{toki_1995}, NL3 \cite{lalazissis_1997}, and DD2 \cite{typel_2010,kumar_2024} RMF models.
    The former two possess nonlinearity via meson-meson interactions and the latter one contains density-dependent meson-nucleon couplings, fitted to Dirac-Brueckner-Hartree-Fock calculations (further details are given in Ref.~\cite{typel_1999}). All of these RMF parametrizations employ $\sigma$, $\omega$, and $\rho$ meson interaction channels, and omit the presence of nuclear clusters at subsaturation density~\cite{Hempel2010NuPhA837,Fischer2020PhRvC102}. 
    Bulk nuclear matter properties for RHF, TMA, NL3, and DD2  
    are provided in Table~\ref{tab:Nuclear_properties}. 
    We note here that these RMF nuclear EOSs have been commonly employed in astrophysical studies of core-collapse supernovae~\cite{Hempel12,Steiner2013ApJ765L,Fischer17} and binary neutron star mergers (cf. Refs.~\cite{Bauswein:2014PhRvD90,Bauswein19,Bauswein:2020PhRvL125} and references therein).
    We also develop TMA$\delta$ \cite{Greco:2002PhRvC65_delta-meson} and DD2$\delta$, which differ from the original TMA and DD2 parametrizations by the inclusion of $\delta$ mesons. We will discuss these models further in Sec.~\ref{sec:RMF_delta} and Appendix~\ref{app:H_approx} below, when investigating the impact of $\delta$ mesons.
    
    For further evaluation and implementation of the RHF and RMF models (cf. conditions A and B in Tables~\ref{tab:AB} and \ref{tab:cond}), in addition to temperature, we ensure identical baryon density and proton fraction, hence the different values of the neutron and proton chemical potentials and the different values of the self-energies of RHF, TMA, NL3, and DD2. We note in particular the different magnitudes of $\Sigma_n^0-\Sigma_p^0$. We will return to this aspect further below, when discussing the neutrino opacity.
    We note further that within these selected RMF models, neutron and proton scalar self-energies are identical, which is attributed to the absence of the isovector-scalar $\delta$ meson \cite{kubis_1997}. 
    Models with $\delta$ mesons feature isospin splitting and hence different neutron and proton scalar self-energies, which will be discussed in Sec.~\ref{sec:RMF_delta} and Appendix~\ref{app:H_approx}. 
    Within the RHF, by construction, the neutron and proton scalar self-energies are different, as detailed in Appendix~\ref{app:Hartree-Fock}.
    
    We emphasize here that the neutron and proton chemical potentials, as well as the self-energies, differ from the values used in Ref.~\cite{Guo2020PhRvD102}, due to the different nuclear matter model. Nevertheless, these conditions are taken from simulations of self-consistent neutrino-driven core-collapse supernova explosions from the Garching group~\cite{bollig17}, featuring the nuclear EOS of Lattimer and Swesty~\cite{LSEOS}, and where muonic neutrino interactions are taken into account self-consistently, with the treatment of charged-current muonic processes at the mean-field level. 

    \begin{table}[t!]
    \caption{Two selected conditions \cite{Guo2020PhRvD102}, listing temperature $T$, baryon density $n_{\rm B}$, lepton fractions, and chemical potentials for electrons and muons, $Y_e$, $Y_\mu$, $\mu_e$, and $\mu_\mu$, respectively. Except the lepton fractions and $n_{\rm B}$, where the latter is in units of the saturation density $n_0$, all quantities are given in units of MeV.}
    \begin{ruledtabular}
    \begin{tabular}{lcccccc}
        Condition & $T$ & $n_{\rm B}/n_0$ & $Y_e$ & $Y_\mu$ & $\mu_e$ & $\mu_\mu$ \\
        \colrule
        A & 38.3 & 0.406 & 0.13 & 0.040 & 83 & 67 \\
        B & 15.2 & 0.053 & 0.11 & 0.002 & 45 & 38 
    \end{tabular}
    \label{tab:AB}
    \end{ruledtabular}
    \end{table}

\begin{figure*}[t]
    \subfigure[~Condition~A]{\includegraphics[width=0.495\textwidth]{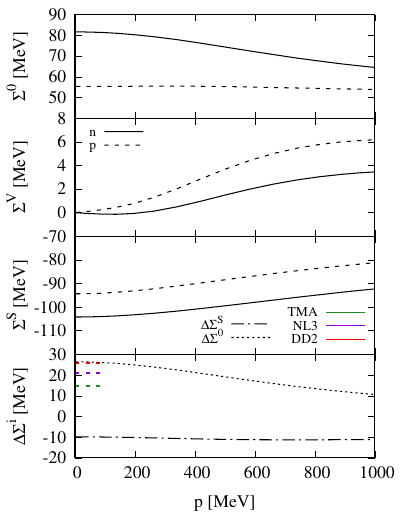}}
    \subfigure[~Condition~B]{\includegraphics[width=0.495\textwidth]{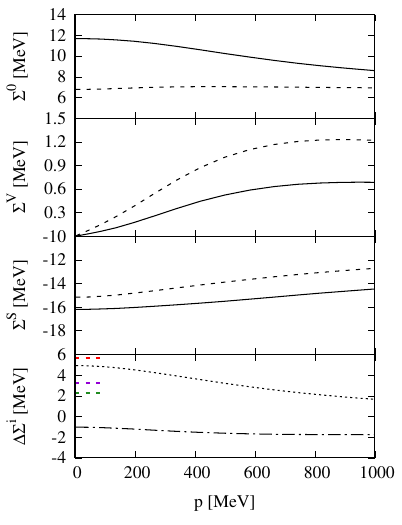}}
    \caption{
    Hartree-Fock nucleon self-energy components (black lines): scalar $\Sigma^S$, timelike $\Sigma^0$ and spacelike (vector) $\Sigma^V$, with respect to momentum $p$ for neutrons (solid lines) and protons (dashed lines), at two representative conditions A (left panel) and B (right panel) listed in Tables~\ref{tab:AB} and \ref{tab:cond}. 
    In the bottom panels we show the difference between neutron and proton self-energies, $\Delta\Sigma^S=\Sigma^S_n-\Sigma^S_p$ (dash-dotted lines) and $\Delta\Sigma^0=\Sigma^0_n-\Sigma^0_p$ (dotted lines). 
    For comparison, short horizontal lines include $\Delta\Sigma^0$ for TMA (green), NL3 (violet), and DD2 (red). We note that $\Delta\Sigma^S=0$ for all RMF models employed here.}
    \label{fig:HF-selfenergies}
    \end{figure*}

    Furthermore, we introduce another RMF model, which is the reduction of the RHF model, henceforth denoted as RMF*, based on the same mesons: $\sigma$, $\omega$, $\rho$, and the same nonlinear terms as the RHF, except pions and tensor terms, which vanish at the mean-field level, as explained in Appendix~\ref{app:Hartree-Fock}, and are omitted by design within all RMF models. In order to achieve this reduction, we take all the same rest masses and coupling constants as given in Appendix~\ref{app:Hartree-Fock} and refit the $\sigma$ and $\omega$ coupling constants to obtain at $n_0^\text{RHF}=0.148$~fm$^{-3}$ the binding energy $E_B=-15.73$~MeV, which, after the normalization, are given by $g_\sigma^2/4\pi=9.155$ and $g_\omega^2/4\pi=10.190$. 
    The resulting neutron and proton chemical potentials, together with the self-energies, are listed in Table~\ref{tab:cond}. Details about the comparison between RHF and RMF* are given in Sec.~\ref{sec:RHF_vs_reduced_RHF}, while details about the RHF model used here are given in Appendix~\ref{app:Hartree-Fock}, together with the numerical evaluation of the momentum-dependent self-energies, as shown in Fig.~\ref{fig:HF-selfenergies} for both conditions A (left panel) and B (right panel). 
    The numerical values are listed in Tables~\ref{tab:HF_Sigma_A} and \ref{tab:HF_Sigma_B} in Appendix~\ref{app:appendix_HF-tables}. 
    Moreover, in Fig.~\ref{fig:HF-selfenergies}, in the lowest panel, we include the differences between neutron and proton components of the self-energies for our nonlinear RHF model, in comparison to $\Delta\Sigma^0$ for the RMF models (note, there is no vector self-energy in RMF models and $\Delta\Sigma^S=0$ for the RMF models employed here). 
    From this analysis, it becomes evident that the values obtained for RHF  resemble closest those of DD2, at $p=0$, but this starts to differ with increasing momenta. 
    In addition, RHF exhibits the impact from the nonzero $\Delta\Sigma^S$, which turns out to be large, on the same order as $\Delta\Sigma^0$. 
    This becomes important for the calculations of the neutrino opacity, since both $\Delta\Sigma^S$ and $\Delta\Sigma^0$ appear explicitly in the expressions of the rates within the RHF approach as well as in the modified $Q$-value. These features are absent in the RMF treatment. We will return to this aspect further below in Sec.~\ref{sec:RHF_vs_RMF_results}.

    \begin{table}[t!]
    \caption{Thermodynamic neutron and proton chemical potentials, $\mu_n$ and $\mu_p$, as well as scalar and timelike components of the self-energy, $\Sigma^S$ and $\Sigma^0$, for RHF and a selection of RMF models, for the two representative conditions A and B (see Table~\ref{tab:AB}). All quantities are given in units of MeV. We note here the different saturation densities given in Table~\ref{tab:Nuclear_properties} (note RHF and RMF* have the same value of the saturation density).}
    \begin{ruledtabular}
    \begin{tabular}{ccccddd}
        & Model & $\mu_n$ & $\mu_p$ &  \multicolumn{1}{c}{$\Sigma^S$} & \multicolumn{1}{c}{$\Sigma^0_n$} & \multicolumn{1}{c}{$\Sigma^0_p$} \\
        \colrule
        A & RHF & 894.3 & 814.9 & \multicolumn{1}{c}{$\cdots^\text{a}$} & \multicolumn{1}{c}{$\cdots^\text{a}$} & \multicolumn{1}{c}{$\cdots^\text{a}$} \\ 
        & TMA & 896.4 & 814.4 & -153.58 & 128.43 & 113.61 \\
        & NL3 & 900.0 & 811.7 &  -157.57 & 135.33 & 114.38 \\
        & DD2 & 904.8 & 811.3 &  -202.40 & 180.83 & 155.14 \\ 
        & RMF* & 888.7 & 814.4 & -126.58 & 95.52 & 88.30 \\[5pt]
        B & RHF & 914.9 & 877.8 & \multicolumn{1}{c}{$\cdots^\text{a}$} & \multicolumn{1}{c}{$\cdots^\text{a}$} & \multicolumn{1}{c}{$\cdots^\text{a}$} \\
        & TMA & 915.0 & 878.8 & -21.86 & 17.28 & 15.00 \\
        & NL3 & 914.3 & 877.2 &  -23.33 & 17.91 & 14.69 \\
        & DD2 & 915.7 & 876.1 & -33.43 & 29.05 & 23.37 \\
        & RMF* & 911.5 & 876.5 & -21.36 & 13.14 & 12.04 
    \end{tabular}
    \label{tab:cond}
    \end{ruledtabular}
    $^\text{a}$~Momentum-dependent RHF self-energies are listed in Tables~\ref{tab:HF_Sigma_A} and \ref{tab:HF_Sigma_B} for conditions A and B (see also Fig.~\ref{fig:HF-selfenergies}). \\
    \end{table}
    \begin{table}[b]
    \centering
    \caption{\label{tab:Nuclear_properties}
    Nuclear matter properties of all models under investigation, at nuclear saturation and for zero temperature, listing saturation density, $n_0$, normalized effective mass with respect to the averaged nucleon rest mass, $m^*/M_N$, binding energy per nucleon, $E_B$, compression modulus, $K$, symmetry energy, $E_{sym}$, and its slope, $L$. 
    For RHF, we take $m^*=m^*(p_F)$ at the Fermi momentum $p_F$. 
    The bottom row contains the current experimental constraints.}
    \begin{ruledtabular}
    \begin{tabular}{ccccccd}
    \multirow{2}{*}{Model} & $n_0$ & $E_B$ & $m^*/M_N$ & $K$ & $E_{sym}$ & \multicolumn{1}{c}{$L$} \\
    & $[\textrm{fm}^{-3}]$ & [MeV] & & [MeV] & [MeV] & \multicolumn{1}{c}{[MeV]} \\
    \colrule
    RHF         & 0.148 & $-$15.77 & 0.76 & 252 & 29.6 & 63.5 \\
    TMA         & 0.147 & $-$16.03 & 0.64 & 317 & 30.6 & 90.0 \\
    TMA$\delta$ & 0.147 & $-$16.03 & 0.64 & 317 & 30.6 & 113.7 \\
    NL3         & 0.148 & $-$16.24 & 0.60 & 272 & 37.4 & 118.6 \\
    DD2         & 0.149 & $-$16.02 & 0.63 & 243 & 32.7 & 57.9 \\
    DD2$\delta$ & 0.149 & $-$16.02 & 0.63 & 243 & 32.7 & 58.0 \\[5pt]
    Exp. & $\sim$0.15 & $-$16$\pm$1$^\text{b}$ & 0.8$\pm$0.1$^\text{a}$ & 230$\pm$20$^\text{b}$ & 29$-$34$^\text{b}$ & \multicolumn{1}{c}{43$-$75$^\text{b}$}
    \end{tabular}
    \end{ruledtabular}
    \raggedright $^\text{a}$~Ref.~\cite{li_2018} \\
    $^\text{b}$~Ref.~\cite{harris_2025}
    \end{table}

    One of the complications that arises within the RHF treatment is the Dirac delta in expressions~\eqref{eq:diff_absorption_HF} and \eqref{eq:total_absorption_HF}. Note that within the RMF approximation, the Dirac delta functions can be removed by applying the appropriate identity and correct integration limits (for details, cf. Ref.~\cite{roberts_2017} and references therein), which we closely follow, as detailed in Appendix~\ref{app:H_approx}. 
    It is important to emphasize that this is no longer possible at the RHF level, due to the explicit momentum dependence of the nucleon self-energies. Hence, in our numerical procedure, we apply the property of the Dirac delta given by Eq.~\eqref{delta_property} and remove the integral over $\mu_{24}$ by making a substitution $\mu_{24}=\mu^0_{24}$, where $\mu^0_{24}$ is the root of $g(\mu^0_{24})=0$ found via the numerical root-finding procedure.
    
    In the numerical calculations of Eqs.~\eqref{eq:diff_absorption_HF} and \eqref{eq:total_absorption_HF}, we also evaluate the Dirac delta in the RMF approximation, where we note that within this approach, one can easily integrate out $\phi_{13}$ and $\phi_{24}$ to obtain $(2\pi)^2$. All integrals are evaluated using the Gauss-Legendre quadrature. To find the angle root of the Dirac delta satisfying $g(\mu^0_{24})=0$ in Eq.~\eqref{delta_property}, we tested both a brute-force search [we created a fine grid from $-1$ to 1, checked when the Dirac delta's argument, $g(\mu_{24})$, is equal to zero, and then performed a linear interpolation to obtain $\mu_{24}=\mu^0_{24}$] and a globally convergent Newton-Raphson method, which yielded quantitatively comparable results. Only results that are in the range $-1\le\mu\le1$ are accepted, otherwise, the result is equal to zero\footnote{If the Newton-Raphson fails to converge within a finite number of steps, the Dirac delta is assumed to be zero.}. 
    The lower limit of the $q_0$ integral is set to --$400$~MeV and the upper limit of the $p_2$ integration is in the range of $1.0$--$1.5$~GeV.

    \subsection{Electronic charged-current processes}
    \begin{figure*}[t]
    \subfigure[condition A, no medium effects]{\includegraphics[width=0.495\textwidth]{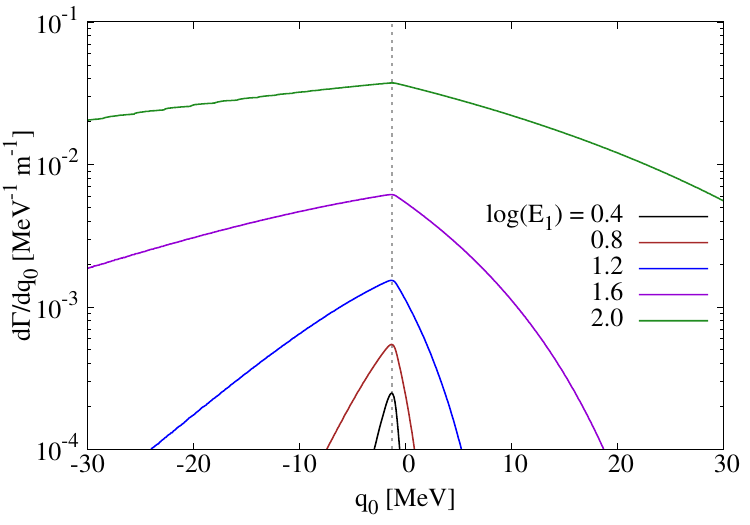}}
    \subfigure[condition A, medium effects]{\includegraphics[width=0.495\textwidth]{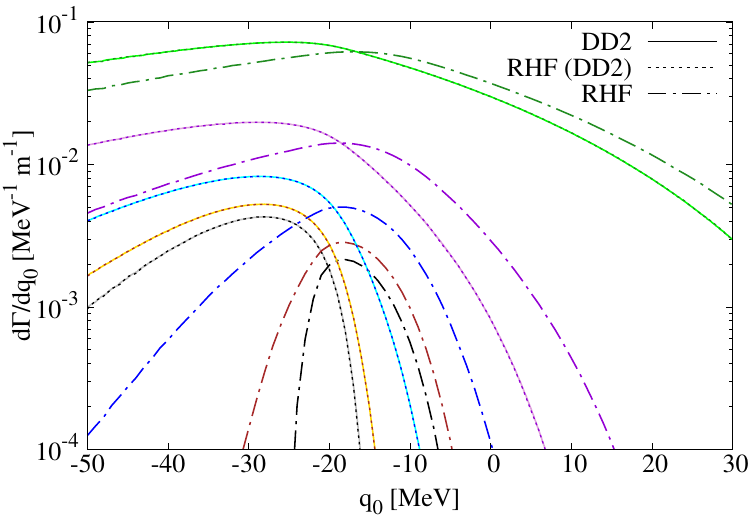}}
    \\
    \subfigure[condition B, no medium effects]{\includegraphics[width=0.495\textwidth]{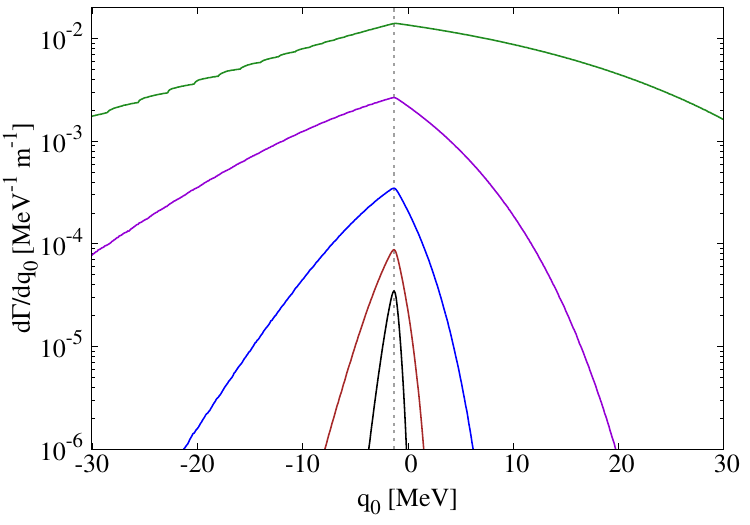}}
    \subfigure[condition B, medium effects]{\includegraphics[width=0.495\textwidth]{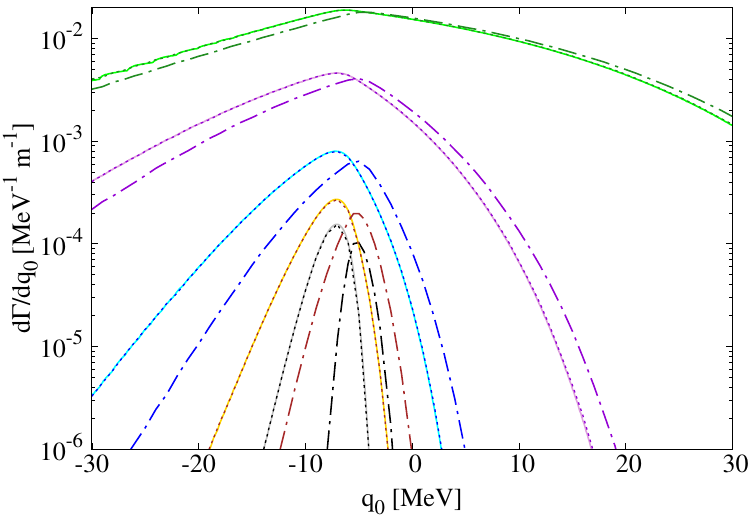}}
    \caption{Angle integrated neutrino absorption rate, $d\Gamma/dq_0$, for the process $\nu_e + n \rightarrow e^- + p$, as a function of energy transfer $q_0$, for a selection of incoming neutrino energies $E_1$ (in log scale), for $\log(E_1/{\rm MeV})=0.4$--$2.0$, for the two conditions A (top panels) and B (bottom panels), given in Tables~\ref{tab:AB} and \ref{tab:cond}, for the LO hadronic vertex \eqref{eq:cc_current}, Eq.~\eqref{eq:LO}.
    Results are shown for the vacuum rates (left panels), with zero nucleon self-energies, and with the medium effects included (right panels), where we distinguish the RMF approach (we select DD2 as the representative RMF model here), denoted as DD2 (solid lines), and two RHF treatments.
    The latter employ the full momentum-dependent self-energies RHF (dash-dotted lines) and the reduction to DD2, RHF(DD2), by choosing the constant scalar and timelike component of the self-energies as well as zero vector self-energies (see the text for details).
    }
    \label{fig:diff_absor_electrons}
    \end{figure*}
    In the following, we evaluate the angle integrated neutrino absorption rate, $d\Gamma/dq_0$ [see Eq.~\eqref{eq:total_absorption_HF}], for electron neutrino absorption on neutrons,
    \begin{eqnarray}
        \nu_e+n\rightarrow e^- + p~,
    \end{eqnarray}
    for the LO hadronic vertex \eqref{eq:cc_current} without form factors, see Eq.~\eqref{eq:LO}. This is shown in Fig.~\ref{fig:diff_absor_electrons} for conditions A (top panels) and B (bottom panels), for selected incoming neutrino energies, $E_1$ (in log scale) from $\log(E_1/{\rm MeV})=0.4$--$2.0$. First, we validate our numerical procedure considering the vacuum differential absorption rates (left panels), i.e., all medium effects are set to zero and we apply the corresponding RHF neutron and proton chemical potentials. These results agree qualitatively with Ref.~\cite{roberts_2017} (see their Fig.~2). 
    Unfortunately, a quantitative comparison is not possible here due to the lack of certain data, e.g., neutron and proton chemical potentials.
    
    Second, we evaluate the rates at the RMF level, shown in the right panels of Fig.~\ref{fig:diff_absor_electrons}.
    Therefore, we select DD2 as a representative RMF parametrization (solid lines), since its nuclear properties are the most similar to the ones for the nonlinear RHF. For the further validation of the rate expressions at the RHF level, we distinguish the full approach, implementing the momentum-dependent medium modifications (dash-dotted lines) and the semireduction of the RHF approach to DD2, by choosing constant DD2 values for the neutron and proton scalar and timelike components of the self-energies (dotted lines) as well as zero vector self-energies, denoted as RHF(DD2). The values of the neutron and proton chemical potentials as well as nonzero self-energies are listed in Table~\ref{tab:cond}.
    The results of the angle integrated neutrino absorption rates agree quantitatively with those of the actual RMF treatment at all energies $E_1$ (see the dotted lines, which lie exactly on top of solid lines). This confirms the correct RMF limit of the corresponding numerical implementation of the momentum-dependent RHF model.
    \begin{figure*}[t]
    \subfigure[condition A, no medium effects]{\includegraphics[width=0.495\textwidth]{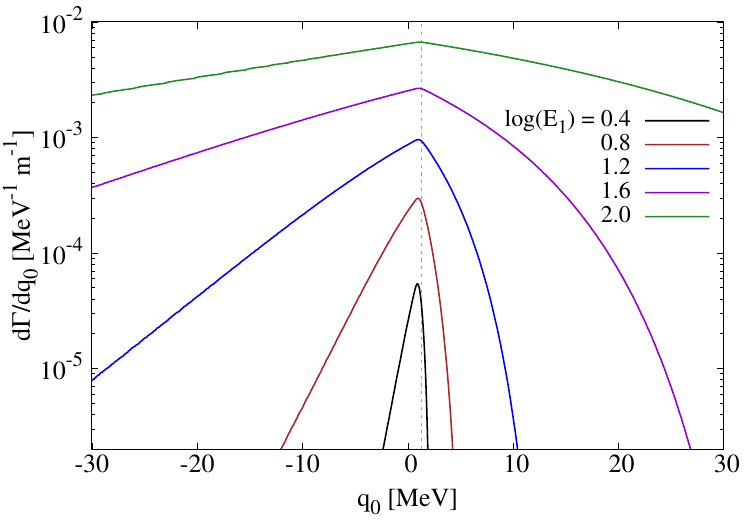}}
    \subfigure[condition A, medium effects]{\includegraphics[width=0.495\textwidth]{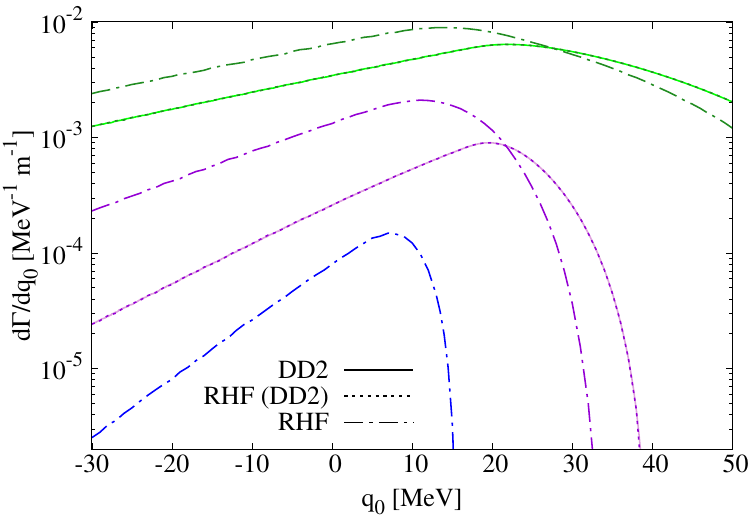}}
    \\
    \subfigure[condition B, no medium effects]{\includegraphics[width=0.495\textwidth]{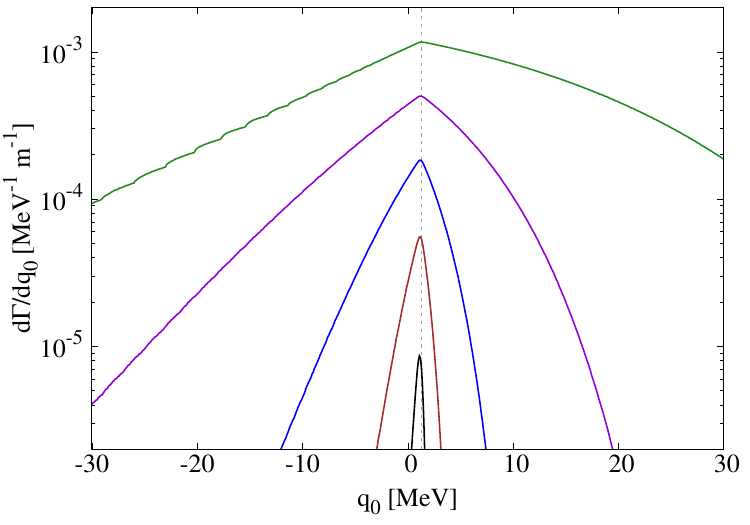}}
    \subfigure[condition B, medium effects]{\includegraphics[width=0.495\textwidth]{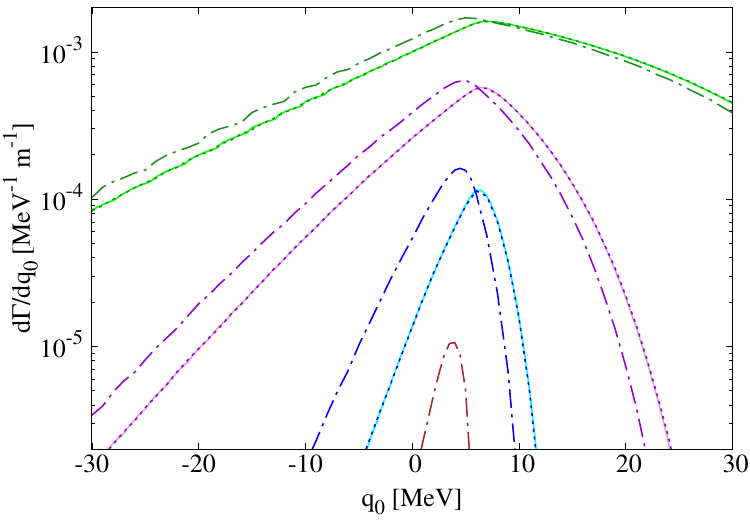}}
    \caption{The same as Fig.~\ref{fig:diff_absor_electrons} but for the charged-current process involving electron antineutrinos, $\bar\nu_e + p \rightarrow e^+ + n$.}
    \label{fig:diff_absor_anti_electrons}
    \end{figure*}

    Third, moving to the comparison of RMF---again at the example of DD2---and RHF treatments, shown in Fig.~\ref{fig:diff_absor_electrons} (solid and dash-dotted lines), under the same conditions, it becomes evident that both exhibit qualitatively different $q_0$ dependencies, especially at low neutrino energies. We note here that, if no medium effects are considered, the rates peak at $q_0=m_p-m_n=-1.293$~MeV, i.e., neutron-proton rest mass difference (gray, dashed vertical lines in the left panels of Fig.~\ref{fig:diff_absor_electrons}). 
    The peaks of the angle integrated neutrino absorption rates shift to larger negative values, if the medium modifications are taken into account at the RMF level, which is associated with the presence of nonzero self-energies (right panels)---the magnitude of the rate's shifts at the RMF level is given by the effective (medium) $Q$-value~\cite{roberts_2012,martinez_2012}, $Q^*=m^*_p-m^*_n + (\Sigma^0_p - \Sigma^0_n)$ [apply Eq.~\eqref{E_eff_HF} for the difference in the single-particle energies, $E_p-E_n$, at $\boldsymbol{p}=0$], with neutron and proton effective masses and timelike self-energies, $m^*_{n/p}$ and $\Sigma^0_{n/p}$, respectively, which is generally negative under neutron-rich conditions. 
    Therefore, especially at low neutrino energies, the peaks of the rates are located at $Q^*$, which for DD2 has values of $Q^*\approx-27$~MeV for condition A and $Q^*\approx-7$~MeV for condition B.
    We note further that for all RMF models under investigation here, we have $m^*_p-m^*_n=m_p-m_n$, since both neutrons and protons have identical scalar self-energies (see the RMF rate expressions in Appendix~\ref{app:H_approx}). 

    \begin{figure*}[t]
    \subfigure[Condition A, ($\nu_e,\bar\nu_e$)]{\includegraphics[width=0.495\textwidth]{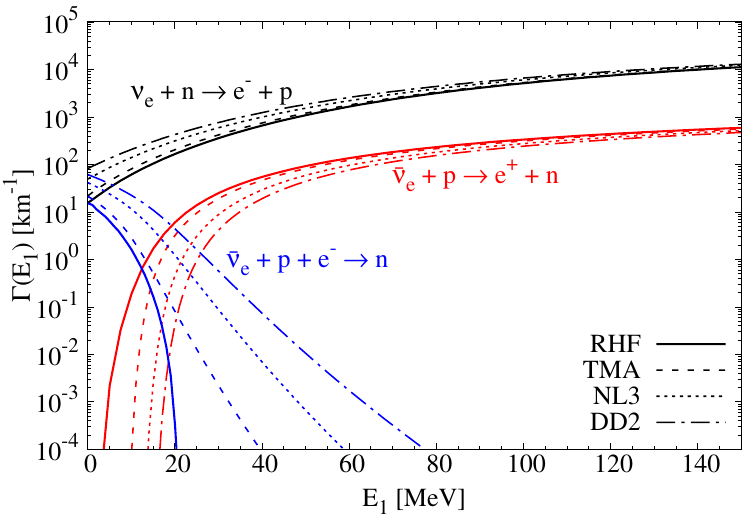}\label{fig:neutrino_opacity_vs_E1_a}}
    \subfigure[Condition A, ($\nu_\mu,\bar\nu_\mu$)]{\includegraphics[width=0.495\textwidth]{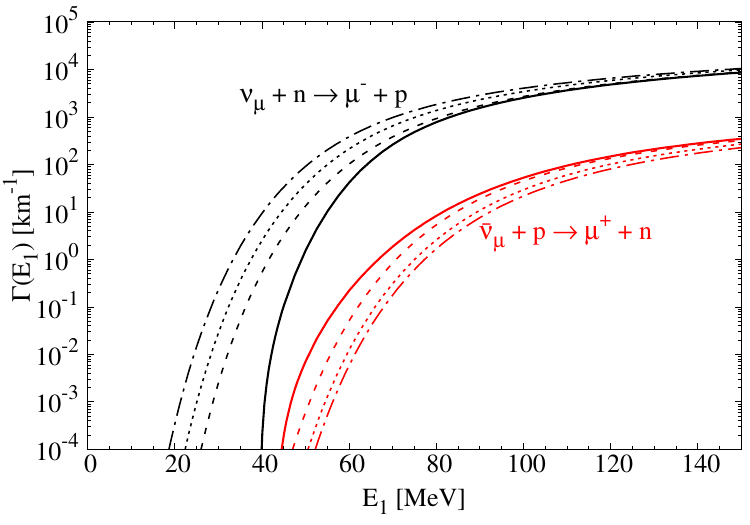}\label{fig:neutrino_opacity_vs_E1_b}}
    \\
    \subfigure[Condition B, ($\nu_e,\bar\nu_e$)]{\includegraphics[width=0.495\textwidth]{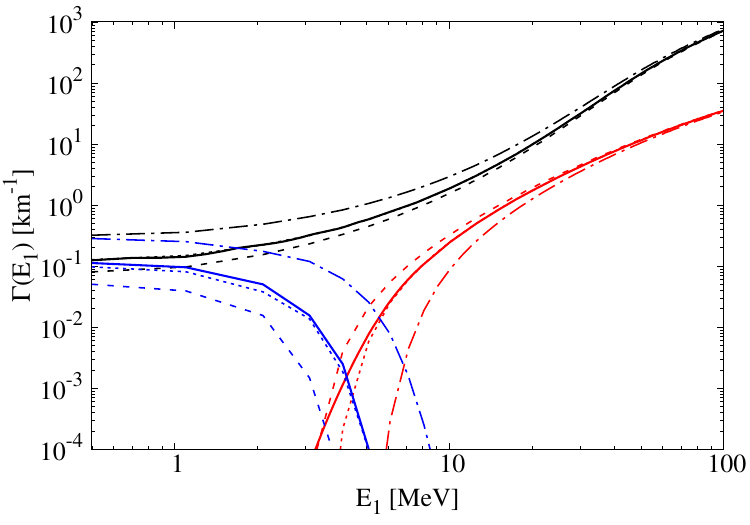}}
    \subfigure[Condition B, ($\nu_\mu,\bar\nu_\mu$)]{\includegraphics[width=0.495\textwidth]{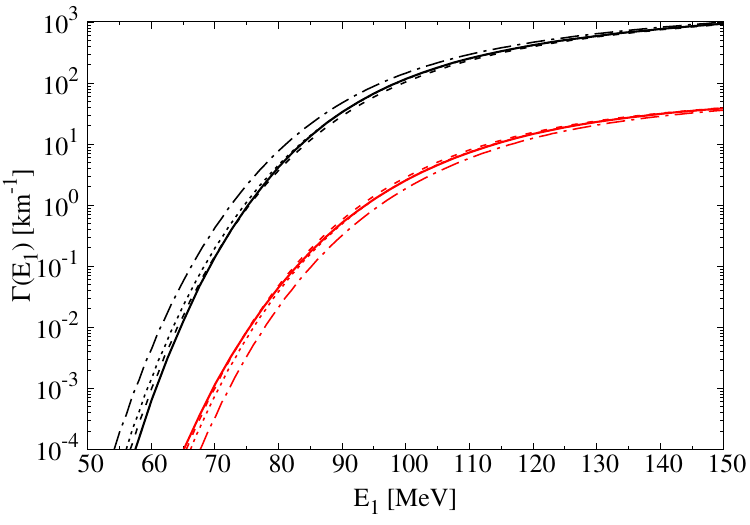}}
    \caption{Neutrino opacity as a function of the incoming (anti)neutrino energy, $E_1$, for conditions A (top panels) and B (bottom panels), see Tables~\ref{tab:AB} and \ref{tab:cond}, comparing RHF (solid lines) and our selection of representative RMF models: TMA (dashed lines), NL3 (dotted lines), and DD2 (dash-dotted lines), employing here the full hadronic vertex~\eqref{eq:cc_current}, i.e., weak magnetism, pseudoscalar terms, and form factors, see expressions~\eqref{eq:LO}--\eqref{form_factors}. 
    Black lines correspond to reactions involving neutrinos, red lines those for antineutrinos, and blue lines for the inverse neutron decay, where we distinguish electronic charged-current processes (left panels) and muonic charged-current processes (right panels).}
    \label{fig:neutrino_opacity_vs_E1}
    \end{figure*}

    Fourth, investigating the rates within the RHF approach (dash-dotted lines), we observe a shift of the rate's peaks back, toward less negative values of $q_0$, in comparison to DD2 (solid lines). 
    This finding is confirmed by calculating the modified $Q$-value within RHF, i.e., $Q^*=m_p-m_n+[\Sigma^S_p(0)-\Sigma^S_n(0)] + [\Sigma^0_p(0) - \Sigma^0_n(0)]$, assuming here constant values of the self-energies at the zero momentum and neglecting the generally small vector parts of the self-energies (cf. Fig.~\ref{fig:HF-selfenergies}), i.e., $\Sigma^V_{n/p}=0$ here.
    For condition A, we obtain $Q^*\approx-18$~MeV and for condition B, $Q^*\approx-5$~MeV.
    The differences of the rates between RHF and DD2 treatments, shown in Fig.~\ref{fig:diff_absor_electrons}, are attributed to the fact that neutrons and protons have naturally different scalar self-energies within the RHF approach [see expression~\eqref{Sigma_S_HF_full} in Appendix~\ref{app:Hartree-Fock}], even without the explicit inclusion of the isovector-scalar $\delta$ meson, and that the nucleon self-energies are momentum dependent. 
    In fact, the $\delta$-meson contributions to the EOS can be large, depending on the RMF parametrization~\cite{Greco:2002PhRvC65_delta-meson,Greco:2003PhRvC67_delta-meson,Roca-Maza:2011PhRvC84_delta-meson}, which we further elaborate on in Sec.~\ref{sec:RMF_delta}.
    Now, enforcing $\Sigma^S_p(0)-\Sigma^S_n(0)=0$, within RHF, we obtain $Q^*\approx-28$~MeV for condition~A and $Q^*\approx-6$~MeV for condition~B. These values resemble closely those of DD2. 
    This shift in the $q_0$ dependence of the angle integrated neutrino absorption rates has important consequences for the energy dependence of the neutrino opacity. We will return to this aspect later.
    
    In the following, we repeat the calculations for electron antineutrino absorption on protons, 
    \begin{eqnarray}
        \bar\nu_e + p \rightarrow e^+ + n~,
    \end{eqnarray}
    also here at the level of the leading-order hadronic vertex function without form factors. This can be done straightforwardly, with the proper replacements in expressions~\eqref{eq:diff_absorption_HF} and \eqref{eq:total_absorption_HF}, i.e., replacing electron with positron, swapping neutron and proton propagators, and changing signs in several places in the lepton tensor components, as is detailed in Sec.~\ref{sec:HF_approx}. The results are shown in Fig.~\ref{fig:diff_absor_anti_electrons}, for the same conditions A (top panels) and B (bottom panels) and for the same selection of incoming electron antineutrino energies, $E_1$. Equivalently to Fig.~\ref{fig:diff_absor_electrons}, the left panels evaluate the rates without medium modifications, from which it becomes evident that the rates peak at $q_0=m_n-m_p=1.2935$~MeV (gray, dashed vertical lines in the left panels of Fig.~\ref{fig:diff_absor_anti_electrons}), exactly as for the $\nu_e$ rates but with opposite sign. The reason, therefore, is well known, while for $\nu_e$ the conversion of neutrons to protons is exoenergetic, the reverse conversion of protons into neutrons is endoenergetic, in particular under neutron-rich conditions. This has long been explored at the level of the zero-momentum transfer approximation~\cite{Bruenn85} and at the RMF level \cite{reddy_1998}.

    As a consequence, in the case of medium-modified rates (right panels in Fig.~\ref{fig:diff_absor_anti_electrons}), opposite to the behavior for $\nu_e$ (see Fig.~\ref{fig:diff_absor_electrons}), the peaks of the rates are shifted toward larger positive values of $q_0$ at higher densities (condition A) and lower positive values of $q_0$ at lower densities (condition B). However, at the RMF and RHF levels, these shifts are directly proportional to the corresponding medium $Q$-values, $Q^*=m^*_n-m^*_p + (\Sigma^0_n - \Sigma^0_p)$ and $Q^*=m_n-m_p+[\Sigma^S_n(0)-\Sigma^S_p(0)] + [\Sigma^0_n(0) - \Sigma^0_p(0)]$, respectively, only at low densities (condition B) with the values given above for $Q^*$ for $\nu_e$ with the extra minus sign, i.e., $Q^*_{\bar\nu_e}=-Q^*_{\nu_e}$, while at higher density (condition A), $Q^*$-values for $\nu_e$ and $\bar\nu_e$ do not give the same values with opposite sign. The reason for this is such that the expression for the antineutrino opacity is not entirely symmetric to the one for neutrino opacity both at the RMF and RHF levels. 
    Moreover, the behavior of the magnitudes of RHF in comparison to DD2 is opposite to those of $\nu_e$ in which they are suppressed, whereas the rates for $\bar\nu_e$ are enhanced. 
    And finally, since the treatment of antineutrinos is not exactly symmetric to the one for neutrinos within RMF and RHF, we observe that we have a much stronger suppression of the $\bar\nu_e$ rate magnitudes at lower $E_1$ within DD2 compared to RHF.
    
    In order to further highlight the impact of the RHF treatment, in comparison to the RMF approach, we now turn to the discussion of the neutrino opacity~\eqref{eq:total_absorption_HF}, i.e., $\Gamma(E_1)=1/\lambda_1$, for which we integrate the angle integrated electron neutrino absorption rate, $d\Gamma/dq_0$, over $q_0$. 
    Further below, we will return to the discussion of the impact of the different contributions of the hadronic vertex~\eqref{eq:cc_current}, such as weak magnetism, pseudoscalar contributions, and form factors. 
    Here, we employ the full vertex for the discussion of the opacity for both RMF and RHF approaches. 
    Figure~\ref{fig:neutrino_opacity_vs_E1} shows the opacity as a function of the incoming neutrino energy $E_1$, comparing the RHF (solid line) and our selected RMF models: TMA (dashed lines), NL3 (dotted lines), and DD2 (dash-dotted lines), for the two conditions A (top panels) and  B (bottom panels). 
    The left panels evaluate the electronic charged-current rates for $\nu_e$ (black lines) and $\bar\nu_e$, distinguishing between absorption (red lines) and inverse neutron decay (blue lines), and the right panels evaluate the muonic charged-current rates for $\nu_\mu$ (black lines) and $\bar\nu_\mu$ (red lines).

    \begin{figure*}[t]
    \subfigure[$\nu_e+n\rightarrow e^-+p$, comparing RHF and RMF models
    ]{\includegraphics[width=0.4925\textwidth]{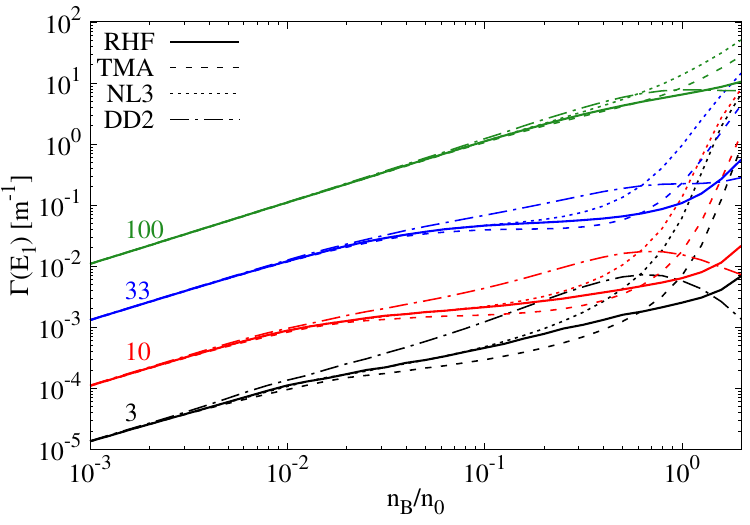}
    \label{fig:neutrino_opacity_vs_nB_B_a}}
    \hfill
    \subfigure[RHF, hadronic vertex contributions]{\includegraphics[width=0.4925\textwidth]{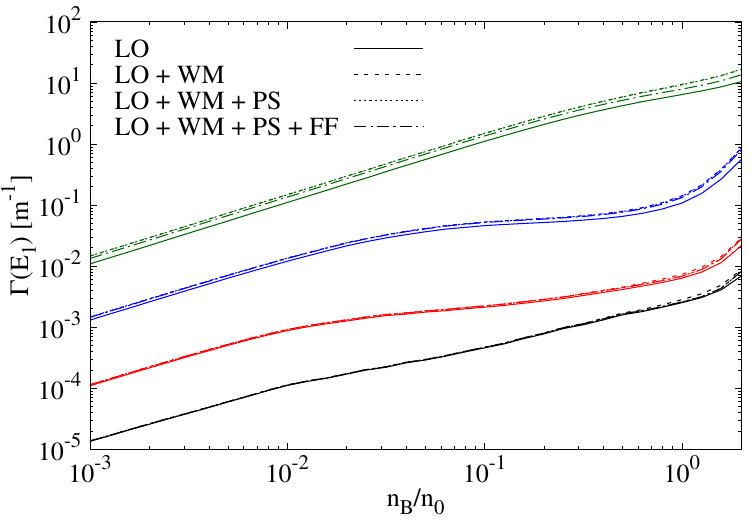}}
    \\
    \subfigure[$\bar\nu_e+p\rightarrow e^++n$, comparing RHF and RMF models]{\includegraphics[width=0.4925\textwidth]{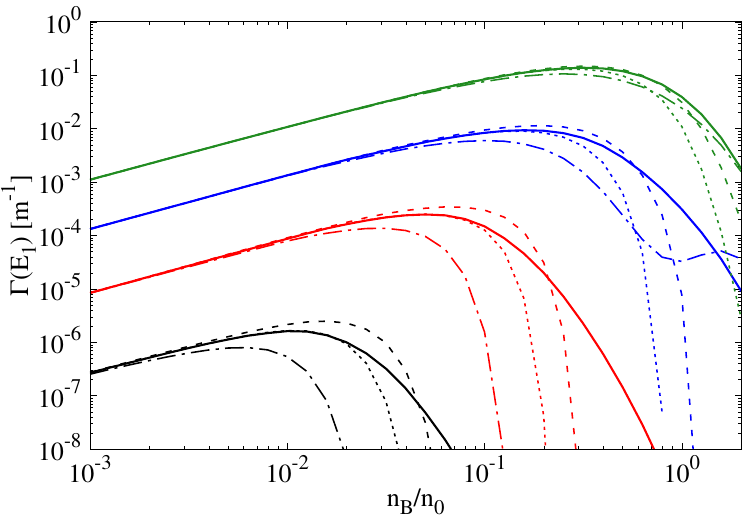}
    \label{fig:neutrino_opacity_vs_nB_B_c}}
    \hfill
    \subfigure[RHF, hadronic vertex contributions]{\includegraphics[width=0.4925\textwidth]{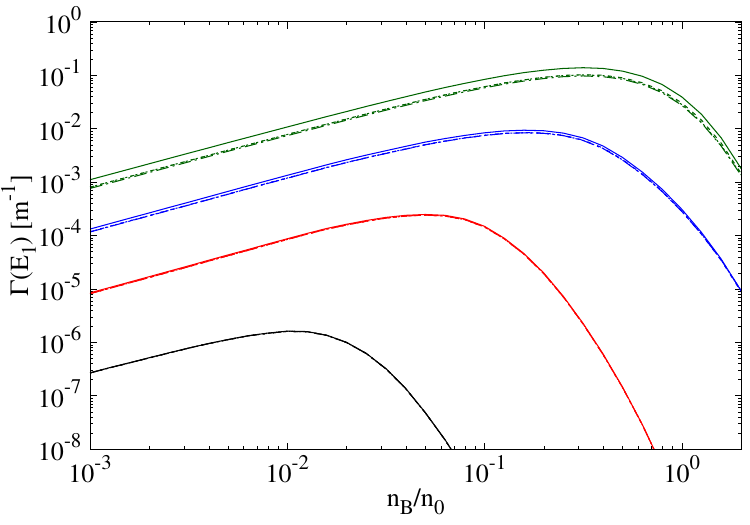}}
    \caption{Neutrino opacity as a function of the normalized baryon density, $n_{\rm B}$, in units of the saturation density, $n_0$, for a selection of (anti)neutrino energies of $E_1=3,10,33,100$~MeV, corresponding to condition B (Tables~\ref{tab:AB} and \ref{tab:cond}), for $\nu_e$ (top panels) and $\bar\nu_e$ (bottom panels). 
    Left panel: comparing our selection of representative RMF models: TMA (dashed lines), NL3 (dotted lines), and DD2 (dash-dotted lines), and the RHF approach (solid lines) at the leading-order hadronic vertex.
    Right panel: rates at the RHF level comparing different contributions of the hadronic vertex~\eqref{eq:cc_current}, namely, LO, WM, PS terms, and FFs.}
    \label{fig:neutrino_opacity_vs_nB_B}
    \end{figure*}

    To calculate neutrino opacity, Eq.~\eqref{eq:total_absorption_HF}, for the inverse neutron decay (blue lines)
    \begin{eqnarray}
        \bar\nu_e+p+e^-\rightarrow n~,
    \end{eqnarray}
    the following modifications have to be introduced~\cite{Fischer2020PhRvC101}:
    \begin{enumerate}
        \item[(1)] the expressions of the lepton and hadron tensors are identical to the ones for the antineutrinos, as introduced in Sec.~\ref{sec:HF_approx};
        \item[(2)] replacement of the lepton Pauli blocking, $1-f_3(E_3)\rightarrow f_3(E_3)$;
        \item[(3)] the energy transfer of leptons becomes $q_0=E_1-E_3\rightarrow E_1+E_3$;
        \item[(4)] replacement in Eq.~\eqref{p1_n_tilde_H}, $E_3\rightarrow-E_3$;
        \item[(5)] the integration limits of $q_0$ are from $E_1+m_3$ to $\infty$; 
        \item[(6)] since the process considered is described by Fig.~\ref{fig:neutrino_self_energy} but with an arrow for $p_3$ in the opposite direction, which implies the usage of $G_3^<(p_3)$, it introduces a global minus sign.
    \end{enumerate}
    
    Since the integration over $q_0$ is limited from above by $E_1-m_3$ (only for $\nu_e$ and $\bar\nu_e$ absorptions), the most relevant for the behavior of $\Gamma(E_1)$ is the dependence of $d\Gamma/dq_0$ below this limit. 
    For $\nu_e$ and $\bar\nu_e$ it covers all the peaks in both approximations. The neutrino opacities within the RHF are suppressed for $\nu_e$ and enhanced for $\bar\nu_e$, compared to the RMF rates, especially to DD2, which exhibits the strongest enhancement of the $\nu_e$ opacity and the strongest suppression of the $\bar\nu_e$ opacity, respectively, due to the largest $\Delta\Sigma^0$ (see the lowest panel in Fig.~\ref{fig:HF-selfenergies}) among all RMF models investigated here.

    The magnitudes of enhancement and suppression depend on the conditions. At condition B (bottom panel), at lower density, the effect is smaller, which is associated with the shift of the effective $Q$-value---this can be identified through the drop in $\bar\nu_e$ opacity to zero---from 
    $Q_\text{TMA}^*\approx 3$~MeV, 
    $Q_\text{NL3}^*\approx 4$~MeV, and 
    $Q_\text{DD2}^*\approx 6$~MeV 
    for the RMF models to 
    $Q_{\rm RHF}^*\approx 3$~MeV for RHF, 
    compared to the higher baryon density at condition A (see Tables~\ref{tab:AB} and \ref{tab:cond}), with $Q$-values ranging from 
    $Q_\text{TMA}^*\approx 10$~MeV, 
    $Q_\text{NL3}^*\approx 14$~MeV, and
    $Q_\text{DD2}^*\approx 17$~MeV 
    for RMF models to 
    $Q_{\rm RHF}^*\approx 4$~MeV for RHF. 
    
    Moreover, in the case of the inverse neutron decay [blue lines in the left panels of Fig.~\eqref{fig:neutrino_opacity_vs_E1}], we observe a faster drop to zero of the opacity for condition B compared to condition A, with a much larger suppression within the RHF approximation, compared to the RMF approximations, among which DD2 exhibits the largest enhancement due its large $\Delta\Sigma^0$.

    \subsection{Muonic charged-current processes}
    The corresponding rates for the muonic charged-current reactions\footnote{There is no inverse neutron decay channel for muonic processes due to the large muon rest mass.},
    \begin{eqnarray}
        \nu_\mu + n \rightarrow \mu^- + p~,\qquad \bar\nu_\mu + p \rightarrow \mu^+ + n~,
    \end{eqnarray}
    are obtained by proper replacements of the electron with muon and positron with antimuon contributions in the rate expression ~\eqref{eq:total_absorption_HF}, in particular by replacing the electron rest mass with that of muons, which has the biggest impact on the energetics in terms of modifying the effective $Q$-values, due to the large muon rest mass of $m_\mu=105.658$~MeV~\cite{PDG}, in comparison to the electron rest mass. 
    The corresponding RMF opacities agree quantitatively with the data shown in Figs.~2 and 3 in Ref.~\cite{Guo2020PhRvD102} if we take the same underlying nuclear model. 
    The comparison with the RHF rates shows the same suppression (enhancement) of the $
    \nu_{\mu}(\bar\nu_\mu)$ opacity as discussed above for the electronic charged-current reactions, but the shift of the effective $Q$-value between conditions A and B is much stronger here in case of the RHF approach compared to the left panels in Fig.~\ref{fig:neutrino_opacity_vs_E1}. 
    Also, here, the magnitude depends on the conditions, ranging from 
    $Q_\text{TMA}^*\approx65$~MeV, 
    $Q_\text{NL3}^*\approx66$~MeV, and
    $Q_\text{DD2}^*\approx68$~MeV 
    for RMF models to 
    $Q_{\rm RHF}^*\approx65$ MeV 
    for RHF at condition B and from 
    $Q_\text{TMA}^*\approx47$~MeV, 
    $Q_\text{NL3}^*\approx50$~MeV, and
    $Q_\text{DD2}^*\approx53$~MeV 
    for RMF models to 
    $Q_{\rm RHF}^*\approx45$ MeV for RHF at condition A.

    \subsection{Density dependence and contributions to the hadronic vertex}
    In the following, we will analyze the impact of the different contributions to the hadronic vertex~\eqref{eq:cc_current}, i.e., the LO given by Eq.~\eqref{eq:LO}, and contributions from WM, Eq.~\eqref{eq:WM}, PS terms, Eq.~\eqref{eq:PS}, and FFs, Eq.~\eqref{form_factors}. The resulting opacity for condition B is illustrated in Fig.~\ref{fig:neutrino_opacity_vs_nB_B} as a function of the baryon density for a selection of incoming neutrino energies of $E_1=3,10,33,100$~MeV, for $\nu_e$ (top panels) and $\bar\nu_e$ (bottom panels). 
     
    To begin with, the left panels of Fig.~\ref{fig:neutrino_opacity_vs_nB_B} compare the opacity for the RMF and RHF models at the leading order of the hadronic vertex, confirming the previous analysis at the level of the neutrino energy-dependent opacity, i.e., the suppression (enhancement) of the $\nu_e(\bar\nu_e)$ opacity at low (anti)neutrino energies and intermediate densities, especially compared to DD2. However, at supranuclear densities for $\nu_e$, we see that the RHF approach starts to dominate over DD2, but becomes increasingly suppressed compared to TMA and NL3. We note here that above the saturation density, the uncertainty of the results within all the nuclear models starts to increase due to the lack of relevant experimental data.
    From Fig.~\ref{fig:neutrino_opacity_vs_nB_B}, we see that the different magnitudes of the neutrino opacity that appear with increasing density are attributed to the fact that the medium modifications due to the self-energies become increasingly larger. 
    Especially interesting is the modification of the effective medium modified $Q$-value for the $\bar\nu_e$, which systematically shifts toward higher densities, relevant especially for low and intermediate antineutrino energies. For $\nu_e(\bar\nu_e)$ within DD2, we see that its magnitude starts to decrease (increase) at supranuclear densities, especially at smaller neutrino energies [see Fig.~\ref{fig:neutrino_opacity_vs_nB_B_a} and \ref{fig:neutrino_opacity_vs_nB_B_c}]. 
    This is due to the fact that the density-dependent $\rho$ coupling constant is given in DD2 as
    \begin{equation}
        g_\rho(n_{\rm B})=g_\rho(n_0)\exp\left\{-a_\rho(x-1)\right\} \ , \label{g_rho_DD2}
    \end{equation}
    with coefficient $a_\rho$ and $x=n_{\rm B}/n_0$ \cite{typel_1999}, which, at supranuclear densities, become smaller and eventually approaches zero. This, in turn, results in $\Sigma^0_n$ and $\Sigma^0_p$ converging to one value, whereas in other RMF models, such as TMA, this phenomenon is not present, as $g_\rho$ is a constant and, therefore, the difference between $\Sigma^0_n$ and $\Sigma^0_p$ increases continuously with increasing density.

    We note here that the opacities for RMF and RHF approaches coincide toward low densities for all neutrino energies, for both $\nu_e$ and $\bar\nu_e$, which is due to the fact that the self-energies become negligible and the nuclear matter properties become indistinguishable.
    
    In the right panels of Fig.~\ref{fig:neutrino_opacity_vs_nB_B} we explore the different hadronic current contributions, expressions~\eqref{eq:LO}--\eqref{form_factors}, for varying density, but otherwise condition B (see Tables~\ref{tab:AB} and \ref{tab:cond}). 
    Compared to LO, especially at large neutrino energies, LO+WM+PS+FF increases $\nu_e$ rates, while both LO+WM and LO+WM+PS increase it even further by the same value, which confirms the findings of Ref.~\cite{Guo2020PhRvD102}. Concerning neutrino opacity for $\bar\nu_e$, we observe the largest decrease for LO+WM+PS+FF, while LO+WM and LO+WM+PS exhibit smaller but comparable decreases. Furthermore, the magnitude of the higher-order hadronic current contributions for $\nu_e$ increases with increasing density, in particular above saturation density, while for $\bar\nu_e$, the impact of those contributions becomes smaller, due to medium suppression, for all neutrino energies, especially smaller ones.

    \begin{figure*}[htp]
    \subfigure[Condition A]{\includegraphics[width=0.49\textwidth]{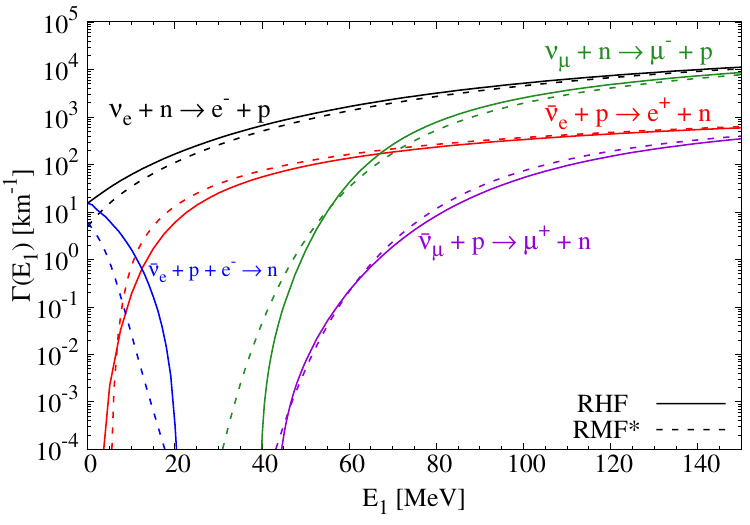}
    \label{fig:neutrino_opacity_B_RHF_a}}
    \hfill
    \subfigure[Condition B]{\includegraphics[width=0.49\textwidth]{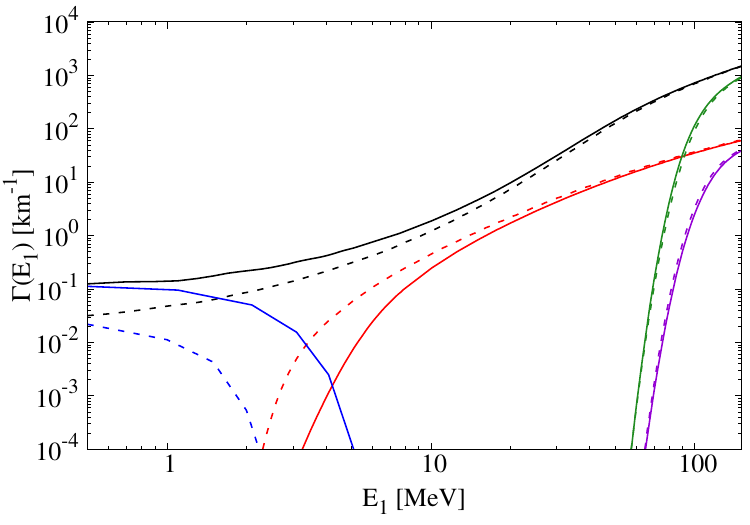}
    \label{fig:neutrino_opacity_B_RHF_b}}
    \\
    \subfigure[$\nu_e+n\rightarrow e^-+p$]{\includegraphics[width=0.49\textwidth]{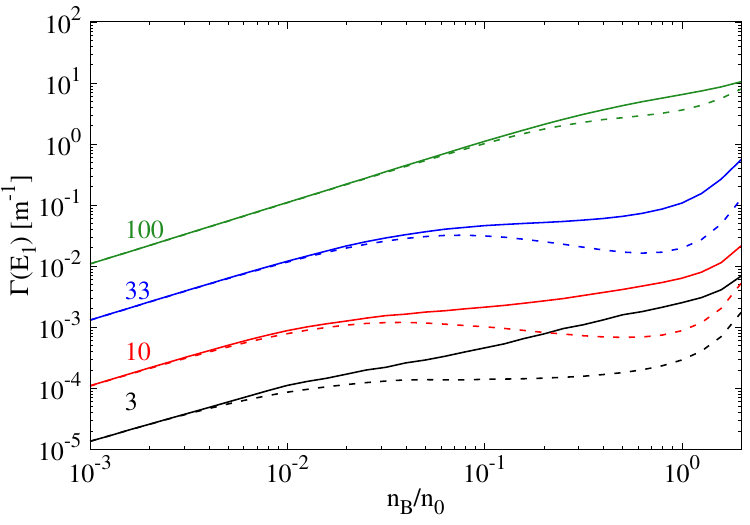}
    \label{fig:neutrino_opacity_B_RHF_c}}
    \hfill
    \subfigure[$\bar\nu_e+p\rightarrow e^++n$]{\includegraphics[width=0.49\textwidth]{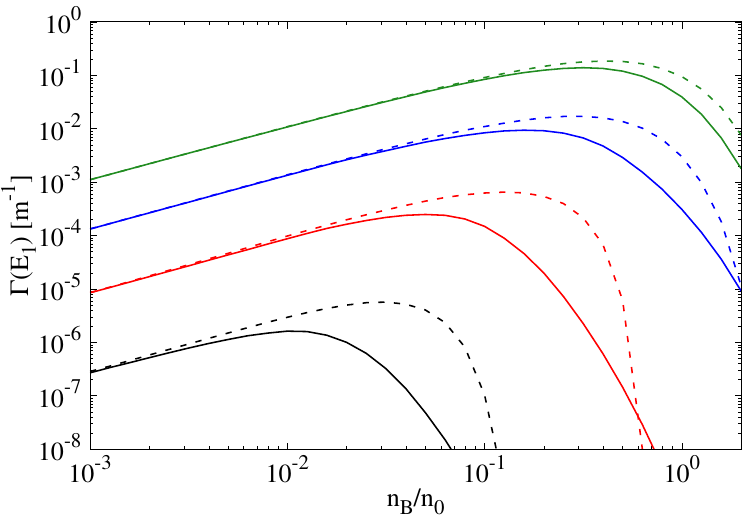}
    \label{fig:neutrino_opacity_B_RHF_d}}
    \caption{Comparison of RHF (solid lines) and the reduced RMF$^*$ (dashed lines) models. Upper panel: same as Fig.~\ref{fig:neutrino_opacity_vs_E1} for condition A in graphs~(a) and (b) and condition B in graphs~(c) and (d).
    Lower panel: same as Fig.~\ref{fig:neutrino_opacity_vs_nB_B} for condition B.}
    \label{fig:neutrino_opacity_B_RHF}
    \end{figure*}

\section{Neutrino opacity for the reduced RMF$^*$}
\label{sec:RHF_vs_reduced_RHF}
    In order to analyze the large opacity differences observed between the RMF models and RHF, we develop another RMF parametrization as a specific reduction of the RHF model. Henceforth denoted as RMF$^*$, it is based on the following mesons: $\sigma$, $\omega$, $\rho$, and the same nonlinear terms with standard meson-nucleon couplings. The resulting nonlinear RMF model contains refitted $\sigma$ and $\omega$ coupling constants in order to obtain the same nuclear saturation density and binding energy as for RHF (the saturation densities $n_0$ are given in Table~\ref{tab:Nuclear_properties} and more details are given in Sec.~\ref{sec:RHF_vs_RMF_results}). In other words, the bulk nuclear matter properties of RHF and RMF$^*$, except $n_0$ and $E_B$ fitted exactly, are not the same, namely, for RMF*, we have $m^*/M_N=0.73$, $K=102$ MeV, $E_{sym}=21.8$ MeV, and $L=58.3$ MeV. The corresponding thermodynamic chemical potentials as well as self-energies are listed in Table~\ref{tab:cond}. 
    
    It becomes evident that, while the thermodynamic neutron and proton chemical potentials are of the same magnitude, for all nuclear models and both conditions A and B, RMF$^*$ has the smallest neutron-proton timelike self-energy difference. We note further that, as for the other RMF models, also for RMF$^*$, we do not consider isospin splitting interactions due to $\delta$ mesons, and hence neutron and proton scalar self-energies are identical.
    
    The different self-energies of RMF$^*$ have important consequences for the kinematics of the neutrino opacity, shown in Fig.~\ref{fig:neutrino_opacity_B_RHF}, comparing our RHF approach (solid lines) and RMF$^*$ (dashed lines). 
    In the Figs.~\ref{fig:neutrino_opacity_B_RHF_a} and \ref{fig:neutrino_opacity_B_RHF_b}, we compare the opacity for conditions A and B, respectively, for different channels as a function of the incoming neutrino energy $E_1$. It becomes evident that the RMF$^*$ results compare qualitatively with those of RHF. In particular, the medium modified $Q$-values and the associated thresholds for $\bar\nu_e$ are very close to those of the rates within the RHF treatment, 
    $Q_{\rm RMF^*}^*\approx5$~MeV
    for condition A and 
    $Q_{\rm RMF^*}^*\approx2$~MeV
    for condition B. 
    A similar trend is observed for the muonic charged-current reactions, especially for condition B. It confirms that the deduced reduction agrees qualitatively with the RHF approach for the chosen conditions, but at the same time obtaining different nuclear properties, which highlights the importance of the momentum-dependent self-energies.
    The differences become negligible toward lower densities.
    This density dependence is illustrated in Figs.~\ref{fig:neutrino_opacity_B_RHF_c} for $\nu_e$ and \ref{fig:neutrino_opacity_B_RHF_d} for $\bar\nu_e$, for the same selection of incoming neutrino energies as before in Fig.~\ref{fig:neutrino_opacity_vs_nB_B}, comparing our RHF approach (solid lines) and RMF$^*$ (dashed lines). We omit here muonic processes for simplicity. And finally, we observe a different behavior compared to other RMF models, in which RMF* with increasing density is more suppressed compared to RHF, which we attribute to the smallest value in $\Sigma^0_n-\Sigma^0_p$ among all chosen RMF models.

\section{Impact of the $\delta$ meson}
\label{sec:RMF_delta}

    In this section, we investigate the impact of the $\delta$[$a_0$(980)] meson on neutrino opacities by incorporating them into the TMA and DD2 RMF models, which we denote as TMA$\delta$ and DD2$\delta$, respectively (cf. Appendix~\ref{app:H_approx}). The inclusion of these mesons leads to a splitting of the scalar self-energies of neutrons and protons, as given by Eq.~\eqref{Sigma_S_H_final}, and therefore we have $\Sigma^S_n\neq\Sigma^S_p$, which, in general, resembles the situations of RHF.
    
    In TMA$\delta$, the $\delta$-meson coupling only affects the slope of the symmetry energy (see Table~\ref{tab:Nuclear_properties}).
    In order to reproduce nuclear saturation properties, i.e., $E_{sym}$ at $n_0$, we refit the $\rho$-meson coupling, which is increased by a factor of about 2 for TMA$\delta$, $g_\rho=6.789$, for the following $\delta$-meson coupling and mass, $g_\delta=7.852$ and $m_\delta=980$~MeV, respectively \cite{Greco:2002PhRvC65_delta-meson}. The remaining coupling constants and rest masses are identical to those in TMA \cite{toki_1995}.
    
    In DD2$\delta$, the mass of the $\delta$ meson is taken to be the same as that used in TMA$\delta$. Only very small changes in the existing parameters were observed after refitting the model to the nuclear properties at saturation density. In particular, the coupling parameter for the $\rho$ meson is modified according to $g_{\rho}(\text{DD2}\delta) = g_\rho(\rm DD2)\times1.0223=3.707821$, while the $\delta$-meson coupling is set to $g_\delta=g_\rho(\rm DD2)\times0.3=1.088082$. The density dependence of the $\delta$-meson coupling is assumed to have the same functional form as that of the $\rho$ meson \cite{typel_1999} given by Eq.~\eqref{g_rho_DD2}, where the parameter $a_{\delta}$ is obtained by constraining the slope of the symmetry energy and is given by 
    $a_\delta= a_\rho(\rm DD2)/1.35=0.3844$, while the $\rho$-meson parameter is kept unchanged,
    $a_\rho(\rm DD2\delta)=a_\rho(\rm DD2)$. These newly fitted parameters are close to those obtained in density-dependent RMF models with tensor couplings, where both nuclear matter and finite-nucleus properties are constrained \cite{2020EPJA...56..160T}.

    \begin{table*}[t!]
    \caption{Same as Table~\ref{tab:cond}, but depicting only TMA$\delta$ and DD2$\delta$ with different neutron and proton scalar self-energies, $\Sigma^S_n$ and $\Sigma^S_p$, respectively, which for the RMF models without $\delta$ mesons are equal to each other.}
    \begin{ruledtabular}
    \begin{tabular}{llccdddd}
        & Model & $\mu_n$ & $\mu_p$ &  \multicolumn{1}{c}{$\Sigma^S_n$} &  \multicolumn{1}{c}{$\Sigma^S_p$} & \multicolumn{1}{c}{$\Sigma^0_n$} & \multicolumn{1}{c}{$\Sigma^0_p$} \\
        \colrule
        \multirow{2}{*}{A} & TMA$\delta$ & 896.4 & 814.9 & -171.30 & -135.49 & 144.66 & 97.37 \\ 
        & DD2$\delta$ & 904.8 & 811.3 & -202.94 & -201.85  & 181.39 & 154.66 \\[5pt]
        \multirow{2}{*}{B} & TMA$\delta$ & 914.6 & 879.2 & -24.77 & -18.95 & 19.77 & 12.51 \\
        & DD2$\delta$ & 915.7 & 876.1 & -33.55 & - 33.32 & 29.17 & 23.25
    \end{tabular}
    \label{tab:cond_delta_mesons}
    \end{ruledtabular}
    \end{table*}
    
    In Table~\ref{tab:cond_delta_mesons}, we provide the chemical potentials and the values of the self-energies for conditions A and B, in which, as opposed to Table~\ref{tab:cond}, we have two different $\Sigma^S_n\neq\Sigma^S_p$. We note that the inclusion of the $\delta$ meson alters the TMA self-energies significantly and introduces a large splitting in the scalar self-energies for both conditions A and B, whereas in DD2$\delta$ the effect is almost negligible.

\begin{figure*}[t!]
    \subfigure[$T=0$]{\includegraphics[width=0.49\textwidth]{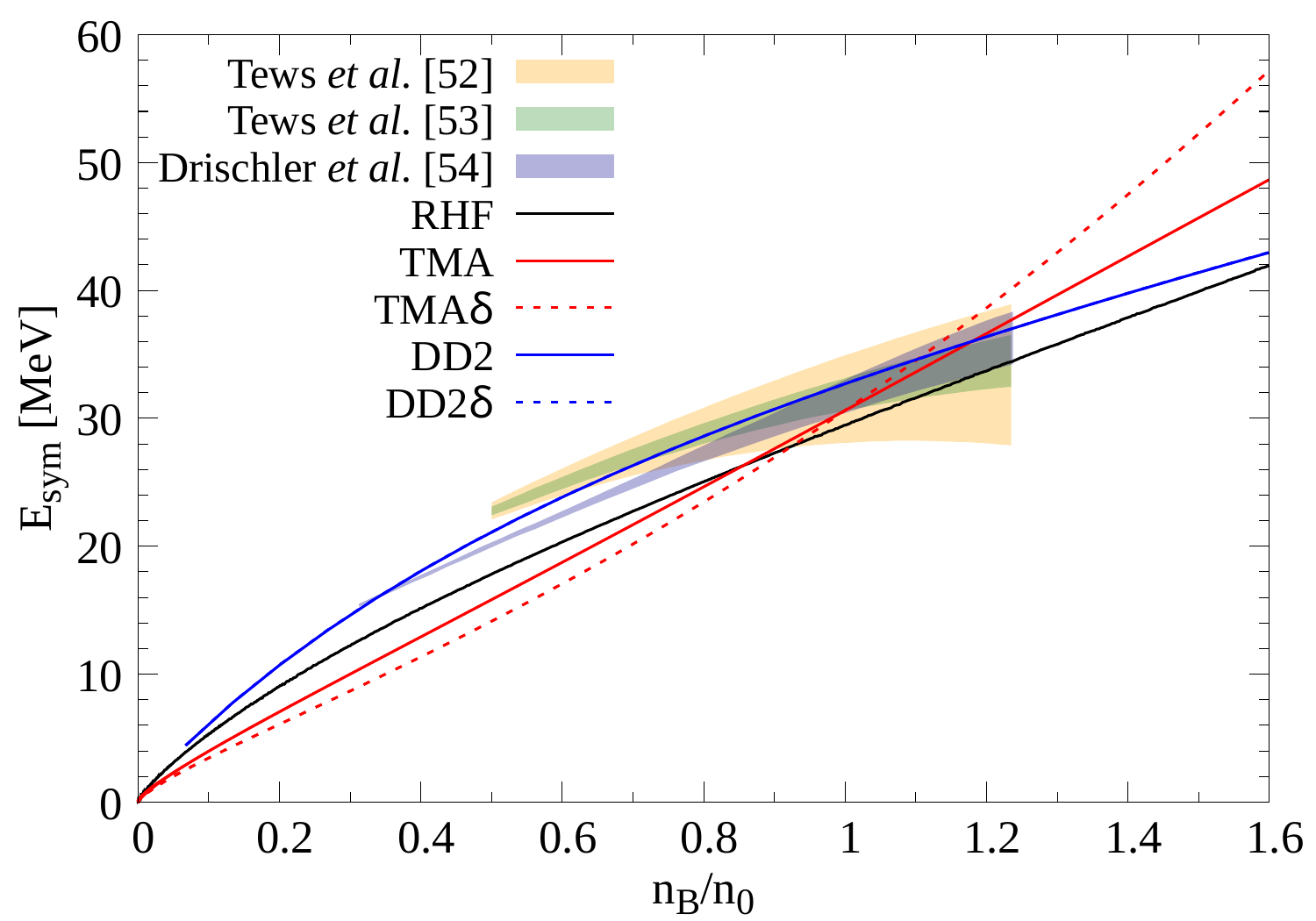}\label{fig:E_sym}}
    \subfigure[Condition A]{\includegraphics[width=0.49\textwidth]{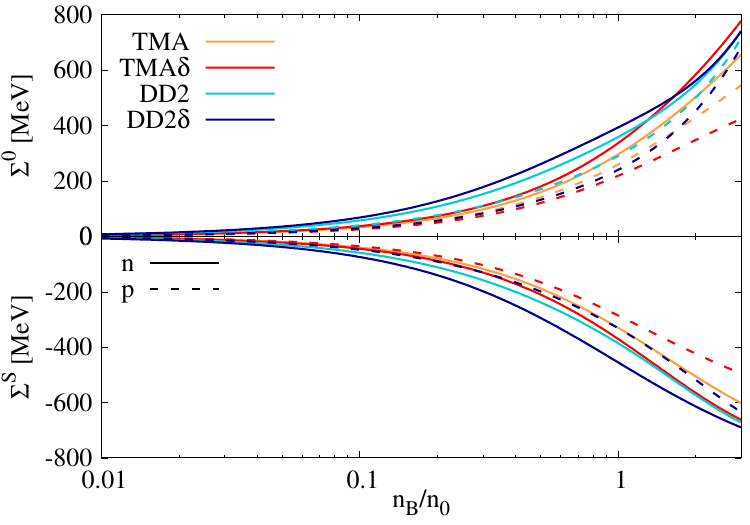}\label{fig:Self_energies_delta}} \\
    \subfigure[Condition A, $(\nu_e,\bar\nu_e)$]{\includegraphics[width=0.49\textwidth]{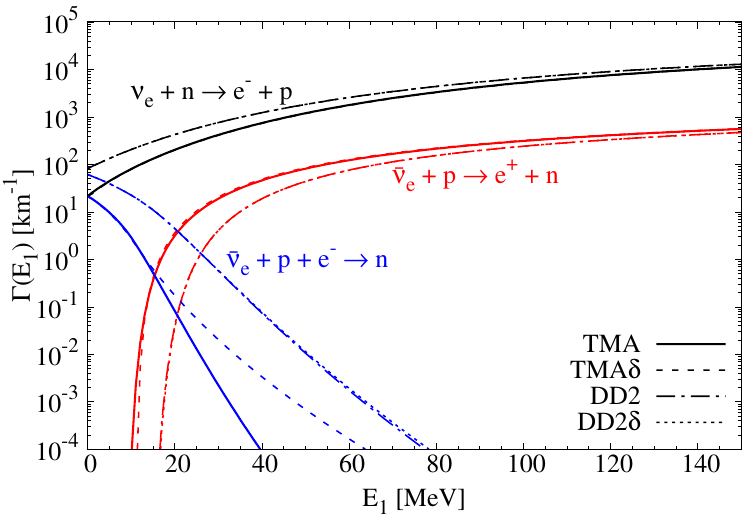}\label{fig:Neutrino_opacity_A_electrons_RMF_delta}}
    \subfigure[Condition A, $(\nu_\mu,\bar\nu_\mu)$]{\includegraphics[width=0.49\textwidth]{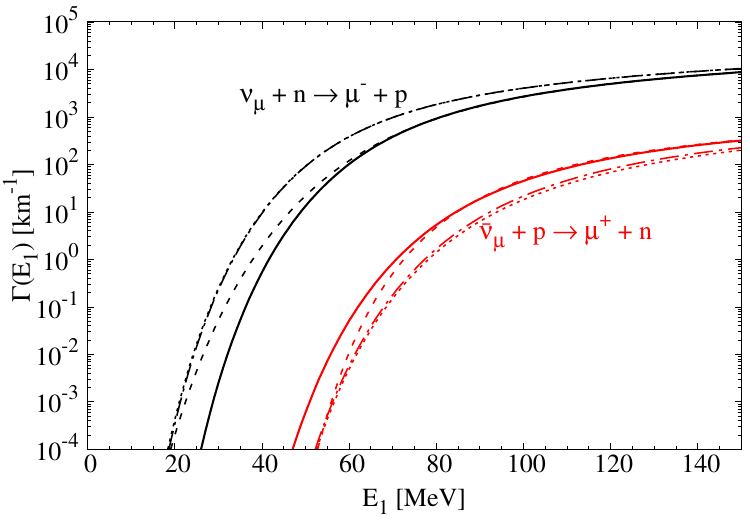}\label{fig:Neutrino_opacity_A_muons_RMF_delta}}
    \caption{ 
    Top panels EOS comparison, showing the nuclear symmetry energy (left panel) as a function of the normalized baryon density, $n_{\rm B}$, in units of the saturation density, $n_0$, comparing RHF (black solid line), TMA (red solid line), TMA$\delta$ (red dashed line), DD2 (blue solid line), and DD2$\delta$ (blue dashed line), with the colorful bands for selected recent constraints from $\chi$-effective field theory at the N$^2$LO with lower (yellow band) and larger cutoffs (green band) \cite{tews_2018,tews_2025}, and at the N$^3$LO (blue band) \cite{drischler_2019}, as well as scalar, $\Sigma^S$, and timelike, $\Sigma^0$, self-energies (right panel) for neutrons (solid lines) and protons (dashed lines), with respect to the normalized baryon density comparing TMA (orange line), TMA$\delta$ (red line), DD2 (light blue line), and DD2$\delta$ (dark blue line). 
    In TMA and DD2, we have only one scalar self-energy, which is denoted by one solid line. 
    Bottom panels: the same as Figs.~\ref{fig:neutrino_opacity_vs_E1_a} and \ref{fig:neutrino_opacity_vs_E1_b} for condition A, comparing here TMA (solid line), TMA$\delta$ (dashed line), DD2 (dash-dotted line), and DD2$\delta$ (dotted line).
    \label{fig:RMF_delta}}
    \end{figure*}

    In Fig.~\ref{fig:E_sym}, we present the nuclear symmetry energy of TMA and DD2 with and without $\delta$ meson, in comparison to the RHF model. Moreover, we include the recent constraints from $\chi$-effective field theory at  N$^2$LO with lower \cite{tews_2018} and larger cutoffs \cite{tews_2025} as well as at N$^3$LO \cite{drischler_2019}. 
    We observe that the inclusion of the $\delta$ meson decreases TMA$\delta$'s $E_{sym}$ below $n_0$ and increases it above $n_0$, compared to TMA, which is attributed to the large change of the slope of $E_{\rm sym}$ (see Table~\ref{tab:Nuclear_properties}). 
    The situation is contrary for DD2$\delta$, for which $E_{sym}$ is almost identical to that of DD2 at all densities, as the $\rho$-meson coupling parameter is only slightly modified by $2.23\%$, and the newly introduced $\delta$ meson affects mostly the slope of the symmetry energy, however, only by about 1~MeV (see Table~\ref{tab:Nuclear_properties}), rather than $E_{\rm sym}$.
    We note here that there might well be other DD2$\delta$ parametrizations possible, fitted to nuclear saturation properties, which can have a larger impact on $E_{\rm sym}$. However, large variations might become problematic since DD2 matches all current nuclear matter constraints quantitatively among RMF models and a large variation of $L$ would potentially conflict those.

    In Fig.~\ref{fig:Self_energies_delta}, we compare the self-energies for TMA (orange line) and DD2 (light blue line) with their $\delta$-meson counterparts, TMA$\delta$ (red line) and DD2$\delta$ (dark blue line). We confirm that the inclusion of the $\delta$ meson splits the scalar self-energies into $\Sigma^S_n$ (solid line) and $\Sigma^S_p$ (dashed line). Compared to the reference TMA and DD2 models, $\Sigma^S_n$ has generally smaller values and $\Sigma^S_p$ has larger values (note that $\Sigma^S<0$ for the chosen sign convention here). 
    Moreover, since the $g_\rho$ and $g_\delta$ are constants in TMA$\delta$, we see that the splittings in both scalar (bottom panel) and timelike self-energies (top panel) consistently increase with increasing density, whereas in DD2$\delta$, they start to converge to one value at supranuclear densities, which is attributed to the fact that both $g_\rho$ and $g_\delta$ are density dependent, expressed as an exponential function, which becomes smaller above $n_0$.

    In Figs.~\ref{fig:Neutrino_opacity_A_electrons_RMF_delta} and \ref{fig:Neutrino_opacity_A_muons_RMF_delta}, we analyze the impact of the inclusion of the $\delta$-meson channel on the neutrino opacity in terms of the incoming (anti)neutrino energies for condition A for electronic charged-current processes [Fig.~\ref{fig:Neutrino_opacity_A_electrons_RMF_delta}] and for the muonic charged-current processes [Fig.~\ref{fig:Neutrino_opacity_A_muons_RMF_delta}]. 
    Surprisingly, the resulting opacities are nearly indistinguishable comparing TMA (solid lines) with TMA$\delta$ (dashed lines) and DD2 (dash-dotted lines) with DD2$\delta$ (dotted lines).  
    For TMA$\delta$, for the muonic charged-current reactions, we find a slight enhancement at smaller neutrino energies and suppression at smaller antineutrino energies. 
    However, in contrast to the RHF model, we see that the inclusion of the $\delta$ meson in TMA and DD2, and hence explicit isospin splitting between neutron and proton scalar self-energies, cannot account for the observed reduction of the medium modifications of the charged-current weak rates. We attribute this phenomenon to the explicit momentum dependence of the RHF expressions, that cannot be captured within RMF models by design.

\section{Summary and Conclusions}
\label{sec:summary}

    In this article, we have performed novel calculations of the neutrino opacity for both electronic and muonic charged-current processes \eqref{eq:cc_process}, with the inclusion of the inverse neutron decay, extending the previous calculations within the relativistic Hartree mean-field models \cite{reddy_1998,roberts_2017,Guo2020PhRvD102} by incorporating momentum-dependent nucleon self-energies within the relativistic Hartree-Fock approach, including contributions from weak magnetism, pseudoscalar terms, and weak form factors. 
    We derived the weak rate expressions from the neutrino self-energy within the RHF framework, for use in numerical calculations. We benchmarked the numerical results against the widely used RMF approach~\cite{roberts_2017,Guo2020PhRvD102}.
    
    The calculations show a consistent suppression of the neutrino opacity for $\nu_e$ and $\nu_\mu$ within the RHF approach, compared to the RMF approximation, for both high and low temperatures and densities.
    For antineutrinos, $\bar\nu_e$ and $\bar\nu_\mu$, we observe the opposite tendency, i.e., an enhancement of the RHF opacity in comparison to the RMF results. Concerning the inverse neutron decay, we see that it is 
    suppressed within RHF compared to RMF, especially DD2. 
    Furthermore, we perform a detailed analysis of different contributions of the hadron current, showing that they play a vital role especially at high (anti)neutrino energies.
    
    In addition, the momentum dependence shows that the observed large differences between RHF and RMF models is indeed non-negligible for a wide range of temperatures and densities, 
    which we tested against the DD2 reference RMF model as well as by implementing a {\em reduced} RMF* model. 
    In the former case, the nuclear matter properties agree with those of RHF, but the neutrino rates differ largely, while in the latter case, the nuclear matter properties differ largely but the neutrino rates are very similar.
    These analyses confirm that the large differences reported here for the neutrino opacity comparing RHF and RMF are not entirely attributed to different nuclear models and nuclear matter properties.
    Besides additional terms in the neutrino rate expressions within RHF, compared to RMF, which have been commonly employed in astrophysical simulations~\cite{Hempel2010NuPhA837,Hempel12,Steiner2013ApJ765L}, we attribute the opacity differences between RHF and RMF to a large extent to momentum-dependent self-energies and the inclusion of the Fock terms. 
    In general, it is well known that besides the $\rho$ meson, the isovector $\delta$-meson interaction channel gives rise to large isovector optical potential (cf. Ref.~\cite{li_2018} for a comprehensive review), which is related to the scalar self-energy differences between neutrons and protons, being on the same order as the timelike self-energy differences between neutrons and protons. It is this very interaction channel that gives rise to effective mass spitting of neutrons and protons, which is naturally included in the RHF model as well as in other quantum hadrodynamics approaches, such as (Dirac-)Brueckner-Hartree-Fock~\cite{Wang:2023PhRvC108_delta-meson,Lenske:2001PhRvC64_delta-meson}.
    To this end, we examined the inclusion of $\delta$ mesons in the TMA and DD2 RMF models, for which we developed the nuclear matter parametrizations here, denoted as TMA$\delta$ and DD2$\delta$, and found, despite largely different scalar self-energies between neutrons and protons, only a negligible impact on the neutrino opacity compared to TMA and DD2, for the conditions explored here. These findings highlight the importance of the momentum dependence and associated Fock terms in the weak rate calculations. 
    We note here that the inclusion of the $\delta$ meson within RHF was recently investigated in Ref.~\cite{li_2026}, which can further contribute to the isospin splitting of the scalar self-energies.
    
    The here reported reduction of the rate differences between $\nu_e$ and $\bar\nu_e$, comparing the RHF treatment with the commonly employed RMF approach, due to the inclusion of neutron contributions to the proton scalar and timelike self-energy and vice versa within the RHF framework, along with the momentum dependence (see Appendix~\ref{app:Hartree-Fock}), might have important consequences on the differences between $\nu_e$ and $\bar\nu_e$ fluxes and spectra in simulations of core-collapse supernovae, and in particular the long-term proto-neutron star deleptonization phase when the neutrinospheres continuously shift to increasing densities~\cite{Fischer12}. We leave this important application of the here developed novel rates in simulations of core-collapse supernovae and binary neutron star mergers for future investigation.

    Lastly, the calculations of the neutrino opacity in asymmetric nuclear matter, within RHF, do not conserve the current  condition of the vector component of the polarization tensor expression, Eq.~\eqref{Pi_free}, i.e., $q^\mu\,\Pi^V_{\mu\nu}=0$. 
    This deficit was pointed out previously in Ref.~\cite{roberts_2017}, based in the RMF approach. 
    We further note that in our present approach, many-body correlations to the neutrino rates are absent. It has been shown that these might have a non-negligible impact (cf. Refs.~\cite{Reddy99,burrows_1999}, employing the random phase approximation).
    We leave this investigation to future work.
 
\begin{acknowledgments}
    We wish to thank Gabriel Mart\'{i}nez-Pinedo for helpful discussions and for carefully reading the manuscript, as well as Pok Man Lo and Micaela Oertel for insightful comments. 
    This work was supported by the Polish National Science Center (NCN) under Grant Nos. 2023/49/B/ST9/03941 (K.S. and T.F.) and 2022/47/D/ST9/03092 (K.S.). 
    The computations were performed at the Wroc\l{}aw Centre for Networking and Supercomputing (WCSS).
\end{acknowledgments}

\section*{DATA AVAILABILITY}
There are no publicly available research data or software supporting this manuscript. Requests for further information or data should be sent to the authors.

\appendix

\section{
\label{app:Hartree-Fock} 
Momentum-dependent self-energies within the RHF framework
}
    The nuclear matter that is considered in the current paper in the calculations of the neutrino opacity is composed of two nucleons---protons ($p$) and neutrons ($n$)---described by the Dirac isodoublet
    $\psi=\begin{pmatrix}
            \psi_p \\
            \psi_n
    \end{pmatrix}$,
    where the strong interaction is mediated by the exchange of the mesons~\cite{bouyssy_1987}: 
    two isoscalar-scalar $\sigma$ and isoscalar-vector $\omega$ mesons, and two isovector-scalar $\pi$ and isovector-vector $\rho$ mesons.  The Lagrangian can be decomposed as follows:
    \begin{equation}\label{L}
    \mathcal{L}=\mathcal{L}_0+\mathcal{L}_I+\mathcal{L}_{NL} \ .
    \end{equation}
    The free part is given by the following Lagrangian:
    \begin{align}\label{L_0}
            \mathcal{L}_0=& \sum_{a=p,n}\bar{\psi}_a(i\gamma^\mu\partial_\mu-m_a)\psi_a+\frac{1}{2}(\partial_\mu\sigma\partial^\mu\sigma-m_\sigma^2\sigma^2) \nonumber \\
            &-\frac{1}{4}\omega_{\mu\nu}\omega^{\mu\nu}+\frac{1}{2}m_\omega^2\omega_\mu\omega^\mu+\frac{1}{2}(\partial_\mu\boldsymbol{\pi}\cdot\partial^\mu\boldsymbol{\pi}-m_\pi^2\boldsymbol{\pi}^2) \nonumber \\
            &-\frac{1}{4}\boldsymbol{\rho}_{\mu\nu}\cdot\boldsymbol{\rho}^{\mu\nu}+\frac{1}{2}m_\rho^2\boldsymbol{\rho}_\mu\cdot\boldsymbol{\rho}^\mu \ ,
    \end{align}
    where the index $a=\{p,n\}$ refers to isospin and the field tensors are $\omega_{\mu\nu}=\partial_\mu\omega_\nu-\partial_\nu\omega_\mu$ and $\boldsymbol{\rho}_{\mu\nu}=\partial_\mu\boldsymbol{\rho}_\nu-\partial_\nu\boldsymbol{\rho}_\mu$. 
    The parameters in this model are the rest masses $m_p$, $m_n$, $m_\sigma$, $m_\omega$, $m_\pi$, and $m_\rho$, of the fields $\psi_p$, $\psi_n$, $\sigma$, $\omega_\mu$, $\pi^i$, and $\rho^i_\mu$, respectively. Note that there are three isovector mesons, i.e., $\pi^i=\{\pi^0,\pi^+,\pi^-\}$ and $\rho^i_\mu=\{\rho^0_\mu,\rho^+_\mu,\rho^-_\mu\}$.
    
    The interacting part of $\mathcal{L}$ is given as follows:
    \begin{align}
        \mathcal{L}_I=&\;\bar\psi\bigg(-g_\sigma\sigma-g_\omega\gamma^\mu\omega_\mu-\frac{f_\pi}{m_\pi}\gamma_5\gamma^\mu\partial_\mu\boldsymbol{\pi}\cdot\boldsymbol{\tau} \nonumber \\
        &-g_\rho\gamma^\mu\boldsymbol{\rho}_\mu\cdot\boldsymbol{\tau}+\frac{f_\rho}{2m_n}\sigma_{\mu\nu}\partial^\nu\boldsymbol{\rho}^\mu\cdot\boldsymbol{\tau}\bigg)\psi \ , \label{L_int}
    \end{align}
    where $\tau^i$ are the regular isospin Pauli matrices. In addition to the masses, here, the parameters are the meson-nucleon coupling constants $g_\sigma$, $g_\omega$, $g_\rho$, as well as $f_\pi$, $f_\rho$. The latter are the tensor coupling constants. We note here that there is an additional tensor coupling, $f_\omega$, which is related to the isoscalar anomalous magnetic moment of the nucleon. However, it possesses a small value, compared to the other tensor couplings, and is, therefore, omitted in the present calculations \cite{bouyssy_1987}. 
    For the pions, we choose the pseudovector representation.
    Despite the resulting Lagrangian being nonrenormalizable, this choice avoids too strong $\pi N$ interactions and nonphysical modifications of the nucleon spectrum at normal densities~\cite{matsui_1982}.

    For the nonlinear part of $\mathcal{L}$, we follow Refs.~\cite{bernardos_1993,savushkin_1997},
    \begin{equation}
        \mathcal{L}_{NL}=\frac{1}{2}m_\sigma^2\sigma^2-\frac{1}{2}m_\omega^2\omega_\mu\omega^\mu-U(\sigma,\omega) \ , 
        \label{L_NL}
    \end{equation}
    with the following potential-energy functional~\cite{boguta_1977,bodmer_1991,sugahara_1994}:
    \begin{align} \label{U_sw}
        U(\sigma,\omega)=&\;\frac{1}{2}m_\sigma^2\sigma^2+\frac{1}{3}g_3\sigma^3+\frac{1}{4}g_4\sigma^4  \nonumber \\
        &-\frac{1}{2}m_\omega^2\omega_\mu\omega^\mu-\frac{1}{4}c_4(\omega_\mu\omega^\mu)^2 \ ,
    \end{align}
    with $g_3$, $g_4$, and $c_4$ additional meson-meson coupling constants.
    It is furthermore common to introduce squared effective masses as follows:
    \begin{align}
        \hat m_\sigma^{*2}&=\frac{1}{\sigma}\frac{\partial U(\sigma,\omega)}{\partial\sigma}=m_\sigma^2+g_3\sigma+g_4\sigma^2 \ , \label{mhat_eff_s} \\
        \hat m_\omega^{*2}&=-\frac{1}{\omega_\alpha}\frac{\partial U(\sigma,\omega)}{\partial\omega^\alpha}=m_\omega^2+c_4\omega_\mu^2 \ , 
        \label{mhat_eff_w}
    \end{align}
    such that the Lagrangian~\eqref{L} features the following equations of motion for $\sigma$ and $\omega$ mesons:
    \begin{align}
        (\partial_\mu^2+\hat m_\sigma^{*2})\sigma&=-g_\sigma\bar\psi\psi \ , \label{eom_sigma} \\
       \partial_\mu\omega^{\mu\nu}+\hat m_\omega^{*2}\omega^\nu&=g_\omega\bar\psi\gamma^\nu\psi \ . 
        \label{eom_omega}
    \end{align}
    For the calculation of the nucleon self-energies, using the newly defined effective meson masses---these are operators---it is common to introduce the mean-field values of the $\sigma$ and $\omega$ mesons, e.g., defined in Ref.~\cite{reinhard_1989},
    \begin{align}
        \sigma\rightarrow\langle\sigma\rangle&\equiv\sigma_0 \ , \label{sigma_0} \\
       \omega_\mu\rightarrow\langle\omega_\mu\rangle&\equiv\delta_{\mu0}\omega_0 \ , \label{omega_0}
    \end{align}
    where $\delta_{\mu0}$ limits the meson fields to the timelike components since the spacelike contributions vanish on account of the rotation invariance. Applying the mean-field values, Eqs.~\eqref{mhat_eff_s} and \eqref{mhat_eff_w} become
    \begin{align}
        m_\sigma^{*2}&=m_\sigma^2+g_3\sigma_0+g_4\sigma_0^2 \ , \label{m_eff_s} \\
        m_\omega^{*2}&=m_\omega^2+c_4\omega_0^2 \ , \label{m_eff_w}
    \end{align}
    where $\sigma_0$ and $\omega_0$ are obtained from the following gap equations:
    \begin{align}
        m_\sigma^{*2}\sigma_0&=-g_\sigma\langle\bar\psi\psi\rangle \ , \label{eom_sigma_0} \\
        m_\omega^{*2}\omega_0&=g_\omega\langle\bar\psi\gamma^0\psi\rangle \ ,\label{eom_omega_0}
    \end{align}
    with scalar density, $\langle\bar\psi\psi\rangle=n_S$, given by Eq.~\eqref{nS_HF}, and baryon density, $\langle\bar\psi\gamma^0\psi\rangle=n_{\rm B}$, given by Eq.~\eqref{nB_HF}.

    Finally, with the effective masses given by Eqs.~\eqref{m_eff_s} and \eqref{m_eff_w}, the Green's functions of the $\sigma$ and $\omega$ fields satisfy the following equations:
    \begin{align}
        &\left(\partial_\mu^2+m_\sigma^{*2}\right)i\Delta^0(x-y)=-i\delta^{(4)}(x-y) \ , 
        \label{Delta_NL} \\
        &\left[\left(\partial_\mu^2+m_\omega^{*2}\right)g^{\mu\nu}-\partial^\mu\partial^\nu\right]iD_{\mu\nu}^0(x-y)=i\delta^{(4)}(x-y) \ ,
        \label{D_NL}
    \end{align}
    with the Green's functions for $\sigma$, $i\Delta^0$, and $\omega$, $iD_{\mu\nu}^0$. We note that these Green's functions are identical to the ones for the standard scalar and vector fields without the self-interacting terms. Therefore, in the final expressions of the nucleon self-energies, the free meson rest masses $m_\sigma$ and $m_\omega$ have to replaced by $m^*_\sigma$ and $m^*_\omega$, respectively.
    
\begin{figure}[t!]
        \includegraphics[width=0.36\textwidth]{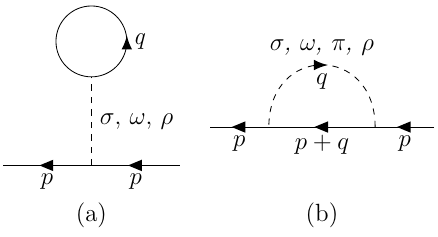}
        \caption{\label{fig:Hartree_and_Fock} Interacting Dirac Green's function at the second order, where a) is the Hartree (tadpole) diagram and b) is the Fock (exchange) diagram. Solid lines are described by the Dirac propagator and the dashed lines by the meson propagators.}
    \end{figure}

    In terms of the Dirac Green's functions, the relativistic Hartree-Fock approximation is based on the two Feynman diagrams at the second order presented in Fig.~\ref{fig:Hartree_and_Fock} \cite{chin_1977,horowitz_1983,serot_1986}. Due to the conservation of momentum at the vertices, the meson momentum is equal to zero in the Hartree diagram, Fig.~\ref{fig:Hartree_and_Fock}(a), hence the $\pi$, the tensor $\rho$, and the vector-tensor $\rho$ terms in the interacting Lagrangian will not contribute. Moreover, when considering the mean-field approximation, both $\pi^\pm$ and $\rho^\pm$ mesons are not present, as they would change the isospin, which is prohibited by charge conservation at the vertex. In the case of the Fock diagram, Fig.~\ref{fig:Hartree_and_Fock}(b), all the mesons are involved. Additionally, in our considerations, we neglect the contribution of the retardation terms, i.e., $\propto E(p)-E(q)$, which describe the exchange of the nucleon energy between two vertices in the Fock diagram, Fig.~\ref{fig:Hartree_and_Fock}(b).

    For both $\rho$ and $\pi$ mesons, we use the following decomposition of the inner products in the Lagrangian~\cite{bjorken_1964}:
    \begin{align}
    \boldsymbol{\tau}\cdot\boldsymbol{\pi}&=\tau^3\pi^0+\sqrt{2}(\tau^+\pi^++\tau^-\pi^-) \ , 
    \label{inner_prod_pi} \\
    \boldsymbol{\tau}\cdot\boldsymbol{\rho}_\mu&=\tau^3\rho^0_\mu+\sqrt{2}(\tau^+\rho^+_\mu+\tau^-\rho^-_\mu) \ , 
    \label{inner_prod_rho}
    \end{align}
    where $\tau^\pm=\frac{1}{2}(\tau^1\pm i\tau^2)$ and  $(\tau^i)^2=(\tau^3)^2+2\tau^+\tau^-+2\tau^-\tau^+$. 
    Note that for the charged mesons exchanged, an additional factor of $\sqrt{2}$ arises at a vertex. 
    If no mixing of the nucleons is considered, as in the case of symmetric nuclear matter, only the first term with $\tau^3$ will contribute to the Feynman rules, with $+1$ for protons and $-1$ for neutrons.

    The momentum-dependent nucleon self-energies are decomposed as in Eq.~\eqref{Sigma_HF} \cite{bouyssy_1987} and their explicit forms for asymmetric nuclear matter \cite{sun_2008} for nonzero temperature \cite{soutome_1990,yang_2019,yang_2021} are given as follows (note the explicit appearance of the effective $\sigma$ and $\omega$ masses, $m_\sigma^*$ and $m_\omega^*$, respectively):
\begin{widetext}
    \begin{align}
        \Sigma_a^S(p)=&-\left(\frac{g_\sigma}{m^*_\sigma}\right)^2n_S+\frac{1}{(4\pi)^2}\frac{1}{p}\Bigg\{\int_0^\infty dqq f_a(E_a)\Bigg[\hat M_{a}(q)\sum_{i\in\mathcal{A}}B_{i,aa}(p,q)+\frac{1}{2}\hat P_{a}(q)D_{aa}(q,p)\Bigg] \nonumber \\
        &+2\int_0^\infty dqq f_b(E_b)\Bigg[\hat M_{b}(q)\sum_{j\in\mathcal{B}}B_{j,ab}(p,q)+\frac{1}{2}\hat P_{b}(q)D_{ab}(q,p)\Bigg]\Bigg\} \ , \label{Sigma_S_HF_full} \\
        \Sigma_a^0(p)=&\left(\frac{g_\omega}{m^*_\omega}\right)^2 n_{\rm B}\pm\left(\frac{g_\rho}{m_\rho}\right)^2n_I+\frac{1}{(4\pi)^2}\frac{1}{p}\Bigg\{\int_0^\infty dqqf_a(E_a)\sum_{i\in\mathcal{A}}A_{i,aa}(p,q) \nonumber \\
        &+2\int_0^\infty dqqf_b(E_b)\sum_{j\in\mathcal{B}}A_{j,ab}(p,q)\Bigg\} \ , \label{Sigma_0_HF_full} \\
        \Sigma_a^V(p)=&\,\frac{1}{(4\pi)^2}\frac{1}{p}\Bigg\{\int_0^\infty dqqf_a(E_a)\Bigg[\hat P_{a}(q)\sum_{i\in\mathcal{A}}C_{i,aa}(p,q)+\frac{1}{2}\hat M_{a}(q)D_{aa}(p,q)\Bigg] \nonumber \\
        &+2\int_0^\infty dqqf_b(E_b)\Bigg[\hat P_{b}(q)\sum_{j\in\mathcal{B}}C_{j,ab}(p,q)+\frac{1}{2}\hat M_{b}(q)D_{ab}(p,q)\Bigg]\Bigg\} \ , \label{Sigma_V_HF_full}
    \end{align}
\end{widetext}
    where $a=\{p,n\}$, $b=\{n \ \text{if} \ a=p \ \text{and} \ p \ \text{if} \ a=n\}$ (which accounts for the nucleon mixing), $p=|\mathbf{p}|$, $q=|\mathbf{q}|$, $+$ is for protons, $-$ for neutrons, $f_{a/b}(E_{a/b})$ is the Fermi-Dirac distribution function with the expression given in the main text, where the single-particle energy $E_{a/b}$ is found via Eq.~\eqref{E_eff_HF}, and we introduce the following sets: $\mathcal{A}=\{\sigma,\omega,\pi,\rho_V,\rho_T\}$ and $\mathcal{B}=\{\pi,\rho_V,\rho_T\}$, with the following notations:
    \begin{align}
        \hat P_a(p)&\equiv\frac{p^*_a(p)}{E^*_a(p)} \ , \label{hat_P} \\ 
        \hat M_a(p)&\equiv\frac{m^*_a(p)}{E^*_a(p)} \ , \label{hat_M}
    \end{align}
    and the baryon, scalar, and isospin densities are expressed as
    \begin{align}
        n_{\rm B} & \equiv\sum_a\frac{1}{\pi^2}\int_0^\infty dqq^2f_a(E_a)=n_p+n_n \ , \label{nB_HF} \\
        n_S&\equiv\sum_a\frac{1}{\pi^2}\int_0^\infty dqq^2f_a(E_a)\hat M_a(q) \label{nS_HF} \ , \\
        n_I&\equiv n_p-n_n \label{nI_HF} \ .
    \end{align}
    All functions $A$, $B$, and $C$ are given explicitly in Table~\ref{tab:HF_functions} with the following functions:
    \begin{align}
        \theta_{i,ab}(p,q)&\equiv\ln\left|\frac{\tilde A_{i,ab}(p,q)+2pq}{\tilde A_{i,ab}(p,q)-2pq}\right| \ , \label{theta_HF} \\
        \Phi_{i,ab}(p,q)&\equiv\frac{\tilde A_{i,ab}(p,q)}{4pq}\theta_{i,ab}(p,q)-1 \ , \label{phi_HF} \\
        \tilde A_{i,ab}(p,q)&\equiv p^2+q^2+m_i^2 \ , \label{A_tilde_HF}
    \end{align}
    where $i=\{\sigma,\omega,\pi,\rho\}$, and with $i=\sigma,\omega$, we have $m^*_\sigma$ and $m^*_\omega$, respectively.

    \begin{table*}[t]
    \caption{\label{tab:HF_functions} RHF functions, which appear in Eqs.~\eqref{Sigma_S_HF_full}--\eqref{Sigma_V_HF_full}, where $p=|\mathbf{p}|$, $q=|\mathbf{q}|$, and the other functions are defined in Eqs.~\eqref{theta_HF}--\eqref{A_tilde_HF}, where all $(p,q)$ and $ab$ dependencies are omitted for clarity.}
    \begin{ruledtabular}
    \begin{tabular}{cccc}
        \textrm{i} & $A_i$ & $B_i$ & $C_i$ \\
        \colrule
        \\[-2pt]
        $\sigma$ & $g_\sigma^2\theta_\sigma$ & $g_\sigma^2\theta_\sigma$ & $-2g_\sigma^2\Phi_\sigma$ \\[10pt]
        $\omega$ & $2g_\omega^2\theta_\omega$ & $-4g_\omega^2\theta_\omega$ & $-4g_\omega^2\Phi_\omega$ \\[10pt]
        $\pi$ & $\left(\frac{f_\pi}{m_\pi}\right)^2\left[-m_\pi^2\theta_\pi+4pq\right]$
        & $-\left(\frac{f_\pi}{m_\pi}\right)^2\left[m_\pi^2\theta_\pi-4pq\right]$ &
        $-2\left(\frac{f_\pi}{m_\pi}\right)^2\big[pq\theta_\pi-(p^2+q^2)\Phi_\pi\big]$  \\[20pt]
        $\rho_V$ & $2g_\rho^2\theta_\rho$ & $-4g_\rho^2\theta_\rho$ & $-4g_\rho^2\Phi_\rho$ \\[10pt]
        $\rho_T$ &
        $\left(\frac{f_\rho}{2m_n}\right)^2\left[-m_\rho^2\theta_\rho+4pq\right]$
        & $-3\left(\frac{f_\rho}{2m_n}\right)^2\left[m_\rho^2\theta_\rho-4pq\right]$ & $-4\left(\frac{f_\rho}{2m_n}\right)^2\big[pq\theta_\rho-(p^2+q^2-m_\rho^2/2)\Phi_\rho\big]$ \\[20pt]
        $\rho_{VT}$ & \multicolumn{3}{c}{$D=12\frac{f_\rho g_\rho}{2m_n}[\theta_\rho p-2\Phi_\rho q]$} \\[10pt]
    \end{tabular}
    \end{ruledtabular}
    \end{table*}

    The nucleon self-energies given by Eqs.~\eqref{Sigma_S_HF_full}--\eqref{Sigma_V_HF_full} and the two gap equations for $\sigma_0$ and $\omega_0$ given by Eqs~\eqref{eom_sigma_0} and \eqref{eom_omega_0}, respectively, form a system of coupled nonlinear equations that must be solved numerically. The method we implement here is based on the globally convergent Newton-Raphson root-finding method for the N-dimensional nonlinear system of equations. The coupled nucleon self-energies, $\sigma_0$ and $\omega_0$, are solved for the given neutron and proton chemical potentials ($\mu_n$ and $\mu_p$, respectively), where we start the numerical procedure at high density corresponding to high chemical potentials with the initial guesses, which for the self-energies are momentum independent. Then, the obtained values of the self-energies, $\sigma_0$ and $\omega_0$, are used to find the next values for a lower chemical potential, until we reach the required subnuclear density. In the implementation of the Newton-Raphson method, we create two 20-point momentum grids, $p_a\in\{p_1,...,p_{20}\}$ for $a=\{p,n\}$, using the 20-point Gauss-Legendre integration method, which we find to be sufficient, where $p_{20}=5 p_F$ is the maximum $p$ taken as five times the Fermi momentum of the same infinite symmetric nuclear matter computed at zero temperature. 
    Here, the proton or neutron chemical potential is compared to values of the chemical potential above the gas-liquid phase transition\footnote{At subsaturation density, the presence of the gas-liquid phase transition at low temperatures features multiple solutions.} and the corresponding $p_F$ is taken, otherwise, if the chemical potential is below the phase transition, we always take the same $p_F$ at the phase transition.
    In total, there are 122 functions that form the system of equations to solve with 122 unknowns, i.e., $20$ scalar self-energies, $\Sigma^S_a$, $20$ timelike self-energies, $\Sigma^0_a$, and $20$ vector self-energies, $\Sigma^V_a$, each of which for protons and neutrons, and 1 for $\sigma_0$ and 1 for $\omega_0$.
    
    In the present paper, we choose parametrization
    BM-A for the model M1 in Ref.~\cite{savushkin_1997}, keeping the $\delta$-force components in the $\pi$ and $\rho$ interactions \cite{bouyssy_1987}, henceforth the explicit appearance of the extra $\pm 4pq$ terms in the functions $A_\pi$, $A_{\rho_T}$, $B_\pi$, and $B_{\rho_T}$ \cite{horowitz_1983} (see Table~\ref{tab:HF_functions}), 
    however, with the modified $\sigma$ and $\omega$ coupling constants, in order to obtain a nuclear binding energy, $E_B=-15.77$ MeV, at saturation density, which we picked to be $n_0=0.148$ fm$^{-3}$. 
    We employ the following model parameters for the rest masses: $m_n=939.565$ MeV, $m_p=938.272$ MeV, $m_\sigma=550$ MeV, $m_\omega=782.6$ MeV, $m_\rho=764$ MeV, and $m_\pi=139$ MeV, as well as the following coupling constants: $g_\sigma^2/4\pi=6.275$, $g_\omega^2/4\pi=4.120$, $g_\rho^2/4\pi=0.55$, $f_\pi^2/4\pi=0.08$, $f_\rho/g_\rho=3.7$, $g_3=-5.992\,m_n\,g_\sigma^3/10^3$, $g_4=2.828\,g_\sigma^4/10^3$, and $c_4=2.844\,g_\omega^4/10^2$.

\section{\label{app:H_approx}Neutrino opacity within the RMF approximation}
    
    In this appendix, we revisit the essential formalism for the calculations of the neutrino absorption rate within the RMF approximation~\cite{serot_1986} and point out a number of differences in the final expressions compared to Ref.~\cite{roberts_2017}. 
    In this approach, the dressed nucleon quantities are those in Eqs.~\eqref{m_eff_HF}--\eqref{E_eff_HF}, except that the nucleon self-energies are momentum independent, given as follows:
    \begin{align}
        \Sigma^S_a&=-\frac{g_\sigma^2}{m^{2}_\sigma} n_S\mp\frac{g_\delta^2}{m^{2}_\delta} n_{I,S} \ , \label{Sigma_S_H_final} \\
        \Sigma_a^0&=\frac{g_\omega^2}{m^{2}_\omega} n_{\rm B} \pm\frac{g_\rho^2}{m_\rho^2}n_I \label{Sigma_0_H_final} \ ,
    \end{align}
    where $n_{\rm B}$, $n_S$, and $n_I$ are given by Eqs.~\eqref{nB_HF}--\eqref{nI_HF}, respectively, $n_{I,S}=n_{S,p}-n_{S,n}$ is the isospin scalar density with the proton, $n_{S,p}$, and neutron, $n_{S,n}$, scalar density, the upper sign is for protons, lower one for neutrons, and the vector part of the self-energy is $\Sigma^V_a=0$, such that $p_a^*=p_a$, due to rotational invariance, which is assumed for nuclear matter. We note here that for the nonlinear RMF models, such as TMA and NL3, in Eqs.~\eqref{Sigma_S_H_final} and \eqref{Sigma_0_H_final}, the following replacements: $m_\sigma\rightarrow m_\sigma^*$ and $m_\omega\rightarrow m_\omega^*$, yield effective $\sigma$ and $\omega$ masses, which are given by the very same Eqs.~\eqref{m_eff_s} and \eqref{m_eff_w}, respectively, accounting for meson-meson interactions.
    We note here further that in TMA, NL3, and DD2 RMF models with $g_\delta=0$, $\Sigma^S_n=\Sigma^S_p$, which is no longer true in the RHF approximation, TMA$\delta$, and DD2$\delta$, and all the meson-nucleon coupling constants are density dependent in the DD2 and DD2$\delta$ RMF models. The latter aspect requires adding the corresponding rearrangement term, $\Sigma_0^R$, in the timelike self-energy Eq.~\eqref{Sigma_0_H_final}, as is given in Refs.~\cite{typel_1999,kumar_2024}.

    In order to incorporate the isovector-scalar $\delta$ mesons in the RMF models, the following terms are added to the Lagrangian~\eqref{L}, i.e., kinetic and rest mass contributions in Eq.~\eqref{L_0}:
    \begin{equation}
    \mathcal{L}_0^{(\delta)} 
    = 
    \frac{1}{2}(\partial_\mu\boldsymbol{\delta}\cdot\partial^\mu\boldsymbol{\delta}
    -m_\delta^2\boldsymbol{\delta}^2)~,
    \end{equation}
    and interaction contribution in Eq.~\eqref{L_int}, 
    \begin{equation}
    \mathcal{L}_I^{(\delta)} = -
    g_\delta\bar\psi\boldsymbol{\tau}\cdot\boldsymbol{\delta}\psi~,
    \end{equation}
    with $\delta$-meson rest mass, $m_\delta$, coupling constant, $g_\delta$, and $\delta^i=\{\delta^0,\delta^+,\delta^-\}$ \cite{Greco:2002PhRvC65_delta-meson}.
    
    In this formalism, the differential absorption rate can be obtained from Eq.~\eqref{eq:diff_absorption_HF} with the modified RMF effective quantities, which depend no longer on momentum but only on bulk properties of the nuclear medium, and the modified $\tilde q^\mu$, as follows:
    \begin{equation}
        \frac{d\,\Gamma(E_1)}{dE_3d\mu_{13}}=\frac{G_F^2C^2}{32\pi^2}\frac{p_3}{E_3}\frac{1-f_3(E_3)}{1-\exp\{-(q_0+\Delta\mu)/T\}}L_{\mu\nu}\mathcal{I}^{\mu\nu} \ , \label{eq:diff_absorption_H}
    \end{equation}
    where $L_{\mu\nu}$ is defined by Eq.~\eqref{lepton_tensor_explicit} and 
    \begin{align}
        \mathcal{I}^{\mu\nu}&=-2\int\frac{d^3p_2}{(2\pi)^3}\text{Im}\,\tilde\Pi^{\mu\nu}(q_0,q) \nonumber \\
        &=\int\frac{d^3p_2}{(2\pi)^2}\frac{f_2-f_4}{4E_2^*E_4^*}\Lambda^{\mu\nu}\delta(\tilde q_0+E_2^*-E_4^*) \ , \label{I_H}
    \end{align}
    which we introduce here to make a connection with the final expressions derived in the integral forms in Ref.~\cite{roberts_2017}, where $\text{Im}\,\tilde\Pi^{\mu\nu}$ and $\Lambda^{\mu\nu}$ are given by Eqs.~\eqref{Im_tilde_Pi} and \eqref{Lambda_HF}, respectively, with the effective quantities taken from the RMF approximation. Here, we introduce the following effective 4-momentum transfer $\tilde q_\mu=(\tilde q_0,\boldsymbol{q})$ with $\tilde q_0=q_0+\Delta U$, where $\Delta U=\Sigma_2^0-\Sigma_4^0$ is the difference of the neutron and proton RMF timelike self-energies.

    Following Eq.~\eqref{eq:L_Pi_Hartree-Fock}, the contracted $L_{\mu\nu}$ and $\mathcal{I}^{\mu\nu}$ tensors in Eq.~\eqref{eq:diff_absorption_H} can be written in general as
    \begin{align}
        L_{\mu\nu}\mathcal{I}^{\mu\nu}= & \sum_{i\in\mathcal{A}}\mathcal{C}_i\big[L_LI_L^i + L_QI_Q^i - 2L_{M+}I_{M+}^i \nonumber \\
        &- 2L_{M-}I_{M-}^i + 2L_{T+}I_{T+}^i + 2L_{T-}I_{T-}^i\big] \ , \label{L_Pi_Hartree}
    \end{align}
    where $\mathcal{A}$ and $\mathcal{C}_i$ are given in \eqref{eq:AC}. All components of the lepton tensor are given by Eq.~\eqref{eq:L_components}, where within the mean-field approximation the 4-momentum contractions are modified as follows:
    \begin{align}
        p_1\cdot q=&\;\frac{q_\mu^2-m_3^2}{2} \ , \label{p1_q_H} \\[5pt]
        p_1\cdot\tilde q=&\;p_1\cdot q+\Delta U E_1 \ , \label{p1_q_tilde_H} \\[5pt]
        p_1\cdot\tilde n=&-\;\frac{q_\mu^2}{2q}\bigg[E_1+E_3-\Delta U + 2\Delta U E_1\frac{q_0}{q_\mu^2} \nonumber \\
        &+({q_0+\Delta U})\frac{m_3^2}{q_\mu^2}\bigg] \ , \label{p1_n_tilde_H}
        \\[5pt]
        \tilde n\cdot q=&-q\Delta U \ , \label{n_tilde_q_H} \\[5pt]
        q\cdot\tilde q=&\;q_\mu^2+q_0\Delta U \label{q_q_tilde_H} \ .
    \end{align}
    We note that, compared to \cite{roberts_2017}, expression~\eqref{p1_n_tilde_H} differs by a sign\footnote{We confirm $q_0+\Delta U$ instead of $q_0-\Delta U$~\cite{roberts_2017}.}.
    
    Concerning $\mathcal{I^{\mu\nu}}$, for the sake of numerical calculations, it is useful to remove the Dirac delta in Eq.~\eqref{I_H}, which we denote for now as $\delta(g(\mu))$, where $\mu\equiv\mu_{24}$. Then, it can be rewritten as
    \begin{equation}
        \delta(g(\mu))=\frac{\delta(\mu-\mu_0)}{|g'(\mu_0)|} \label{delta_property}
    \end{equation}
    by applying the following property of the Dirac delta with $\mu_0$ being the root of $g(\mu)$. In the RMF approach, we have the following definition:
    \begin{equation}
        \mu_0=\frac{1}{2p_2q}\left[\tilde q_\mu^2\beta+2E_2^*\tilde q_0\right] \ , \label{mu0_H}
    \end{equation}
    where $\beta=1+(m_2^{*2}-m_4^{*2})/\tilde q_\mu^2$, and hence we obtain
    \begin{equation}
        \delta(\tilde q_0+E_2^*-E_4^*)=\frac{\delta(\mu-\mu_0)}{p_2q}\left|E_2^*+\tilde q_0\right| \ . \label{delta_property_H}
    \end{equation}
    To ensure that $\mu_0$ is the correct root of $g(\mu)$, it must be in the range of $-1\le\mu_0\le1$, which, after integrating over $\mu$, imposes the new limit $\Theta(E_2^*-e_-)$ for $E_2^*$, where
    \begin{equation}
        e_-=-\beta\frac{\tilde q_0}{2}+\frac{q}{2}\sqrt{\beta^2-\frac{4m_2^{*2}}{\tilde q_\mu^2}} \ .
    \end{equation}
    
    Applying Eqs.~\eqref{delta_property} and \eqref{delta_property_H}, Eq.~\eqref{I_H} can be rewritten as
    \begin{align}
        \mathcal{I}^{\mu\nu}
        &=\frac{1}{(4\pi)^2}\int d\Omega_{24}\int_0^\infty dp_2p_2^2\frac{f_2-f_4}{E_2^*E_4^*} \Lambda^{\mu\nu}\frac{\delta(\mu-\mu_0)}{|g'(\mu_0)|} \label{I_H_delta} \\
        &=\frac{1}{(4\pi)^2}\frac{1}{q}
        \int d\Omega_{24}\int_{m_2^*}^\infty dE_2^*\delta(\mu-\mu_0)
      \nonumber \\
        &\qquad\qquad\times\Lambda^{\mu\nu}\,\Theta(E_2^*-e_-)(f_2-f_4) \ . \label{I_H_full}
    \end{align}
    If the function $g(\mu)$ has a complicated form, we apply Eq.~\eqref{I_H_delta}, which is the case for the RHF approximation, as detailed in Sec.~\ref{sec:HF_approx}.
    In this case, the root $\mu_0$ is obtained numerically, for which we employ the Newton–Raphson method, being retained if $-1 \le \mu_0 \le 1$. 
    Moreover, the derivative $g'(\mu_0)$ must also be evaluated numerically. 
    On the contrary, in the RMF approximation, we instead use Eq.~\eqref{I_H_full}.

    \begin{figure*}[t!]
    \subfigure[$\nu_e+n\rightarrow e^-+p$]{\includegraphics[width=0.49\textwidth]{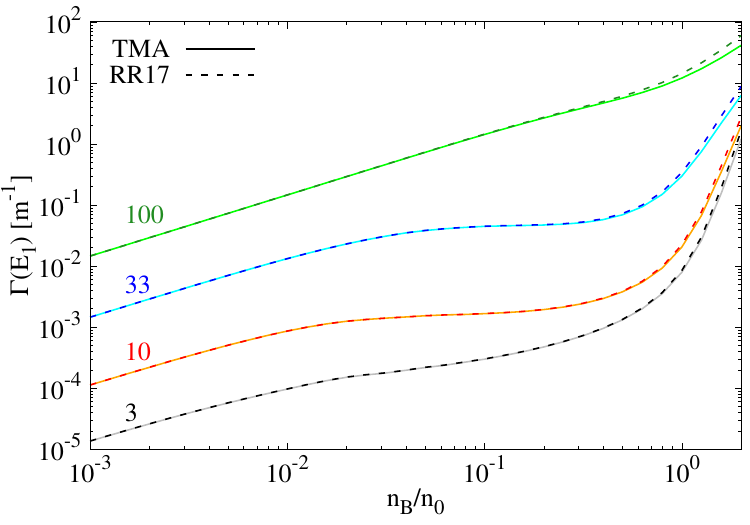}\label{fig:neutrino_opacity_B_RMF_comp_a}}
    \subfigure[$\bar\nu_e+p\rightarrow e^++n$]{\includegraphics[width=0.49\textwidth]{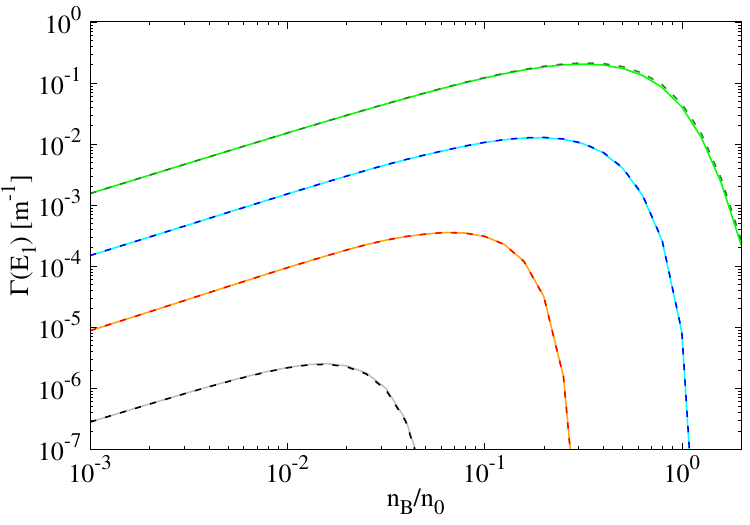}\label{fig:neutrino_opacity_B_RMF_comp_b}}
    \caption{ 
    The same as Fig.~\ref{fig:neutrino_opacity_vs_nB_B}, but comparing two different opacity expressions, Appendix~\ref{app:H_approx} (solid lines) and Ref.~\cite{roberts_2017}, denoted here as RR17 (dashed lines). Both calculations employed the same TMA RMF parametrization.
    \label{fig:RMF_comp}}
    \end{figure*}
    
    The decomposition of $\Lambda^{\mu\nu}$ in Eq.~\eqref{I_H_full} is given by Eqs.~\eqref{Lambda_V}--\eqref{Lambda_AP}, however, the 4-momentum contractions must be replaced by the following expressions:
    \begin{align}
        p_2^*\cdot\tilde q=&\;E_2^*E_4^*-p_2q\mu-E_2^{*2} \xrightarrow[]{\mu=\mu_0} -\frac{1}{2}\beta\tilde q_\mu^2 \ , \label{p2_tilde_q_tilde} \\[5pt]
        p_2^*\cdot\tilde n=&\;E_2^*q-p_2\tilde q_0\mu \xrightarrow[]{\mu=\mu_0} -\frac{\tilde q_\mu^2}{2q}\left[2E_2^*+\tilde q_0\beta\right] \ , \label{p2_tilde_n_tilde}
    \end{align}
    where the arrow indicates the substitution of the angle root, as given in Eq.~\eqref{mu0_H}, which would be the case after imposing the Dirac delta in $\mathcal{I^{\mu\nu}}$, Eq.~\eqref{I_H_full}.
    
    Taking into account the form of $\mathcal{I}$ given by Eq.~\eqref{I_H_full}, we define the following integrals:
    \begin{align}
        I_Q&=\frac{\tilde q_\mu^2}{4\pi q}\int_{e_-}^\infty dE_2^*(f_2-f_4) \ , \\[5pt]
        I_L&=-\frac{\tilde q_\mu^2}{4\pi q^3}\int_{e_-}^\infty dE_2^*(f_2-f_4)[2E_2^*+\tilde q_0\beta]^2 \ , \\[5pt]
        I_M&=-\frac{\tilde q_\mu^2}{4\pi q^2}\int_{e_-}^\infty dE_2^*(f_2-f_4)[2E_2^*+\tilde q_0\beta] \ , \\[5pt]
        I_T&=-\frac{1}{2}I_L+\left(\frac{2m_2^{*2}}{\tilde q_\mu^2}-\frac{\beta^2}{2}\right)I_Q \ .
    \end{align}
    
    Based on these definitions, and applying Eqs.~\eqref{p2_tilde_q_tilde}--\eqref{p2_tilde_n_tilde}, along with imposing $\delta(\mu-\mu_0)$, the nonzero tensor components of Eq.~\eqref{I_H} can be cast into the following integral forms with the use of Eqs.~\eqref{Lambda_V}--\eqref{Lambda_AP}:
    \begin{enumerate}
        \item[(1)] Vector:
        \begin{align}
            I_Q^V&=(\lambda^2+\sigma_--1)I_Q \ , \\[5pt]
            I_L^V&=I_L+\sigma_-I_Q \ , \\[5pt]
            I_{T+}^V&=I_T+\sigma_-I_Q \ , \\[5pt]
            I_{M+}^V&=\lambda I_M \ ,
        \end{align}
        where $\lambda=\beta-1$ and $\sigma_\pm=1-(m_2^*\pm m_4^*)^2/\tilde q_\mu^2$.

        \item[(2)] Axial-vector:
        \begin{align}
            I_Q^A&=(\lambda^2+\sigma_+-1)I_Q \ , \\[5pt]
            I_L^A&=I_L+\sigma_+I_Q \ , \\[5pt]
            I_{T+}^A&=I_T+\sigma_+I_Q \ , \\[5pt]
            I_{M+}^A&=\lambda I_M \ .
        \end{align}

        \item[(3)] Tensor:
        \begin{align}
            I_L^T&=\frac{{\tilde q_\mu^2}}{4{M_N^2}}\left[\left(\sigma_--\beta^2+\frac{4{m_2^{*2}}}{{\tilde q_\mu^2}}\right)I_Q-I_L\right] \ , \\[5pt]
            I_{T+}^T&=\frac{{\tilde q_\mu^2}}{4{M_N^2}}\left[\left(\sigma_--\beta^2+\frac{4{m_2^{*2}}}{{\tilde q_\mu^2}}\right)I_Q-I_T\right] \ ,
        \end{align}
        with small differences arising, compared to Eqs.~(57) and (58) in Ref.~\cite{roberts_2017}\footnote{
        We find $\tilde q_\mu^2$ instead of $q_\mu^2$ and $M_N^2$ instead of $m_2^2$ in the denominator outside the brackets, and $m_2^{*2}$ instead of $m_2^2$ in the numerator~\cite{roberts_2017}.
        }.
        
        \item[(4)] Pseudoscalar:
        \begin{equation}
            I_Q^P=-\frac{q_\mu^2}{M_N^2}I_Q \ .
        \end{equation}

        \item[(5)] Vector axial-vector:
        \begin{equation}
            I_{T-}^{VA}=-2iI_M \ .
        \end{equation}

        \begin{table*}[htp]
        \centering
        \caption{Neutron and proton self-energies for condition A and $l=e$ within the RHF with respect to 20 points of momentum $p$ left-edged obtained from the Gauss-Legendre quadrature.}
        \begin{ruledtabular}
        \begin{tabular}{l|ccd|ccd}
            &\multicolumn{3}{c|}{Neutrons}&\multicolumn{3}{c}{Protons} \\
            \multicolumn{1}{l|}{$p$} & $\Sigma^S$ & $\Sigma^0$ & \multicolumn{1}{c|}{$\Sigma^V$} & $\Sigma^S$ & $\Sigma^0$ & \multicolumn{1}{c}{$\Sigma^V$} \\
            \colrule
               0.000 & -104.163 &  81.684 & -0.006 &  -94.332 &  55.292 &  0.009 \\
               7.436 & -104.161 &  81.680 & -0.012 &  -94.329 &  55.293 &  0.018 \\
              31.553 & -104.133 &  81.642 & -0.051 &  -94.292 &  55.297 &  0.078 \\
              63.425 & -104.053 &  81.529 & -0.094 &  -94.183 &  55.309 &  0.167 \\
             110.679 & -103.846 &  81.233 & -0.127 &  -93.901 &  55.339 &  0.334 \\
             163.835 & -103.496 &  80.723 & -0.103 &  -93.430 &  55.380 &  0.594 \\
             230.019 & -102.908 &  79.849 &  0.033 &  -92.652 &  55.428 &  1.039 \\
             299.309 & -102.148 &  78.684 &  0.300 &  -91.669 &  55.450 &  1.637 \\
             378.451 & -101.143 &  77.111 &  0.724 &  -90.416 &  55.417 &  2.427 \\
             457.220 & -100.045 &  75.378 &  1.211 &  -89.105 &  55.313 &  3.240 \\
             542.140 &  -98.795 &  73.426 &  1.742 &  -87.685 &  55.133 &  4.056 \\
             622.850 &  -97.579 &  71.582 &  2.205 &  -86.368 &  54.916 &  4.719 \\
             705.829 &  -96.331 &  69.768 &  2.613 &  -85.075 &  54.672 &  5.262 \\
             780.761 &  -95.228 &  68.239 &  2.914 &  -83.974 &  54.449 &  5.632 \\
             854.261 &  -94.186 &  66.859 &  3.147 &  -82.963 &  54.238 &  5.894 \\
             916.235 &  -93.347 &  65.792 &  3.301 &  -82.169 &  54.070 &  6.049 \\
             973.602 &  -92.607 &  64.882 &  3.411 &  -81.480 &  53.924 &  6.145 \\
            1016.645 &  -92.077 &  64.247 &  3.475 &  -80.993 &  53.820 &  6.192 \\
            1052.727 &  -91.649 &  63.745 &  3.519 &  -80.604 &  53.738 &  6.216 \\
            1072.635 &  -91.420 &  63.479 &  3.539 &  -80.397 &  53.694 &  6.224 \\
        \end{tabular}
        \label{tab:HF_Sigma_A}
        \end{ruledtabular}
        \end{table*}
        
        \begin{table*}[htp]
        \centering
        \caption{The same as in Table~\ref{tab:HF_Sigma_A} but for condition B.}
        \begin{ruledtabular}
        \begin{tabular}{l|cdc|ccc}
            &\multicolumn{3}{c|}{Neutrons}&\multicolumn{3}{c}{Protons} \\
            \multicolumn{1}{l|}{$p$} & $\Sigma^S$ & \multicolumn{1}{c}{$\Sigma^0$} & $\Sigma^V$ & $\Sigma^S$ & $\Sigma^0$ & $\Sigma^V$ \\
            \colrule
               0.000 &  -16.190 &  11.689 &  0.003 &  -15.156 &   6.774 &  0.006 \\
               7.436 &  -16.189 &  11.689 &  0.005 &  -15.155 &   6.774 &  0.012 \\
              31.553 &  -16.184 &  11.682 &  0.022 &  -15.143 &   6.780 &  0.053 \\
              63.425 &  -16.169 &  11.661 &  0.045 &  -15.111 &   6.795 &  0.109 \\
             110.679 &  -16.130 &  11.604 &  0.085 &  -15.030 &   6.831 &  0.198 \\
             163.835 &  -16.070 &  11.505 &  0.138 &  -14.903 &   6.883 &  0.309 \\
             230.019 &  -15.978 &  11.329 &  0.215 &  -14.712 &   6.947 &  0.461 \\
             299.309 &  -15.867 &  11.091 &  0.302 &  -14.495 &   6.999 &  0.623 \\
             378.451 &  -15.729 &  10.772 &  0.399 &  -14.246 &   7.031 &  0.794 \\
             457.220 &  -15.579 &  10.432 &  0.483 &  -14.005 &   7.040 &  0.936 \\
             542.140 &  -15.405 &  10.067 &  0.557 &  -13.756 &   7.032 &  1.054 \\
             622.850 &  -15.232 &   9.738 &  0.609 &  -13.534 &   7.017 &  1.134 \\
             705.829 &  -15.051 &   9.430 &  0.646 &  -13.321 &   6.997 &  1.187 \\
             780.761 &  -14.890 &   9.179 &  0.668 &  -13.144 &   6.979 &  1.214 \\
             854.261 &  -14.737 &   8.960 &  0.681 &  -12.984 &   6.962 &  1.226 \\
             916.235 &  -14.614 &   8.794 &  0.686 &  -12.861 &   6.948 &  1.228 \\
             973.602 &  -14.507 &   8.654 &  0.686 &  -12.755 &   6.937 &  1.223 \\
            1016.645 &  -14.429 &   8.558 &  0.685 &  -12.681 &   6.929 &  1.216 \\
            1052.727 &  -14.367 &   8.483 &  0.683 &  -12.623 &   6.922 &  1.209 \\
            1072.635 &  -14.334 &   8.444 &  0.681 &  -12.592 &   6.919 &  1.204 \\
        \end{tabular}
        \label{tab:HF_Sigma_B}
        \end{ruledtabular}
        \end{table*}
        
        \item[(6)] Vector tensor:
        \begin{align}
            I_L^{VT}&=\left(2{\frac{m_2^*}{M_N}}+\Delta\beta\right)I_Q \ , \label{I_VT_L} \\[5pt]
            I_{T+}^{VT}&=\left(2{\frac{m_2^*}{M_N}}+\Delta\beta\right)I_Q \ , \label{I_VT_T+} \\[5pt]
            I_{M+}^{VT}&=-\frac{\Delta}{2}I_M \ ,
        \end{align}
        where $\Delta$ is given below Eq.~\eqref{Lambda_VT_M_plus}. 

        \item[(7)] Axial-vector tensor:
        \begin{equation}
            I_{T-}^{AT}=-i\left(2{\frac{m_2^*}{M_N}}+\Delta\right)I_M \ .
            \label{I_AVT}
        \end{equation}
        We note that also here for the vector tensor and axial-vector tensor terms, we find small differences compared to the literature\footnote{In the first terms in Eqs.~\eqref{I_VT_L}, \eqref{I_VT_T+}, and \eqref{I_AVT}, we obtain a factor of $2m_2^*/M_N$ instead of a factor of 2, as in Eqs.~(59), (60), and (63) in Ref.~\cite{roberts_2017}.}.

        \item[(8)] Axial-vector pseudoscalar:
        \begin{align}
            I_Q^{AP}&=-2\left(\Delta\beta+2\frac{m_2^*}{M_N}\right)I_Q \ , \\[5pt]
            I_{M+}^{AP}&=-\Delta I_M \ .
        \end{align}
        
    \end{enumerate}
    
    From the differential absorption rate, Eq.~\eqref{eq:diff_absorption_H}, we can calculate the neutrino opacity by expressing it in terms of the integrals over $q_0$ and $q$, which are connected to $E_3$ via Eq.~\eqref{q0} and through $\mu_{13}$ via Eq.~\eqref{q}, respectively (cf. Ref.~\cite{reddy_1998}),
    \begin{align}
        \Gamma(E_1)=&\frac{G_F^2C^2}{32\pi^2E_1^2}\int_{-\infty}^{E_1-m_3} dq_0\int_{|E_1-p_3|}^{E_1+p_3}dqq \nonumber \\
        &\times\frac{1-f_3(E_3)}{1-\exp\{-(q_0+\Delta\mu)/T\}} L_{\mu\nu}\mathcal{I}^{\mu\nu} \label{total_absorption_H2} \ .
    \end{align}

\section{\label{app:RMF_term_comp} Comparison of the neutrino opacity with Roberts and Reddy~\cite{roberts_2017}}

    In order to explore the impact of the corrected terms of the mean-field opacity expressions of Ref.~\cite{roberts_2017}, which we pointed out in Appendix~\ref{app:H_approx} before, we compute the neutrino opacity here for both cases employing the TMA RMF EOS. 
    In Fig.~\ref{fig:RMF_comp}, we provide the results of the neutrino opacity calculations for our expressions given in Appendix~\ref{app:H_approx} (solid lines) and the full expression given in Ref.~\cite{roberts_2017}, Eq.~(11), denoted here as RR17 (dashed lines). 
    The results show that for $\nu_e$ in Fig.~\ref{fig:neutrino_opacity_B_RMF_comp_a}, there is an enhancement of the opacity for RR17 compared to TMA based on our expressions around and above saturation density, in particular for high neutrino energies. For intermediate and low neutrino energies, the effect becomes negligible. 
    Similarly for $\bar\nu_e$, shown in Fig.~\ref{fig:neutrino_opacity_B_RMF_comp_b}, for which the opacity is medium suppressed toward high densities, the impact of the corrected terms becomes negligible at all neutrino energies.
    We note that the analysis presented here is medium dependent and might alter quantitatively when considering different conditions.

\section{\label{app:appendix_HF-tables} Tables of the Hartree-Fock self-energies}
    Data tables for the momentum-dependent RHF self-energies are given in Table~\ref{tab:HF_Sigma_A} for condition A and in Table~\ref{tab:HF_Sigma_B} for condition B, separated into scalar $\Sigma^S$, timelike $\Sigma^0$, and vector $\Sigma^V$ self-energies. The momentum $p$ grid employs a 20-point Gauss-Legendre quadrature for the numerical integration procedures, ranging from $p=0$ to $p\approx 1.1$~GeV (left-edged values).
    Here, we distinguish between neutrons (left columns) and protons (right columns) due to the consideration of asymmetric nuclear matter. 

\newpage
\bibliographystyle{apsrev4-1}
\bibliography{bibliography}

@PREAMBLE{
 "\providecommand{\noopsort}[1]{}" 
 # "\providecommand{\singleletter}[1]{#1}%" 
}

@ARTICLE{hirata87,
   author = {{Hirata}, K. and others},
    title = "{Observation of a neutrino burst from the supernova SN1987A}",
  journal = {Phys. Rev. Lett.},
     year = 1987,
    month = apr,
   volume = 58,
    pages = {1490-1493},
      doi = {10.1103/PhysRevLett.58.1490},
   adsurl = {http://ads.nao.ac.jp/abs/1987PhRvL..58.1490H}
}

@ARTICLE{bionta87,
   author = {{Bionta}, R.~M. and others},
    title = "{Observation of a neutrino burst in coincidence with supernova 1987A in the Large Magellanic Cloud}",
  journal = {Phys. Rev. Lett.},
     year = 1987,
    month = apr,
   volume = 58,
    pages = {1494-1496},
      doi = {10.1103/PhysRevLett.58.1494},
   adsurl = {http://ads.nao.ac.jp/abs/1987PhRvL..58.1494B}
}

@ARTICLE{IMB_Bratton88,
       author = {{Bratton}, C.~B. and others},
        title = "{Angular distribution of events from SN1987A}",
      journal = {\prd},
         year = 1988,
        month = jun,
       volume = {37},
       number = {12},
        pages = {3361-3363},
          doi = {10.1103/PhysRevD.37.3361},
       adsurl = {https://ui.adsabs.harvard.edu/abs/1988PhRvD..37.3361B}
}

@ARTICLE{BAKSAN1988PhLB205,
       author = {{Alexeyev}, E.~N. and {Alexeyeva}, L.~N. and {Krivosheina}, I.~V. and {Volchenko}, V.~I.},
        title = "{Detection of the neutrino signal from SN 1987A in the LMC using the INR Baksan underground scintillation telescope}",
      journal = {Phys. Lett. B},
         year = 1988,
        month = apr,
       volume = {205},
       number = {2-3},
        pages = {209-214},
          doi = {10.1016/0370-2693(88)91651-6},
       adsurl = {https://ui.adsabs.harvard.edu/abs/1988PhLB..205..209A}
}

@ARTICLE{LSD1987EL3,
       author = {{Aglietta}, M. and others},
        title = "{On the event observed in the Mont Blanc Underground Neutrino Observatory during the occurrence of supernova 1987a.}",
      journal = {Europhys. Lett.},
         year = 1987,
        month = jan,
       volume = {3},
        pages = {1315-1320},
          doi = {10.1209/0295-5075/3/12/011},
       adsurl = {https://ui.adsabs.harvard.edu/abs/1987EL......3.1315A}
}

@ARTICLE{Janka2025ARNPS,
       author = {{Janka}, Hans-Thomas},
        title = "{Long-Term Multidimensional Models of Core-Collapse Supernovae: Progress and Challenges}",
      journal = {Annu. Rev. Nucl. Part. Sci.},
         year = 2025,
        month = sep,
       volume = {75},
       number = {1},
        pages = {425-461},
          doi = {10.1146/annurev-nucl-121423-100945},
       adsurl = {https://ui.adsabs.harvard.edu/abs/2025ARNPS..75..425J}
}

@ARTICLE{Fischer2024PrPNP,
       author = {{Fischer}, Tobias and {Guo}, Gang and {Langanke}, Karlheinz and {Mart{\'\i}nez-Pinedo}, Gabriel and {Qian}, Yong-Zhong and {Wu}, Meng-Ru},
        title = "{Neutrinos and nucleosynthesis of elements}",
      journal = {Prog. Part. Nucl. Phys.},
         year = 2024,
        month = may,
       volume = {137},
          eid = {104107},
        pages = {104107},
          doi = {10.1016/j.ppnp.2024.104107},
       adsurl = {https://ui.adsabs.harvard.edu/abs/2024PrPNP.13704107F}
}

@ARTICLE{Tamborra2025NatRP7,
       author = {{Tamborra}, Irene},
        title = "{Neutrinos from explosive transients at the dawn of multi-messenger astronomy}",
      journal = {Nat. Rev. Phys.},
         year = 2025,
        month = jun,
       volume = {7},
       number = {6},
        pages = {285-298},
          doi = {10.1038/s42254-025-00828-2},
       adsurl = {https://ui.adsabs.harvard.edu/abs/2025NatRP...7..285T}
}

@ARTICLE{reddy_1998,
       author = {{Reddy}, Sanjay and {Prakash}, Madappa and {Lattimer}, James M.},
        title = "{Neutrino interactions in hot and dense matter}",
      journal = {\prd},
         year = 1998,
        month = jul,
       volume = {58},
       number = {1},
          eid = {013009},
        pages = {013009},
          doi = {10.1103/PhysRevD.58.013009},
       adsurl = {https://ui.adsabs.harvard.edu/abs/1998PhRvD..58a3009R}
}

@article{martinez_2012,
       author = {{Mart{\'\i}nez-Pinedo}, G. and {Fischer}, T. and {Lohs}, A. and {Huther}, L.},
        title = "{Charged-Current Weak Interaction Processes in Hot and Dense Matter and its Impact on the Spectra of Neutrinos Emitted from Protoneutron Star Cooling}",
      journal = {\prl},
         year = 2012,
        month = dec,
       volume = {109},
       number = {25},
          eid = {251104},
        pages = {251104},
          doi = {10.1103/PhysRevLett.109.251104},
       adsurl = {https://ui.adsabs.harvard.edu/abs/2012PhRvL.109y1104M}
}

@ARTICLE{Robert:2012PhRvC86,
       author = {{Roberts}, L.~F. and {Reddy}, Sanjay and {Shen}, Gang},
        title = "{Medium modification of the charged-current neutrino opacity and its implications}",
      journal = {\prc},
         year = 2012,
        month = dec,
       volume = {86},
       number = {6},
          eid = {065803},
        pages = {065803},
          doi = {10.1103/PhysRevC.86.065803},
       adsurl = {https://ui.adsabs.harvard.edu/abs/2012PhRvC..86f5803R}
}

@ARTICLE{Fischer10,
       author = {{Fischer}, T. and {Whitehouse}, S.~C. and {Mezzacappa}, A. and {Thielemann}, F. -K. and {Liebend{\"o}rfer}, M.},
        title = "{Protoneutron star evolution and the neutrino-driven wind in general relativistic neutrino radiation hydrodynamics simulations}",
      journal = {\aap},
         year = 2010,
        month = jul,
       volume = {517},
          eid = {A80},
        pages = {A80},
        doi = {10.1051/0004-6361/200913106},
       adsurl = {https://ui.adsabs.harvard.edu/abs/2010A&A...517A..80F}
}

@ARTICLE{Huedepohl10,
       author = {{H{\"u}depohl}, L. and {M{\"u}ller}, B. and {Janka}, H. -T. and {Marek}, A. and {Raffelt}, G.~G.},
        title = "{Neutrino Signal of Electron-Capture Supernovae from Core Collapse to Cooling}",
      journal = {\prl},
         year = 2010,
        month = jun,
       volume = {104},
       number = {25},
          eid = {251101},
        pages = {251101},
          doi = {10.1103/PhysRevLett.104.251101},
       adsurl = {https://ui.adsabs.harvard.edu/abs/2010PhRvL.104y1101H}
}

@ARTICLE{Fischer2020PhRvC101,
       author = {{Fischer}, Tobias and {Guo}, Gang and {Dzhioev}, Alan A. and {Mart{\'\i}nez-Pinedo}, Gabriel and {Wu}, Meng-Ru and {Lohs}, Andreas and {Qian}, Yong-Zhong},
        title = "{Neutrino signal from proto-neutron star evolution: Effects of opacities from charged-current-neutrino interactions and inverse neutron decay}",
      journal = {\prc},
         year = 2020,
        month = feb,
       volume = {101},
       number = {2},
          eid = {025804},
        pages = {025804},
          doi = {10.1103/PhysRevC.101.025804},
       adsurl = {https://ui.adsabs.harvard.edu/abs/2020PhRvC.101b5804F}
}

@ARTICLE{Guo2020PhRvD102,
       author = {{Guo}, Gang and {Mart{\'\i}nez-Pinedo}, Gabriel and {Lohs}, A. and {Fischer}, Tobias},
        title = "{Charged-current muonic reactions in core-collapse supernovae}",
      journal = {\prd},
         year = 2020,
        month = jul,
       volume = {102},
       number = {2},
          eid = {023037},
        pages = {023037},
          doi = {10.1103/PhysRevD.102.023037},
       adsurl = {https://ui.adsabs.harvard.edu/abs/2020PhRvD.102b3037G}
}

@ARTICLE{roberts_2012,
       author = {{Roberts}, L.~F. and {Shen}, G. and {Cirigliano}, V. and {Pons}, J.~A. and {Reddy}, S. and {Woosley}, S.~E.},
        title = "{Protoneutron Star Cooling with Convection: The Effect of the Symmetry Energy}",
      journal = {\prl},
         year = 2012,
        month = feb,
       volume = {108},
       number = {6},
          eid = {061103},
        pages = {061103},
          doi = {10.1103/PhysRevLett.108.061103},
       adsurl = {https://ui.adsabs.harvard.edu/abs/2012PhRvL.108f1103R}
}

@ARTICLE{mirizzi16,
   author = {{Mirizzi}, A. and {Tamborra}, I. and {Janka}, H.-T. and {Saviano}, N. and 
	{Scholberg}, K. and {Bollig}, R. and {H{\"u}depohl}, L. and 
	{Chakraborty}, S.},
    title = "{Supernova neutrinos: production, oscillations and detection}",
  journal = {Riv. Nuovo Cim.},
     year = 2016,
   volume = 39,
    pages = {1-112},
      doi = {10.1393/ncr/i2016-10120-8},
   adsurl = {http://ads.nao.ac.jp/abs/2016NCimR..39....1M}
}

@ARTICLE{Burrows2026arXiv260209025R,
    doi = {10.3847/1538-4357/ae61ac},
    url = {https://doi.org/10.3847/1538-4357/ae61ac},
    year = {2026},
    month = {may},
    publisher = {The American Astronomical Society},
    volume = {1003},
    number = {1},
    pages = {40},
    author = {Rusakov, Aleksandr and Burrows, Adam and Wang, Tianshu and Vartanyan, David},
    title = {An Exploration of the Equation of State Dependence of Core-collapse Supernova Explosion Outcomes and Signatures},
    journal = {Astrophys. J.},
}

@book{serot_1986,
        author = "Serot, Brian D. and Walecka, J. D.",
        title = "The relativistic nuclear many-body problem",
        volume = 16,
        series = {Advances in Nuclear Physics},
        publisher = "Plenum Press",
        address = "New York",
        year = "1986",
        editor = {Negele, J. W. and Vogt, E.},
}

@book{bellac_2000,
  title={Thermal field theory},
  author={Le Bellac, Michel},
  year={2000},
  publisher={Cambridge University Press, Cambridge, England}
}

@article{kobes_1986,
        title = {Discontinuities of green functions in field theory at finite temperature and density (II)},
        journal = {Nucl. Phys. B},
        volume = {272},
        number = {2},
        pages = {329-364},
        year = {1986},
        issn = {0550-3213},
        doi = {https://doi.org/10.1016/0550-3213(86)90006-4},
        url = {https://www.sciencedirect.com/science/article/pii/0550321386900064},
        author = {Randal L. Kobes and Gordon W. Semenoff}
}

@ARTICLE{PDG,
       author = {{Navas}, S. and {Amsler}, C. and {Gutsche}, T. and {Hanhart}, C. and {Hern{\'a}ndez-Rey}, J.~J. and {Louren{\c{c}}o}, C. and {Masoni}, A. and {Mikhasenko}, M. and {Mitchell}, R.~E. and {Patrignani}, C. and {Schwanda}, C. and {Spanier}, S. and {Venanzoni}, G. and {Yuan}, C.~Z. and {Agashe} and others},
        title = "{Review of particle physics$^{*}$}",
      journal = {\prd},
         year = 2024,
        month = aug,
       volume = {110},
       number = {3},
          eid = {030001},
        pages = {030001},
          doi = {10.1103/PhysRevD.110.030001},
       adsurl = {https://ui.adsabs.harvard.edu/abs/2024PhRvD.110c0001N}
}

@article{roberts_2017,
       author = {{Roberts}, Luke F. and {Reddy}, Sanjay},
        title = "{Charged current neutrino interactions in hot and dense matter}",
      journal = {\prc},
         year = 2017,
        month = apr,
       volume = {95},
       number = {4},
          eid = {045807},
        pages = {045807},
          doi = {10.1103/PhysRevC.95.045807},
       adsurl = {https://ui.adsabs.harvard.edu/abs/2017PhRvC..95d5807R}
}

@article{chin_1977,
        title = {A relativistic many-body theory of high density matter},
        journal = {Ann. Phys. (N.Y.)},
        volume = {108},
        number = {2},
        pages = {301-367},
        year = {1977},
        issn = {0003-4916},
        doi = {https://doi.org/10.1016/0003-4916(77)90016-1},
        url = {https://www.sciencedirect.com/science/article/pii/0003491677900161},
        author = {S.A. Chin},
}

@article{horowitz_1983,
        title = {Properties of nuclear and neutron matter in a relativistic Hartree-Fock theory},
        journal = {\nphysa},
        volume = {399},
        number = {2},
        pages = {529-562},
        year = {1983},
        issn = {0375-9474},
        doi = {https://doi.org/10.1016/0375-9474(83)90262-2},
        url = {https://www.sciencedirect.com/science/article/pii/0375947483902622},
        author = {C.J. Horowitz and Brian D. Serot},
}

@article{bouyssy_1987,
        title = {Relativistic description of nuclear systems in the Hartree-Fock approximation},
        author = {Bouyssy, A. and Mathiot, J.-F. and Van Giai, Nguyen and Marcos, S.},
        journal = {\prc},
        volume = {36},
        issue = {1},
        pages = {380--401},
        numpages = {0},
        year = {1987},
        month = {Jul},
        publisher = {American Physical Society},
        doi = {10.1103/PhysRevC.36.380},
        url = {https://link.aps.org/doi/10.1103/PhysRevC.36.380}
}

@ARTICLE{bernardos_1993,
       author = {{Bernardos}, P. and {Fomenko}, V.~N. and {Giai}, Nguyen Van and {Quelle}, M.~L. and {Marcos}, S. and {Niembro}, R. and {Savushkin}, L.~N.},
        title = "{Relativistic Hartree-Fock approximation in a nonlinear model for nuclear matter and finite nuclei}",
      journal = {\prc},
         year = 1993,
        month = dec,
       volume = {48},
       number = {6},
        pages = {2665-2672},
          doi = {10.1103/PhysRevC.48.2665},
       adsurl = {https://ui.adsabs.harvard.edu/abs/1993PhRvC..48.2665B}
}

@ARTICLE{savushkin_1997,
       author = {{Savushkin}, L.~N. and {Marcos}, S. and {Quelle}, M.~L. and {Bernardos}, P. and {Fomenko}, V.~N. and {Niembro}, R.},
        title = "{Effective interaction for relativistic theory of nuclear structure}",
      journal = {\prc},
         year = 1997,
        month = jan,
       volume = {55},
       number = {1},
        pages = {167-178},
          doi = {10.1103/PhysRevC.55.167},
       adsurl = {https://ui.adsabs.harvard.edu/abs/1997PhRvC..55..167S}
}

@article{toki_1995,
        title = {Relativistic many body approach for unstable nuclei and supernova},
        journal = {\nphysa},
        volume = {588},
        number = {1},
        pages = {c357-c363},
        year = {1995},
        issn = {0375-9474},
        doi = {https://doi.org/10.1016/0375-9474(95)00161-S},
        url = {https://www.sciencedirect.com/science/article/pii/037594749500161S},
        author = {H. Toki and D. Hirata and Y. Sugahara and K. Sumiyoshi and I. Tanihata},
}

@article{lalazissis_1997,
        title = {New parametrization for the Lagrangian density of relativistic mean field theory},
        author = {Lalazissis, G. A. and K\"onig, J. and Ring, P.},
        journal = {\prc},
        volume = {55},
        issue = {1},
        pages = {540--543},
        numpages = {0},
        year = {1997},
        month = {Jan},
        publisher = {American Physical Society},
        doi = {10.1103/PhysRevC.55.540},
        url = {https://link.aps.org/doi/10.1103/PhysRevC.55.540}
}

@ARTICLE{typel_2010,
       author = {{Typel}, S. and {R{\"o}pke}, G. and {Kl{\"a}hn}, T. and {Blaschke}, D. and {Wolter}, H.~H.},
        title = "{Composition and thermodynamics of nuclear matter with light clusters}",
      journal = {\prc},
         year = 2010,
        month = jan,
       volume = {81},
       number = {1},
          eid = {015803},
        pages = {015803},
          doi = {10.1103/PhysRevC.81.015803},
       adsurl = {https://ui.adsabs.harvard.edu/abs/2010PhRvC..81a5803T}
}

@ARTICLE{kumar_2024,
       author = {{Kumar}, Anil and {Thakur}, Pratik and {Sinha}, Monika},
        title = "{Non-radial oscillations in newly born compact star considering effects of phase transition}",
      journal = {Mon. Not. R. Astron. Soc.},
         year = 2024,
        month = may,
       volume = {530},
       number = {1},
        pages = {501-513},
          doi = {10.1093/mnras/stae834},
       adsurl = {https://ui.adsabs.harvard.edu/abs/2024MNRAS.530..501K}
}

@ARTICLE{typel_1999,
       author = {{Typel}, S. and {Wolter}, H.~H.},
        title = "{Relativistic mean field calculations with density-dependent meson-nucleon coupling}",
      journal = {\nphysa},
         year = 1999,
        month = sep,
       volume = {656},
       number = {3},
        pages = {331-364},
          doi = {10.1016/S0375-9474(99)00310-3},
       adsurl = {https://ui.adsabs.harvard.edu/abs/1999NuPhA.656..331T}
}

@ARTICLE{Hempel2010NuPhA837,
       author = {{Hempel}, Matthias and {Schaffner-Bielich}, J{\"u}rgen},
        title = "{A statistical model for a complete supernova equation of state}",
      journal = {\nphysa},
         year = 2010,
        month = jun,
       volume = {837},
       number = {3-4},
        pages = {210-254},
          doi = {10.1016/j.nuclphysa.2010.02.010},
       adsurl = {https://ui.adsabs.harvard.edu/abs/2010NuPhA.837..210H}
}

@ARTICLE{Fischer2020PhRvC102,
       author = {{Fischer}, Tobias and {Typel}, Stefan and {R{\"o}pke}, Gerd and {Bastian}, Niels-Uwe F. and {Mart{\'\i}nez-Pinedo}, Gabriel},
        title = "{Medium modifications for light and heavy nuclear clusters in simulations of core collapse supernovae: Impact on equation of state and weak interactions}",
      journal = {\prc},
         year = 2020,
        month = nov,
       volume = {102},
       number = {5},
          eid = {055807},
        pages = {055807},
          doi = {10.1103/PhysRevC.102.055807},
       adsurl = {https://ui.adsabs.harvard.edu/abs/2020PhRvC.102e5807F}
}

@ARTICLE{Hempel12,
   author = {{Hempel}, M. and {Fischer}, T. and {Schaffner-Bielich}, J. and 
	{Liebend{\"o}rfer}, M.},
    title = "{New Equations of State in Simulations of Core-collapse Supernovae}",
  journal = {\apj},
     year = 2012,
    month = mar,
   volume = 748,
      eid = {70},
    pages = {70},
      doi = {10.1088/0004-637X/748/1/70},
   adsurl = {http://cdsads.u-strasbg.fr/abs/2012ApJ...748...70H}
}

@ARTICLE{Steiner2013ApJ765L,
       author = {{Steiner}, Andrew W. and {Lattimer}, James M. and {Brown}, Edward F.},
        title = "{The Neutron Star Mass-Radius Relation and the Equation of State of Dense Matter}",
      journal = {Astrophys. J. Lett.},
         year = 2013,
        month = mar,
       volume = {765},
       number = {1},
          eid = {L5},
        pages = {L5},
          doi = {10.1088/2041-8205/765/1/L5},
       adsurl = {https://ui.adsabs.harvard.edu/abs/2013ApJ...765L...5S}
}

@ARTICLE{Fischer17,
       author = {{Fischer}, Tobias and {Bastian}, Niels-Uwe and {Blaschke}, David and {Cierniak}, Mateusz and {Hempel}, Matthias and {Kl{\"a}hn}, Thomas and {Mart{\'\i}nez-Pinedo}, Gabriel and {Newton}, William G. and {R{\"o}pke}, Gerd and {Typel}, Stefan},
        title = "{The State of Matter in Simulations of Core-Collapse supernovae{\textemdash}Reflections and Recent Developments}",
      journal = {Pub. Astron. Soc. Aust.},
         year = 2017,
        month = dec,
       volume = {34},
          eid = {e067},
        pages = {e067},
          doi = {10.1017/pasa.2017.63},
       adsurl = {https://ui.adsabs.harvard.edu/abs/2017PASA...34...67F}
}

@ARTICLE{Bauswein:2014PhRvD90,
       author = {{Bauswein}, A. and {Stergioulas}, N. and {Janka}, H.-T.},
        title = "{Revealing the high-density equation of state through binary neutron star mergers}",
      journal = {\prd},
         year = 2014,
        month = jul,
       volume = {90},
       number = {2},
          eid = {023002},
        pages = {023002},
          doi = {10.1103/PhysRevD.90.023002},
       adsurl = {https://ui.adsabs.harvard.edu/abs/2014PhRvD..90b3002B}
}

@ARTICLE{Bauswein19,
       author = {{Bauswein}, Andreas and {Bastian}, Niels-Uwe F. and {Blaschke}, David B. and {Chatziioannou}, Katerina and {Clark}, James A. and {Fischer}, Tobias and {Oertel}, Micaela},
        title = "{Identifying a First-Order Phase Transition in Neutron-Star Mergers through Gravitational Waves}",
      journal = {\prl},
         year = 2019,
        month = feb,
       volume = {122},
       number = {6},
          eid = {061102},
        pages = {061102},
          doi = {10.1103/PhysRevLett.122.061102},
       adsurl = {https://ui.adsabs.harvard.edu/abs/2019PhRvL.122f1102B}
}

@ARTICLE{Bauswein:2020PhRvL125,
       author = {{Bauswein}, Andreas and {Blacker}, Sebastian and {Vijayan}, Vimal and {Stergioulas}, Nikolaos and {Chatziioannou}, Katerina and {Clark}, James A. and {Bastian}, Niels-Uwe F. and {Blaschke}, David B. and {Cierniak}, Mateusz and {Fischer}, Tobias},
        title = "{Equation of State Constraints from the Threshold Binary Mass for Prompt Collapse of Neutron Star Mergers}",
      journal = {\prl},
         year = 2020,
        month = oct,
       volume = {125},
       number = {14},
          eid = {141103},
        pages = {141103},
          doi = {10.1103/PhysRevLett.125.141103},
       adsurl = {https://ui.adsabs.harvard.edu/abs/2020PhRvL.125n1103B}
}

@ARTICLE{Greco:2002PhRvC65_delta-meson,
       author = {{Liu}, B. and {Greco}, V. and {Baran}, V. and {Colonna}, M. and {di Toro}, M.},
        title = "{Asymmetric nuclear matter: The role of the isovector scalar channel}",
      journal = {\prc},
         year = 2002,
        month = apr,
       volume = {65},
       number = {4},
          eid = {045201},
        pages = {045201},
          doi = {10.1103/PhysRevC.65.045201},
       adsurl = {https://ui.adsabs.harvard.edu/abs/2002PhRvC..65d5201L}
}

@ARTICLE{kubis_1997,
       author = {{Kubis}, S. and {Kutschera}, M.},
        title = "{Nuclear matter in relativistic mean field theory with isovector scalar meson}",
      journal = {Phys. Lett. B},
         year = 1997,
        month = feb,
       volume = {399},
        pages = {191-195},
          doi = {10.1016/S0370-2693(97)00306-7},
       adsurl = {https://ui.adsabs.harvard.edu/abs/1997PhLB..399..191K}
}

@ARTICLE{bollig17,
       author = {{Bollig}, R. and {Janka}, H.-T. and {Lohs}, A. and {Mart{\'\i}nez-Pinedo}, G. and {Horowitz}, C.~J. and {Melson}, T.},
        title = "{Muon Creation in Supernova Matter Facilitates Neutrino-Driven Explosions}",
      journal = {\prl},
         year = 2017,
        month = dec,
       volume = {119},
       number = {24},
          eid = {242702},
        pages = {242702},
          doi = {10.1103/PhysRevLett.119.242702},
       adsurl = {https://ui.adsabs.harvard.edu/abs/2017PhRvL.119x2702B}
}

@ARTICLE{LSEOS,
   author = {{Lattimer}, J.~M. and {Swesty}, F.},
    title = "{A generalized equation of state for hot, dense matter}",
  journal = {Nucl. Phys. A},
     year = 1991,
    month = dec,
   volume = 535,
    pages = {331-376},
      doi = {10.1016/0375-9474(91)90452-C},
   adsurl = {http://cdsads.u-strasbg.fr/abs/1991NuPhA.535..331L}
}

@ARTICLE{li_2018,
       author = {{Li}, Bao-An and {Cai}, Bao-Jun and {Chen}, Lie-Wen and {Xu}, Jun},
        title = "{Nucleon effective masses in neutron-rich matter}",
      journal = {Prog. Part. Nucl. Phys.},
         year = 2018,
        month = mar,
       volume = {99},
        pages = {29-119},
          doi = {10.1016/j.ppnp.2018.01.001},
       adsurl = {https://ui.adsabs.harvard.edu/abs/2018PrPNP..99...29L}
}

@ARTICLE{harris_2025,
       author = {{Harris}, Steven P.},
        title = "{Symmetry energy expansion and the peak value of the bulk viscosity}",
      journal = {\prc},
         year = 2025,
        month = sep,
       volume = {112},
       number = {3},
          eid = {035806},
        pages = {035806},
          doi = {10.1103/r28j-pvpm},
       adsurl = {https://ui.adsabs.harvard.edu/abs/2025PhRvC.112c5806H}
}

@ARTICLE{Greco:2003PhRvC67_delta-meson,
       author = {{Greco}, V. and {Colonna}, M. and {di Toro}, M. and {Matera}, F.},
        title = "{Collective modes of asymmetric nuclear matter in quantum hadrodynamics}",
      journal = {\prc},
         year = 2003,
        month = jan,
       volume = {67},
       number = {1},
          eid = {015203},
        pages = {015203},
          doi = {10.1103/PhysRevC.67.015203},
       adsurl = {https://ui.adsabs.harvard.edu/abs/2003PhRvC..67a5203G}
}

@ARTICLE{Roca-Maza:2011PhRvC84_delta-meson,
       author = {{Roca-Maza}, X. and {Vi{\~n}as}, X. and {Centelles}, M. and {Ring}, P. and {Schuck}, P.},
        title = "{Relativistic mean-field interaction with density-dependent meson-nucleon vertices based on microscopical calculations}",
      journal = {\prc},
         year = 2011,
        month = nov,
       volume = {84},
       number = {5},
          eid = {054309},
        pages = {054309},
          doi = {10.1103/PhysRevC.84.054309},
       adsurl = {https://ui.adsabs.harvard.edu/abs/2011PhRvC..84e4309R}
}

@ARTICLE{Bruenn85,
   author = {{Bruenn}, S.~W.},
    title = "{Stellar core collapse - Numerical model and infall epoch}",
  journal = {Astrophys. J. Suppl. Ser.},
     year = 1985,
    month = aug,
   volume = 58,
    pages = {771-841},
      doi = {10.1086/191056},
   adsurl = {http://cdsads.u-strasbg.fr/abs/1985ApJS...58..771B}
}

@ARTICLE{2020EPJA...56..160T,
       author = {{Typel}, Stefan and {Alvear Terrero}, Diana},
        title = "{Parametrisations of relativistic energy density functionals with tensor couplings}",
      journal = {Eur. Phys. J. A},
         year = 2020,
        month = jun,
       volume = {56},
       number = {6},
          eid = {160},
        pages = {160},
          doi = {10.1140/epja/s10050-020-00172-2},
       adsurl = {https://ui.adsabs.harvard.edu/abs/2020EPJA...56..160T}
}

@ARTICLE{tews_2018,
       author = {{Tews}, I. and {Carlson}, J. and {Gandolfi}, S. and {Reddy}, S.},
        title = "{Constraining the Speed of Sound inside Neutron Stars with Chiral Effective Field Theory Interactions and Observations}",
      journal = {\apj},
         year = 2018,
        month = jun,
       volume = {860},
       number = {2},
          eid = {149},
        pages = {149},
          doi = {10.3847/1538-4357/aac267},
       adsurl = {https://ui.adsabs.harvard.edu/abs/2018ApJ...860..149T}
}

@article{tews_2025,
  title = {Neutron matter from local chiral effective field theory interactions at large cutoffs},
  author = {Tews, Ingo and Somasundaram, Rahul and Lonardoni, Diego and G\"ottling, Hannah and Seutin, Rodric and Carlson, Joseph and Gandolfi, Stefano and Hebeler, Kai and Schwenk, Achim},
  journal = {Phys. Rev. Res.},
  volume = {7},
  issue = {3},
  pages = {033024},
  numpages = {7},
  year = {2025},
  month = {Jul},
  publisher = {American Physical Society},
  doi = {10.1103/r314-6r62},
  url = {https://link.aps.org/doi/10.1103/r314-6r62}
}

@article{drischler_2019,
  title = {Chiral Interactions up to Next-to-Next-to-Next-to-Leading Order and Nuclear Saturation},
  author = {Drischler, C. and Hebeler, K. and Schwenk, A.},
  journal = {Phys. Rev. Lett.},
  volume = {122},
  issue = {4},
  pages = {042501},
  numpages = {6},
  year = {2019},
  month = {Jan},
  publisher = {American Physical Society},
  doi = {10.1103/PhysRevLett.122.042501},
  url = {https://link.aps.org/doi/10.1103/PhysRevLett.122.042501}
}

@ARTICLE{Wang:2023PhRvC108_delta-meson,
       author = {{Wang}, Sibo and {Tong}, Hui and {Zhao}, Qiang and {Wang}, Chencan and {Ring}, Peter and {Meng}, Jie},
        title = "{Neutron-proton effective mass splitting in neutron-rich matter}",
      journal = {\prc},
         year = 2023,
        month = sep,
       volume = {108},
       number = {3},
          eid = {L031303},
        pages = {L031303},
          doi = {10.1103/PhysRevC.108.L031303},
       adsurl = {https://ui.adsabs.harvard.edu/abs/2023PhRvC.108c1303W}
}

@ARTICLE{Lenske:2001PhRvC64_delta-meson,
       author = {{Hofmann}, F. and {Keil}, C.~M. and {Lenske}, H.},
        title = "{Density dependent hadron field theory for asymmetric nuclear matter and exotic nuclei}",
      journal = {\prc},
         year = 2001,
        month = sep,
       volume = {64},
       number = {3},
          eid = {034314},
        pages = {034314},
          doi = {10.1103/PhysRevC.64.034314},
       adsurl = {https://ui.adsabs.harvard.edu/abs/2001PhRvC..64c4314H}
}

@article{li_2026,
    title = {Role of the δ meson in softening the symmetry energy within the DDRHF model},
    journal = {Phys. Lett. B},
    volume = {880},
    pages = {140860},
    year = {2026},
    issn = {0370-2693},
    doi = {https://doi.org/10.1016/j.physletb.2026.140860},
    url = {https://www.sciencedirect.com/science/article/pii/S0370269326007112},
    author = {Qirui Li and Jinniu Hu and Ying Zhang and Hong Shen},
}

@ARTICLE{Fischer12,
       author = {{Fischer}, T. and {Mart{\'\i}nez-Pinedo}, G. and {Hempel}, M. and {Liebend{\"o}rfer}, M.},
        title = "{Neutrino spectra evolution during protoneutron star deleptonization}",
      journal = {\prd},
         year = 2012,
        month = apr,
       volume = {85},
       number = {8},
          eid = {083003},
        pages = {083003},
          doi = {10.1103/PhysRevD.85.083003},
       adsurl = {https://ui.adsabs.harvard.edu/abs/2012PhRvD..85h3003F}
}

@ARTICLE{Reddy99,
       author = {{Reddy}, Sanjay and {Prakash}, Madappa and {Lattimer}, James M. and
         {Pons}, Jose A.},
        title = "{Effects of strong and electromagnetic correlations on neutrino interactions in dense matter}",
      journal = {\prc},
         year = 1999,
        month = may,
       volume = {59},
       number = {5},
        pages = {2888-2918},
          doi = {10.1103/PhysRevC.59.2888},
       adsurl = {https://ui.adsabs.harvard.edu/abs/1999PhRvC..59.2888R}
}

@ARTICLE{burrows_1999,
       author = {{Burrows}, Adam and {Sawyer}, R.~F.},
        title = "{Many-body corrections to charged-current neutrino absorption rates in nuclear matter}",
      journal = {\prc},
         year = 1999,
        month = jan,
       volume = {59},
       number = {1},
        pages = {510-514},
          doi = {10.1103/PhysRevC.59.510},
       adsurl = {https://ui.adsabs.harvard.edu/abs/1999PhRvC..59..510B}
}

@article{matsui_1982,
        title = {The pion propagator in relativistic quantum field theories of the nuclear many-body problem},
        journal = {Ann. Phys. (N.Y.)},
        volume = {144},
        number = {1},
        pages = {107-167},
        year = {1982},
        issn = {0003-4916},
        doi = {https://doi.org/10.1016/0003-4916(82)90106-3},
        url = {https://www.sciencedirect.com/science/article/pii/0003491682901063},
        author = {T Matsui and Brian D Serot},
}

@article{boguta_1977,
        title = {Relativistic calculation of nuclear matter and the nuclear surface},
        journal = {\nphysa},
        volume = {292},
        number = {3},
        pages = {413-428},
        year = {1977},
        issn = {0375-9474},
        doi = {https://doi.org/10.1016/0375-9474(77)90626-1},
        url = {https://www.sciencedirect.com/science/article/pii/0375947477906261},
        author = {J. Boguta and A.R. Bodmer},
}

@article{bodmer_1991,
        title = {Relativistic mean field theory of nuclei with a vector meson self-interaction},
        journal = {\nphysa},
        volume = {526},
        number = {3},
        pages = {703-721},
        year = {1991},
        issn = {0375-9474},
        doi = {https://doi.org/10.1016/0375-9474(91)90439-D},
        url = {https://www.sciencedirect.com/science/article/pii/037594749190439D},
        author = {A.R. Bodmer},
}

@article{sugahara_1994,
        title = {Relativistic mean-field theory for unstable nuclei with non-linear σ and ω terms},
        journal = {\nphysa},
        volume = {579},
        number = {3},
        pages = {557-572},
        year = {1994},
        issn = {0375-9474},
        doi = {https://doi.org/10.1016/0375-9474(94)90923-7},
        url = {https://www.sciencedirect.com/science/article/pii/0375947494909237},
        author = {Y. Sugahara and H. Toki},
}

@article{reinhard_1989,
        doi = {10.1088/0034-4885/52/4/002},
        url = {https://dx.doi.org/10.1088/0034-4885/52/4/002},
        year = {1989},
        month = {apr},
        publisher = {},
        volume = {52},
        number = {4},
        pages = {439},
        author = {P -G Reinhard},
        title = {The relativistic mean-field description of nuclei and nuclear dynamics},
        journal = {\reppphys},
}

@book{bjorken_1964,
        author = "Bjorken, James D. and Drell, Sidney D.",
        title = "Relativistic quantum mechanics",
        series = "International Series in Pure and Applied Physics",
        publisher = "McGraw-Hill",
        address = "New York",
        year = "1964",
        editor = {Schiff, Leonard I.},
}

@article{sun_2008,
  title = {Neutron star properties in density-dependent relativistic Hartree-Fock theory},
  author = {Sun, Bao Yuan and Long, Wen Hui and Meng, Jie and Lombardo, U.},
  journal = {Phys. Rev. C},
  volume = {78},
  issue = {6},
  pages = {065805},
  numpages = {14},
  year = {2008},
  month = {Dec},
  publisher = {American Physical Society},
  doi = {10.1103/PhysRevC.78.065805},
  url = {https://link.aps.org/doi/10.1103/PhysRevC.78.065805}
}

@ARTICLE{soutome_1990,
       author = {{Soutome}, Kouichi and {Tomoyuki}, Maruyama and {Koichi}, Saito},
        title = "{Relativistic {\ensuremath{\sigma}}-{\ensuremath{\omega}} model at finite temperature in thermo field dynamics}",
      journal = {\nphysa},
         year = 1990,
        month = feb,
       volume = {507},
       number = {3},
        pages = {731-760},
          doi = {10.1016/0375-9474(90)90179-P},
       adsurl = {https://ui.adsabs.harvard.edu/abs/1990NuPhA.507..731S}
}

@article{yang_2019,
  title = {Critical parameters of the liquid-gas phase transition in thermal symmetric and asymmetric nuclear matter},
  author = {Yang, Shen and Zhang, Bo Nan and Sun, Bao Yuan},
  journal = {Phys. Rev. C},
  volume = {100},
  issue = {5},
  pages = {054314},
  numpages = {12},
  year = {2019},
  month = {Nov},
  publisher = {American Physical Society},
  doi = {10.1103/PhysRevC.100.054314},
  url = {https://link.aps.org/doi/10.1103/PhysRevC.100.054314}
}

@article{yang_2021,
  title = {Liquid-gas phase transition of thermal nuclear matter and the in-medium balance between nuclear attraction and repulsion},
  author = {Yang, Shen and Sun, Xiang Dong and Geng, Jing and Sun, Bao Yuan and Long, Wen Hui},
  journal = {Phys. Rev. C},
  volume = {103},
  issue = {1},
  pages = {014304},
  numpages = {8},
  year = {2021},
  month = {Jan},
  publisher = {American Physical Society},
  doi = {10.1103/PhysRevC.103.014304},
  url = {https://link.aps.org/doi/10.1103/PhysRevC.103.014304}
}

\end{document}